\documentclass[11pt, a4paper]{article}
\usepackage{fancyhdr}
\usepackage[latin1]{inputenc}
\usepackage{graphicx,epsfig,color}
\usepackage{wrapfig,rotating}
\usepackage{amssymb,amsmath,array}
\usepackage{epstopdf}
\usepackage{lscape}
\usepackage{url}
\usepackage[]{lineno}
\begin{document}
%                     FEYNMAN(34).TEX 
%  CALLING ROUTINE FOR DRAWING FEYNMAN DIAGRAMS IN LATEX.
%  DOCUMENTATION IN "FEYNMAN - A LaTeX Routine for Generating Feynman Diagrams"
%  Cavendish-HEP 88/11  (Cavendish Labs, Cambridge, UK).
%  See also: Levine, M.J.S., A LaTeX Graphics Routine for Drawing Feynman
%  Diagrams, Cavendish - HEP 89/4.
%  USES THE FOLLOWING TEX FILES:
%   GLUONSETUP(31), PHOTONSETUP(28), FERMIONSETUP(7), SCALARSETUP(9)
%   VERTEX(25), GLUONLINKS, LOOPS(1) 
%
%   THIS PROGRAM PACKAGE NOT TO BE ALTERED WITHOUT THE EXPRESS WRITTEN
%   PERMISSION OF THE AUTHOR.
%
%**************************************************************************
%
%                           SAMPLE USAGE
%  
%  \documentstyle[12pt]{article}
%  \begin{document}
%  \input feynman
%  \textheight 800pt \textwidth 450pt
%  \begin{picture}(10000,18000)
%  \drawline\gluon[\S\REG](0,16000)[8]
%  \drawline\fermion[\SW\REG](\gluonbackx,\gluonbacky)[2000]
%  \drawline\fermion[\SE\REG](\gluonbackx,\gluonbacky)[2000]
%  \end{picture}
%  \end{document}
%
%**************************************************************************
%
\message{FEYNMAN:  For generating Feynman Diagrams in LaTex}
\message{Mark 1.0 Last Altered by MJSL 2/89}
\textheight 650pt \textwidth 400pt  % Page size set.
\setlength{\unitlength}{0.01pt}
\gdef\Feynmanlength{\setlength{\unitlength}{0.01pt}}  % Say \Feynmanlength
\gdef\unlock{\catcode`\@=11}
%  Allows use of "@" in macro names, like PLAIN.TEX does.
\gdef\lock{\catcode`\@=12}%  Change @'s back to their normal category code.
\global\newcount\LINETYPE                     
\global\newcount\LINEDIRECTION
\global\newcount\LINECONFIGURATION
\newcommand{\LTYPE}{\LINETYPE}
\newcommand{\LDIR}{\LINEDIRECTION}
\newcommand{\LCONFIG}{\LINECONFIGURATION}
%DEFAULTS:  Horizontal fermion.   
\global\LINETYPE=1  \global\LINEDIRECTION=0  \global\LINECONFIGURATION=0
%  The parametric code names.  Don't change these.
\global\newcount\fermion    \fermion=1
\global\newcount\scalar     \scalar=2
\global\newcount\photon     \photon=3
\global\newcount\gluon      \gluon=4
\global\newcount\especial   \especial=5
\gdef\N{0}  \gdef\NE{1}  \gdef\E{2}   \gdef\SE{3}
\gdef\S{4}  \gdef\SW{5}  \gdef\W{6}   \gdef\NW{7}
\global\newcount\REG            \global\REG=0
\global\newcount\FLIPPED        \global\FLIPPED=1
\global\newcount\CURLY          \global\CURLY=2
\global\newcount\FLIPPEDCURLY   \global\FLIPPEDCURLY=3
\global\newcount\FLAT           \global\FLAT=4
\global\newcount\FLIPPEDFLAT    \global\FLIPPEDFLAT=5
\global\newcount\CENTRAL        \global\CENTRAL=6
\global\newcount\FLIPPEDCENTRAL \global\FLIPPEDCENTRAL=7
\gdef\LONGPHOTON{6}             \gdef\FLIPPEDLONG{7}
\global\newcount\SQUASHEDGLUON  \global\SQUASHEDGLUON=8
\gdef\SQUASHED{\SQUASHEDGLUON}
%\global\newcount\FLIPPEDSQUASHEDGLUON  \FLIPPEDSQUASHEDGLUON=9
%
%%%%%%%%%%%%%%%%%%%%%%%%%%%%%%%%%%%%%%%%%%%%%%%%%%%%%%%%%%%%%%%%%%%%%%%%%%%%%%%
% SOME COUNTERS AND DEFINITIONS FOR POSITIONS AND LENGTHS OF LINES & FEATURES %
%%%%%%%%%%%%%%%%%%%%%%%%%%%%%%%%%%%%%%%%%%%%%%%%%%%%%%%%%%%%%%%%%%%%%%%%%%%%%%%
\newcount\adjx \adjx=0
\newcount\adjy \adjy=0
\global\newdimen\BIGPHOTONS     \BIGPHOTONS=0pt  %  DEFAULT:  10 & 11-PT PHOTONS
\gdef\bigphotons{\global\BIGPHOTONS=12pt}%FOR 12-PT DOCS DRAWING E-W PHOTONS.
\global\newdimen\THICKPHOTONS     \THICKPHOTONS=0pt  %  FOR E-W PHOTONS 
\global\newdimen\THICKPHOTONSWITCH    \THICKPHOTONSWITCH=0pt
\gdef\THICKPHOTONTEST{
\THICKPHOTONSWITCH=0pt
\ifdim\THICKPHOTONS=0pt \relax
  \else \ifnum\LTYPE=3
           \ifnum\LDIR=2 \THICKPHOTONSWITCH=1pt \fi % THICK \E PHOTON
           \ifnum\LDIR=6 \THICKPHOTONSWITCH=1pt \fi % THICK \W PHOTON
        \fi
\fi
}  % end of THICKPHOTONTEST
\gdef\THICKLINES{\thicklines  \THICKPHOTONS=1pt}
\gdef\THINLINES{\thinlines  \THICKPHOTONS=0pt}
\global\newcount\phantomswitch   \global\phantomswitch=0
\global\newcount\stemlength   \global\stemlength=275   % Default STEM length.
\global\newcount\absstemlength        % A copy of STEM length.
\global\newcount\stemlengthx          % FOR STEMS on particle lines
\global\newcount\stemlengthy          % FOR STEMS on particle lines
\newdimen\FRONTSTEM  \FRONTSTEM=0pt   % FOR STEMS
\newdimen\BACKSTEM   \BACKSTEM=0pt    % FOR STEMS
\newdimen\EITHERSTEM \EITHERSTEM=0pt  % FOR STEMS
\gdef\frontstemmed{\FRONTSTEM=1pt}            % FOR STEMS
\gdef\backstemmed{\BACKSTEM=1pt}              % FOR STEMS
\gdef\stemmed{\FRONTSTEM=1pt  \BACKSTEM=1pt}    % FOR STEMS
\global\newcount\arrowlength                % FOR ARROWS
\global\newdimen\ATTIP   \global\ATTIP=0pt  % FOR ARROWS
\global\newdimen\ATBASE  \global\ATBASE=1pt % FOR ARROWS
\global\newcount\unitboxnumber  % SHOWS THE NUMBER OF `UNIT BOXES' IN LINE
\global\newcount\unitboxnumberpo  % One more than \unitboxnumber (in GLUONSETUP)
\global\newcount\particlelengthx  % THE X-LENGTH OF THE PARTICLE LINE
\gdef\plengthx{\particlelengthx}
\global\newcount\particlelengthy  % THE Y-LENGTH OF THE PARTICLE LINE
\gdef\plengthy{\particlelengthy}  
\global\newcount\boxlengthx  % THE X-LENGTH OF THE BOX:  abs(plengthx) usually
\global\newcount\boxlengthy  % THE y-LENGTH OF THE box:  abs(plengthy) usually
\global\newcount\particleadjustx  % Replaces \gluonadjustx, \scalaradjustx etc.
\global\newcount\particleadjusty  % Replaces \gluonadjusty, \scalaradjusty etc.
\global\newcount\particlelength   % The LENGTH of a particle line BOX (x)
\global\newcount\particlefrontx
\gdef\pfrontx{\particlefrontx}
\global\newcount\PFRONTx
\global\newcount\particlefronty
\gdef\pfronty{\particlefronty}
\global\newcount\PFRONTy
\global\newcount\particlebackx
\gdef\pbackx{\particlebackx}
\global\newcount\particlebacky
\gdef\pbacky{\particlebacky}
\global\newcount\particlemidx
\gdef\pmidx{\particlemidx}
\global\newcount\particlemidy
\gdef\pmidy{\particlemidy}
% SOME SPECIAL DEFS FOR \SCALARs:
\global\newcount\seglength  \global\newcount\gaplength
\global\gaplength=850  %default
\global\seglength=1416  % Length of each seg not including `ends' for attachment
% Now some storage locations for the user:
\global\newcount\Xone    \global\newcount\Yone    % user co-ords (\Xone,\Yone)
\global\newcount\Xtwo    \global\newcount\Ytwo    % user co-ords (\Xtwo,\Ytwo)
\global\newcount\Xthree  \global\newcount\Ythree  % user's (\Xthree,\Ythree)
\global\newcount\Xfour   \global\newcount\Yfour   % user co-ords (\Xfour,\Yfour)
\global\newcount\Xfive   \global\newcount\Yfive   % user co-ords (\Xfive,\Yfive)
\global\newcount\Xsix    \global\newcount\Ysix    % user co-ords (\Xsix,\Ysix)
\global\newcount\Xseven  \global\newcount\Yseven  % user's (\Xseven,\Yseven)
\global\newcount\Xeight  \global\newcount\Yeight  % user's (\Xeight,\Yeight)
%
%  SOME COUNTERS IDENTIFYING VARIOUS LINE PORTIONS AND DIMENSIONS:
%
\newsavebox{\lastline}  %  Default name for an unnamed particle line.
\global\newcount\numlineparts   % Num of pieces each unitbox of the line needs
\global\newcount\upperlineadjx  \upperlineadjx=0  %Default
\global\newcount\upperlineadjy  \upperlineadjy=0  %Default
\global\newcount\lowerlineadjx  \lowerlineadjx=0  %Default
\global\newcount\lowerlineadjy  \lowerlineadjy=0  %Default
\global\newcount\thirdlineadjx  \thirdlineadjx=0  %Default
\global\newcount\thirdlineadjy  \thirdlineadjy=0  %Default
\global\newcount\fourthlineadjx \fourthlineadjx=0  %Default
\global\newcount\fourthlineadjy \fourthlineadjy=0  %Default
\global\newcount\unitboxwidth   \unitboxwidth=1000%Default
\global\newcount\unitboxheight  \unitboxheight=0  %Default
\global\newcount\numupperunits  \numupperunits=8  %Default
\global\newcount\numlowerunits  \numlowerunits=8  %Default
\global\newcount\numthirdunits  \numthirdunits=8  %Default
\global\newcount\numfourthunits \numfourthunits=8  %Default
%  Some counters.  =0 until a line-type is drawn. Then=1.
\global\newcount\fermioncount   \global\fermioncount=0    
\global\newcount\scalarcount    \global\scalarcount=0    
\global\newcount\photoncount    \global\photoncount=0    
\global\newcount\gluoncount     \global\gluoncount=0    
\global\newcount\especialcount  \global\especialcount=0    
\global\newcount\vertexcount    \global\vertexcount=-1
%
%%%%%%%%%%%%%%%%%%%%%%%%%%%%%%%%%%%%%%%%%%%%%%%%%%%%%%%%%%%%%%
%     AUXILIARY ROUTINES FOR SETTING PARTICLE DIRECTIONS     %
%%%%%%%%%%%%%%%%%%%%%%%%%%%%%%%%%%%%%%%%%%%%%%%%%%%%%%%%%%%%%%
\global\newcount\XDIR
\global\newcount\YDIR
\gdef\SETDIR{  % SETS THE DIRECTIONS
\ifcase\LDIR 
     \global\XDIR=0  \global\YDIR=1   %\N  case.
\or  \global\XDIR=1  \global\YDIR=1   %\NE case.
\or  \global\XDIR=1  \global\YDIR=0   %\E  case.
\or  \global\XDIR=1  \global\YDIR=-1  %\SE case.
\or  \global\XDIR=0  \global\YDIR=-1  %\S  case.
\or  \global\XDIR=-1 \global\YDIR=-1  %\SW case.
\or  \global\XDIR=-1 \global\YDIR=0   %\W  case.
\or  \global\XDIR=-1 \global\YDIR=1   %\NW case.
\else\DIRECTERROR 
\fi}  % END OF \SETDIR
\gdef\moduloeight#1{
\ifnum#1>7 \global\advance #1 by -8 
\relax
\moduloeight#1 
\relax
\else \relax  
\fi}
\gdef\multroothalf#1{\global\multiply #1 by 7071 \global\divide #1 by 10000}
\gdef\negate#1{\global\multiply #1 by -1}
\gdef\double#1{\global\multiply #1 by 2}
\gdef\slanttest(#1,#2){ 
\ifodd\LDIR
\multiply #1 by 7071  \divide #1 by 10000
\multiply #2 by 7071  \divide #2 by 10000
\fi
}
\gdef\gslanttest(#1,#2){
\ifodd\LDIR
\multroothalf#1
\multroothalf#2
\fi
}
%
%%%%%%%%%%%%%%%%%%%%%%%%%%%%%%%%%%%%%%%%%%%%%%%%%%%%%%%%%%%%%%
% AUXILIARY ROUTINES FOR SETTING PARTICLE LENGTHS & POSTIONS %
%%%%%%%%%%%%%%%%%%%%%%%%%%%%%%%%%%%%%%%%%%%%%%%%%%%%%%%%%%%%%%
%
\gdef\setplength{ % calcs length of particle line
\global\particlelengthx=\unitboxwidth
\global\particlelengthy=\unitboxheight
\global\multiply \particlelengthx by \unitboxnumber
\global\multiply \particlelengthy by \unitboxnumber
\global\advance \particlelengthx by \particleadjustx
\global\advance \particlelengthy by \particleadjusty
}
\gdef\boxlengthdefault{  % DEFAULT FOR BOX SIZES IN \drawas
\global\boxlengthx=\plengthx
\global\boxlengthy=\plengthy
\ifnum\plengthx<0 \global\multiply\boxlengthx by -1 \fi
\ifnum\plengthy<0 \global\multiply\boxlengthy by -1 \fi
}
\gdef\rearcoords{  %  CALCULATES THE CO-ORDINATES OF THE BACK OF PARTICLE LINE
\global\particlebacky=\particlefronty 
\global\particlebackx=\particlefrontx 
\global\advance \particlebackx by \particlelengthx
\global\advance \particlebacky by \particlelengthy
}
\gdef\midcoords{  %  CALCULATES THE CO-ORDINATES OF THE MID OF PARTICLE LINE
\global\particlemidy=\particlefronty
\global\particlemidx=\particlefrontx
\global\stemlengthx=\particlelengthx  % Convenient variables not being used
\global\stemlengthy=\particlelengthy  
\global\divide\stemlengthx by 2
\global\divide\stemlengthy by 2
\global\advance \particlemidx by \stemlengthx
\global\advance \particlemidy by \stemlengthy
}
\gdef\setparticle{\setplength\rearcoords\midcoords\boxlengthdefault}  %sets line
\gdef\setcoords(#1,#2,#3)(#4,#5,#6)[#7,#8]{  
% Sets co-ords of first 3 line-parts of a line and the unitbox height and width
% Used by photons and gluons.
\global\upperlineadjx=#1
\global\lowerlineadjx=#2
\global\thirdlineadjx=#3
\global\upperlineadjy=#4
\global\lowerlineadjy=#5
\global\thirdlineadjy=#6
\global\unitboxwidth=#7
\global\unitboxheight=#8
}
%
%%%%%%%%%%%%%%%%%%%%%%%%%%%%%%%%%%%%%%%%%%%%%%%%%%%%%%%%%%%%%%%%%%%%%%%%%%%%%%%
%                                                                             %
%                     ROUTINES FOR DRAWING LINES                              %
%                                                                             %
%%%%%%%%%%%%%%%%%%%%%%%%%%%%%%%%%%%%%%%%%%%%%%%%%%%%%%%%%%%%%%%%%%%%%%%%%%%%%%%
%
% **************   ROUTINE FOR DRAWING STORED LINES AND PICTURES   *************
%
\gdef\drawoldpic#1(#2,#3){  % DRAWS PRE-SAVED PICTURE
\global\particlefrontx=#2
\global\particlefronty=#3
\rearcoords  
\midcoords
\put(#2,#3){\usebox{#1}}
}
\gdef\drawsavedline`#1' as #2[#3#4](#5,#6)[#7]{
\global\LINETYPE=#2
\global\LINEDIRECTION=#3
\global\LINECONFIGURATION=#4
\global\particlefrontx=#5
\global\particlefronty=#6
\global\unitboxnumber=#7  
% Formerly called \numhalfwiggles,\numdashes, \numloops, \fermionlength
% #1 is saved linename;   #2 is \LINETYPE;    #3 is \LINEDIRECTION
% #4 is \LINECONFIGURATION (#5,#6)=(x,y) co-ords;  #7 is linelength (eg#wiggles)
\selectcase
\rearcoords% moved from before selectcase.
\midcoords
\ifnum\phantomswitch=0 \drawas{#1}\fi
% if \phantomswitch=1 then just set the line up and don't draw it.
}

\gdef\startphantom{\phantomswitch=1} % BEGIN PHANTOM MODE.
\gdef\stopphantom{\phantomswitch=0}  % END PHANTOM MODE.
% USE AS: 
% \startphantom...\drawline\gluon[...]...\drawvertex\photon...\stopphantom

\gdef\drawas#1{
\global\savebox{#1}(\boxlengthx,\boxlengthy){
\setlength{\unitlength}{0.01pt}
\begin{picture}(\boxlengthx,\boxlengthy)
\multiput(\upperlineadjx,\upperlineadjy)(\unitboxwidth,\unitboxheight)
{\numupperunits}{\upperunitbox}
\ifnum\numlineparts > 1  %  If the line needs 2 parts per unit or more
\multiput(\lowerlineadjx,\lowerlineadjy)(\unitboxwidth,\unitboxheight)
{\numlowerunits}{\lowerunitbox}  
\fi
\ifnum\numlineparts > 2  %  If the line needs 3 parts per unit or more
\multiput(\thirdlineadjx,\thirdlineadjy)(\unitboxwidth,\unitboxheight)
{\numthirdunits}{\thirdunitbox}  
\fi
\ifnum\numlineparts > 3  %  If the line needs 4 parts per unit or more
\multiput(\fourthlineadjx,\fourthlineadjy)(\unitboxwidth,\unitboxheight)
{\numfourthunits}{\lowerunitbox}  
\fi
\end{picture} }
% CHECK STEMS
\global\PFRONTx=\pfrontx  \global\PFRONTy=\pfronty   %save this value
\SETFRONTSTEM
% Now take into account the possibility of THICK E-W photons (drawn twice)
\THICKPHOTONTEST
\ifdim\THICKPHOTONSWITCH=1pt\global\advance\PFRONTy by 20  \fi
\put(\PFRONTx,\PFRONTy) {\usebox{#1}}   %\pfrontX,Y=\particlefrontx,y
%\put(\particlefrontx,\particlefronty) {\usebox{#1}}
\ifdim\THICKPHOTONSWITCH=1pt
\global\advance\PFRONTy by -40
\put(\PFRONTx,\PFRONTy) {\usebox{#1}}   % The second \E or \W photon ->thicker
\global\advance \PFRONTy by 20  %re-adjust:  advanced by -20 in total above.
\fi  %End of \ifdim\THICKPHOTONSWITCH=1
\SETBACKSTEM
\seglength=1416   \gaplength=850   % Re-set \SCALR defaults.
}
%
% *********   ROUTINES FOR STORING LINES  *******
%

\gdef\drawandsaveline`#1' as #2[#3#4](#5,#6)[#7]{
% #1 is saved linename;   #2 is \LINETYPE;    #3 is \LINEDIRECTION
% #4 is \LINECONFIGURATION (#5,#6)=(x,y) co-ords;  #7 is linelength (eg#wiggles)
\global\newsavebox{#1}
\drawsavedline`#1' as #2[#3#4](#5,#6)[#7]
}

\gdef\drawline#1[#2#3](#4,#5)[#6]{   % Draw line but don't name it.
\drawsavedline`\lastline' as #1[#2#3](#4,#5)[#6]}

\gdef\saveas#1{  %  For saving a line after the fact.
\global\newsavebox#1
\drawas#1}
%
%
%%%%%%%%%%%%%%%%%%%%%%%%%%%%%%%%%%%%%%%%%%%%%%%%%%%%%%%%%%%%%%%%%%%%%%%%
%                                                                      %
%                           C A S E S                                  %
%                           ---------                                  % 
%                                                                      %       
%%%%%%%%%%%%%%%%%%%%%%%%%%%%%%%%%%%%%%%%%%%%%%%%%%%%%%%%%%%%%%%%%%%%%%%%
%
% ERROR MESSAGES FOR INCORRECT CASE SPECIFICATION:
\gdef\TYPEERROR{\message{*** ERROR IN PARTICLE TYPE SELECTION ***}
\message{+++ Try with line type \fermion,\scalar,\photon,\gluon 
(see manual) +++}\SETERR}
\gdef\DIRECTERROR{\SETERR\message{*** ERROR IN PARTICLE DIRECTION SELECTION ***}
\message{+++ Try again with direction N, NE, E, SE  etc. or see manual +++}}
\gdef\UNIMPERROR{\message{*** ERROR IN PARTICLE OPTIONS SELECTION ***}
\message{
+++ The requested options combination has not yet been implemented +++}\SETERR}
\gdef\SETERR{\gdef\upperunitbox{{\tiny Error}}  % PRINTS `error' in diagram.
\gdef\lowerunitbox{\relax}
\gdef\thirdunitbox{\relax}
}
\gdef\neglengthcheck{\ifnum\unitboxnumber < 1 
\message{   *** ERROR:  PARTICLE OF NEGATIVE OR ZERO LENGTH REQUESTED. ***   }
\message{   ***         TAKING ABSOLUTE VALUE. ***   }\negate\unitboxnumber \fi}
%%%%%%%%%%%%%%%%%%%%%%%%%%%%%%%%%%%%%%%%%%%%%%%%%%%%%%%%%%%%%%%%%%%%%%%%%%%%%%%%
\gdef\selectcase{  
\neglengthcheck   %  check for particles of negative length.
% select PARTICLE alignment:
\SETDIR  
%  Select particle type
\ifcase\LINETYPE
\TYPEERROR  % \LINETYPE=0 case.
\or \selectfermion  % \LINETYPE=1 case.
\or \selectscalar   % \LINETYPE=2 case.
\or \selectphoton   % \LINETYPE=3 case.
\or \selectgluon    % \LINETYPE=4 case.
\or \selectespecial % \LINETYPE=5 case.
\else \TYPEERROR \fi  }
%%%%%%%%%%% (1) FERMIONS %%%%%%%%%%% 
\gdef\selectfermion{
% Input fermion-setup stuff ONLY IF HAVE NOT DONE SO YET.
% This avoids having to process a fermion if none are drawn.
\ifnum\fermioncount=0 %                        FERMIONSETUP(7).TEX
%  CALLED BY FEYNMAN(34).TEX.
% USED FOR GENERATING FERMION LINES IN FEYNMAN DIAGRAMS IN LATEX.
\global\newcount\fermionlength  %  THE TOTAL FERMION LINE LENGTH.
\global\newcount\fermionlengthx
\global\newcount\fermionlengthy
\global\newcount\fermionfrontx  %}(x,y) co-ord of left of fermion
\global\newcount\fermionfronty  %}
\global\newcount\fermionbackx
\global\newcount\fermionbacky
%%%%%%%%%%%%%%%%%%%%%%%%%%%%%%%%%%%%%%%%%%%%%%%%%%%%%%%%%%%%%%%%%%%%%%%%%%%
\gdef\ALLfermion{  % READ IN FROM FEYNMAN \selectfermion
\global\fermionfrontx=\particlefrontx \global\fermionfronty=\particlefronty
% Error messages for overly-long lines.  See FEYNMAN for negative-lengths.
\ifnum\unitboxnumber > 50000
\message{   *** WARNING *** Fermion of length
\the\unitboxnumber\space requested ***   }
\ifnum\unitboxnumber > 80000
\message{   *** Reducing fermion length to 30000 (max 80000) ***   }
\global\unitboxnumber=30000 \fi \fi  % end of length error
\global\fermionlength=\unitboxnumber % The TOTAL line length
\global\particleadjustx=0   \global\particleadjusty=0 %Default
\global\numlineparts = 1    \global\numupperunits=1
\global\upperlineadjx=-200  \global\upperlineadjy=0
\global\fermionlengthx=\fermionlength    \global\fermionlengthy=\fermionlength
\gslanttest(\fermionlengthx,\fermionlengthy)  % See FEYNMAN22.TEX (FOR \XDIR).
\global\multiply\fermionlengthx by \XDIR  %  In keeping with photons and gluons.
\global\multiply\fermionlengthy by \YDIR  %  In keeping with photons and gluons.
\global\unitboxheight=\fermionlengthy   \global\unitboxwidth=\fermionlengthx   
\global\advance \fermionlengthx by \particleadjustx
\global\advance \fermionlengthy by \particleadjusty
\global\particlelengthx=\fermionlengthx
\global\particlelengthy=\fermionlengthy  
\boxlengthdefault    \rearcoords    \midcoords
\global\fermionbackx=\particlebackx     \global\fermionbacky=\particlebacky
\ifcase\LINECONFIGURATION  %\REG case
\ifnum\XDIR=0 
\gdef\upperunitbox{\line(\XDIR,\YDIR){\boxlengthy}} %\N or \S
\else
\gdef\upperunitbox{\line(\XDIR,\YDIR){\boxlengthx}}
\fi
\else \UNIMPERROR
\fi
}
 \fi   
%                  CONTAINS fermion DEFINITIONS.
\global\advance\fermioncount by 1  % Counts number of fermions drawn. 
\ALLfermion   
}
%%%%%%%%%%% (2) SCALARS %%%%%%%%%%% 
\gdef\selectscalar{
% Input scalar-setup stuff ONLY IF HAVE NOT DONE SO YET.
% This avoids having to process a scalar if none are drawn.
\ifnum\scalarcount=0 %                   SCALARSETUP(9).TEX
% CALLED BY FEYNMAN(34).
% USED FOR GENERATING SCALAR LINES IN  FEYNMAN DIAGRAMS IN LATEX.
\newcount\scalarlength
\newcount\scalarlengthx
\newcount\scalarlengthy
\newcount\scalarfrontx  %}(x,y) co-ord of left of scalar
\newcount\scalarfronty  %}
\newcount\scalarbackx
\newcount\scalarbacky
%%%%%%%%%%%%%%%%%%%%%%%%%%%%%%%%%%%%%%%%%%%%%%%%%%%%%%%%%%%%%%%%%%%%%%%%%%%%
\gdef\ALLscalar{
\global\scalarfrontx=\particlefrontx   % READ IN FROM FEYNMAN \selectscalar
\global\scalarfronty=\particlefronty   % READ IN FROM FEYNMAN \selectscalar
% \gaplength=850  \seglength=1416  % Default defined in FEYNMAN.TEX.
\numlineparts = 1      \numupperunits=\unitboxnumber
\ifcase\LINECONFIGURATION
\global\upperlineadjx=-200     \global\upperlineadjy=0 
\slanttest(\seglength,\gaplength)   %SEE FEYNMAN22.TEX.
\gdef\upperunitbox{\line(\XDIR,\YDIR){\seglength}}
\else \UNIMPERROR % etc.
\fi
\global\unitboxwidth=\seglength  \global\advance\unitboxwidth by \gaplength
\global\multiply \unitboxwidth by \XDIR
\global\unitboxheight=\seglength  \global\advance\unitboxheight by \gaplength
\global\multiply \unitboxheight by \YDIR
\global\particleadjustx=\gaplength \global\multiply\particleadjustx by \XDIR 
\global\particleadjusty=\gaplength \global\multiply\particleadjusty by \YDIR
\negate\particleadjustx   \negate\particleadjusty   % SUBTRACT from linelength
\setparticle  %SCALAR8 
\global\scalarlengthx=\particlelengthx  %SCALAR8 
\global\scalarlengthy=\particlelengthy  %SCALAR8 
% Warning message for overly-long lines.  See FEYNMAN for negative-lengths.
\ifnum\boxlengthx > 50000
\message{   *** WARNING *** Scalar of length in excess of 50000cp requested!}\fi
\ifnum\boxlengthy > 50000
\message{   *** WARNING *** Scalar of length in excess of 50000cp requested!}\fi
\global\scalarbackx=\pbackx      \global\scalarbacky=\pbacky   %SCALAR8 
}
 \fi   
%                 CONTAINS scalar DEFINITIONS.
\global\advance\scalarcount by 1  % Counts number of scalars drawn. 
\ALLscalar
}
%%%%%%%%%%% (3) PHOTONS %%%%%%%%%%% 
\gdef\selectphoton{   % RECURSIVELY RE-DEFINED IN PHOTONSETUP(23+).TEX.
% Input photon-setup stuff ONLY IF HAVE NOT DONE SO YET.
% This avoids having to process a photon if none are drawn.
\ifnum\photoncount=0 %                            PHOTONSETUP(28).TEX
% CALLED BY FEYNMAN(34).TEX.
% USED FOR GENERATING PHOTON LINES IN FEYNMAN DIAGRAMS IN LATEX.
\newcount\numwiggles    \newcount\numwigglespo
\global\newcount\photonlengthx
\global\newcount\photonlengthy
\global\newcount\photonfrontx  %}(x,y) co-ord of left of photon
\global\newcount\photonfronty  %}
\global\newcount\photonbackx
\global\newcount\photonbacky
\newcount\halfwigglelength
\global\font\Twelverom=cmr12
\global\font\Tenrom=cmr10
\gdef\Lbr{{\Twelverom(}}   \gdef\Rbr{{\Twelverom)}}
\gdef\SLbr{{\Tenrom(}}     \gdef\SRbr{{\Tenrom)}}
%  Want \smile,\frown to always be 12-point but won't work!
\gdef\Smile{{\large$\smile$}}  % Default for 10 and 11-point documents.
\gdef\Frown{{\large$\frown$}}  % Default for 10 and 11-point documents.
\ifdim\BIGPHOTONS>0pt  \gdef\Smile{$\smile$} \gdef\Frown{$\frown$} \fi
%  For use with 12-point documents only.  Invoked by saying \bigphotons.
%
%%%%%%%%%%%%%%%%%%%%%%%%%%%%%%%%%%%%%%%%%%%%%%%%%%%%%%%%%%%%%%%%%%%%%%%%%%%
\gdef\selectphoton{   % RECURSIVELY RE-DEFINED HERE.  Define in FEYNMAN.
\global\advance\photoncount by 1  % Counts number of photons drawn. 
\global\photonfrontx=\particlefrontx   % READ IN FROM FEYNMAN \selectphoton
\global\photonfronty=\particlefronty   % READ IN FROM FEYNMAN \selectphoton
% Error messages for overly-long lines.  See FEYNMAN for negative-lengths.
\ifnum\unitboxnumber > 50
\message{   *** WARNING *** Photon with 
\the\unitboxnumber\space half-wiggles requested ***   }
\ifnum\unitboxnumber > 150
\message{   *** Reducing photon length to 10 half-wiggles (max 150) ***   }
\ifnum\unitboxnumber > 1000
\message{   *** Probable Cause:  Photon selected instead of Fermion ***   }
\fi \global\unitboxnumber=10 \fi \fi  % end of length error
\numwiggles=\unitboxnumber
\divide\numwiggles by 2
\global\unitboxnumberpo=\numwiggles % here \unitboxnumberpo is an unused counter
\global\multiply \unitboxnumberpo by -1
\numwigglespo=\unitboxnumber
\advance\numwigglespo by \unitboxnumberpo %\numwigglespo is one greater than 
\global\numlineparts = 2  % DEFAULT                %\numwiggles in this case.
\global\numupperunits=\numwigglespo  % DEFAULT
\global\numlowerunits=\numwiggles  % DEFAULT
\particleadjustx=0  %DEFAULT
\particleadjusty=0  %DEFAULT
% select photon alignment:
\ifcase\LINEDIRECTION
     \Nphoton    %\LINEDIRECTION=0 (NORTH) CASE
\or  \NEphoton   % 1 case
\or  \Ephoton    % 2 case...horizontal photon.
\or  \SEphoton   % .
\or  \Sphoton    % .
\or  \SWphoton   % .
\or  \Wphoton    % .
\or  \NWphoton   % 7 case
\else\DIRECTERROR \fi
\setplength
\global\divide\plengthx by 2  \global\divide\plengthy by 2
\rearcoords  \boxlengthdefault   \midcoords
\global\photonbackx=\pbackx  %PHOTONSETUP26
\global\photonbacky=\pbacky  %PHOTONSETUP26
\global\photonlengthx=\plengthx  %PHOTONSETUP26
\global\photonlengthy=\plengthy  %PHOTONSETUP26
}
%%%%%%%%%%%%%%%%%%%%%%%%%%%%%%%%%%%%%%%%%%%%%%%%%%%%%%%%%%%%%%%%%%%%%%%%%%%
\gdef\SETUNITBOX(#1)[#2][#3]{ % For slanted photons only.
\gdef\upperunitbox{\oval(#1,#1)[#2]}
\gdef\lowerunitbox{\oval(#1,#1)[#3]}
}
%%%%%%%%%%%%%%%%%%%%%%%%%%%%%%%%%%%%%%%%%%%%%%%%%%%%%%%%%%%%%%%%%%%%%%%%%%%
\gdef\Nphoton{  % VERTICAL PHOTONS
\ifcase\LINECONFIGURATION  %\REG case
\setcoords(-490,-250,0)(260,1250,0)[0,2000]
\gdef\upperunitbox{\SLbr}   \gdef\lowerunitbox{\SRbr}
\particleadjusty=10
\or % \FLIPPED case
\setcoords(-271,-501,0)(250,1250,0)[0,2000]   
\gdef\upperunitbox{\SRbr}   \gdef\lowerunitbox{\SLbr}
\or %\CURLY case (a bit shorter).
\particleadjusty=0
\setcoords(-501,-351,0)(300,1400,0)[0,2200]
\gdef\upperunitbox{\Lbr}   \gdef\lowerunitbox{\Rbr}
\or %\FLIPPEDCURLY case.
\setcoords(-353,-499,0)(300,1400,0)[0,2200]
\gdef\upperunitbox{\Rbr}   \gdef\lowerunitbox{\Lbr}
\or % \FLAT case.  Flatter and shorter than \CURLY.
\setcoords(-481,-371,0)(280,1300,0)[0,2000]
\gdef\upperunitbox{\Lbr}   \gdef\lowerunitbox{\Rbr}
\particleadjusty=150
\ifnum\numwiggles=\number\numwigglespo \particleadjustx=-50 \fi
\or %\FLIPPEDFLAT case.  \LINECONFIGURATION=5.
\setcoords(-321,-391,0)(280,1300,0)[0,2000]
\gdef\upperunitbox{\Rbr}   \gdef\lowerunitbox{\Lbr}
\particleadjusty=150
\ifnum\numwiggles=\number\numwigglespo \particleadjustx=80 \fi
\or % \LONGPHOTON
\setcoords(-490,-260,0)(300,1500,0)[0,2400]
\gdef\upperunitbox{\Lbr}   \gdef\lowerunitbox{\Rbr}
\or % \FLIPPEDLONGPHOTON
\setcoords(-301,-531,0)(300,1500,0)[0,2400]
\gdef\upperunitbox{\Rbr}   \gdef\lowerunitbox{\Lbr}
\else \UNIMPERROR
\fi
}
%%%%%%%%%%%%%%%%%%%%%%%%%%%%%%%%%%%%%%%%%%%%%%%%%%%%%%%%%%%%%%%%%%%%%%%%%%%
\gdef\NEphoton{    % NE   SLANTED PHOTONS:  RE-ORDERED IN PHOTONSETUP27
\ifcase\LINECONFIGURATION  %\REG case
\setcoords(425,425,0)(1250,0,0)[1250,1250]       \SETUNITBOX(1250)[br][tl]  
\ifnum\numwigglespo > \number \numwiggles \particleadjustx=15 \fi
\or % \FLIPPED case
\setcoords(1050,-200,0)(625,625,0)[1250,1250]    \SETUNITBOX(1250)[tl][br]
\ifnum\numwigglespo > \number \numwiggles \particleadjustx=25 \fi
\or % \CURLY case.
\setcoords(500,500,0)(1400,0,0)[1400,1400]       \SETUNITBOX(1400)[br][tl]
\or % \FLIPPEDCURLY case
\setcoords(1200,-200,0)(700,700,0)[1400,1400]    \SETUNITBOX(1400)[tl][br]  
\or % \FLAT case
\setcoords(400,400,0)(1200,0,0)[1200,1200]       \SETUNITBOX(1200)[br][tl]  
\or % \FLIPPEDFLAT case
\setcoords(1000,-200,0)(600,600,0)[1200,1200]    \SETUNITBOX(1200)[tl][br]
\else \UNIMPERROR
\fi
\numupperunits=\numwiggles   \numlowerunits=\numwigglespo
}
%%%%%%%%%%%%%%%%%%%%%%%%%%%%%%%%%%%%%%%%%%%%%%%%%%%%%%%%%%%%%%%%%%%%%%%%%%%
\gdef\Ephoton{    %  EASTWARD  HORIZONTAL PHOTONS
\ifcase\LINECONFIGURATION  % REG case
\setcoords(-285,715,0)(-150,-400,0)[2005,0]
\gdef\upperunitbox{\Frown}   \gdef\lowerunitbox{\Smile}
\or  % \FLIPPED case
\setcoords(-285,715,0)(-420,-170,0)[2005,0]
\gdef\upperunitbox{\Smile}   \gdef\lowerunitbox{\Frown}
\else \UNIMPERROR
\fi
\particleadjustx=-15 % Lengths are in centipoints.
}
%%%%%%%%%%%%%%%%%%%%%%%%%%%%%%%%%%%%%%%%%%%%%%%%%%%%%%%%%%%%%%%%%%%%%%%%%%%
\gdef\SEphoton{   % SE   SLANTED PHOTONS:  RE-ORDERED IN PHOTONSETUP27
\ifcase\LINECONFIGURATION  %\REG case
\setcoords(-200,1050,0)(-625,-625,0)[1250,-1250] \SETUNITBOX(1250)[tr][bl]
\ifnum\numwigglespo > \number \numwiggles \particleadjustx=25 \fi
\or % \FLIPPED case
\setcoords(425,425,0)(0,-1250,0)[1250,-1250]     \SETUNITBOX(1250)[bl][tr]
\ifnum\numwigglespo > \number \numwiggles \particleadjustx=15 \fi
\or % \CURLY case.
\setcoords(-200,1200,0)(-700,-700,0)[1400,-1400] \SETUNITBOX(1400)[tr][bl]  
\or % \FLIPPEDCURLY case
\setcoords(500,500,0)(0,-1400,0)[1400,-1400]     \SETUNITBOX(1400)[bl][tr]  
\or % \FLAT case
\setcoords(-200,1000,0)(-600,-600,0)[1200,-1200] \SETUNITBOX(1200)[tr][bl]
\particleadjustx=-20
\or % \FLIPPEDFLAT case
\setcoords(420,420,0)(0,-1200,0)[1200,-1200]     \SETUNITBOX(1200)[bl][tr]
\particleadjustx=40
\else \UNIMPERROR
\fi
}
%%%%%%%%%%%%%%%%%%%%%%%%%%%%%%%%%%%%%%%%%%%%%%%%%%%%%%%%%%%%%%%%%%%%%%%%%%%
\gdef\Sphoton{  % DOWN, DOWN VERTICAL PHOTONS
\ifcase\LINECONFIGURATION  %\REG case
\setcoords(-252,-490,0)(-740,-1740,0)[0,-2000]
\gdef\upperunitbox{\SRbr}   \gdef\lowerunitbox{\SLbr}
\or % \FLIPPED case
\setcoords(-490,-260,0)(-740,-1740,0)[0,-2002]
\gdef\upperunitbox{\SLbr}   \gdef\lowerunitbox{\SRbr}
\or %\CURLY case (a bit shorter).
\setcoords(-299,-449,0)(-870,-1970,0)[0,-2200]
\gdef\upperunitbox{\Rbr}    \gdef\lowerunitbox{\Lbr}
\particleadjusty=-95
\or %\FLIPPEDCURLY case.
\setcoords(-517,-371,0)(-900,-2000,0)[0,-2200]
\gdef\upperunitbox{\Lbr}    \gdef\lowerunitbox{\Rbr}
\particleadjusty=-165
\or % \FLAT case.  Flatter and shorter than \CURLY.  \LINECONFIGURATION=4.
\setcoords(-299,-409,0)(-885,-1905,0)[0,-2000]
\gdef\upperunitbox{\Rbr}   \gdef\lowerunitbox{\Lbr}
\particleadjustx=50     \particleadjusty=-380
\ifodd\unitboxnumber\relax\else\particleadjustx=250 \particleadjusty=-400 \fi
\or %\FLIPPEDFLAT case.  \LINECONFIGURATION=5.
\setcoords(-519,-449,0)(-900,-1920,0)[0,-2000]
\gdef\upperunitbox{\Lbr}   \gdef\lowerunitbox{\Rbr}
\particleadjusty=-370
\ifodd\unitboxnumber\relax\else\particleadjustx=-240 \particleadjusty=-400 \fi
\or % \LONGPHOTON
\gdef\upperunitbox{\Rbr}   \gdef\lowerunitbox{\Lbr}
\setcoords(-325,-555,0)(-900,-2100,0)[0,-2400]
\particleadjusty=-40
\or % \FLIPPEDLONG
\setcoords(-505,-275,0)(-900,-2100,0)[0,-2400]
\gdef\upperunitbox{\Lbr}   \gdef\lowerunitbox{\Rbr}
\particleadjusty=-30  % Lengths are in centipoints.
\else \UNIMPERROR
\fi
}
%%%%%%%%%%%%%%%%%%%%%%%%%%%%%%%%%%%%%%%%%%%%%%%%%%%%%%%%%%%%%%%%%%%%%%%%%%%
\gdef\SWphoton{  % SW SLANTED PHOTONS:  RE-ORDERED IN PHOTONSETUP27
\ifcase\LINECONFIGURATION  %\REG case
\setcoords(-825,-825,0)(0,-1250,0)[-1250,-1250]     \SETUNITBOX(1250)[br][tl]  
\or % \FLIPPED case
\setcoords(-175,-1425,0)(-625,-625,0)[-1250,-1250]  \SETUNITBOX(1250)[tl][br]  
\or % \CURLY case.
\setcoords(-900,-900,0)(0,-1410,0)[-1400,-1400]     \SETUNITBOX(1400)[br][tl]  
\or % \FLIPPEDCURLY case
\setcoords(-200,-1600,0)(-700,-700,0)[-1400,-1400]  \SETUNITBOX(1400)[tl][br]  
\or % \FLAT case
\setcoords(-800,-800,0)(0,-1200,0)[-1200,-1200]     \SETUNITBOX(1200)[br][tl]  
\or % \FLIPPEDFLAT case
\setcoords(-200,-1400,0)(-600,-600,0)[-1200,-1200]  \SETUNITBOX(1200)[tl][br]  
\else \UNIMPERROR
\fi
}
%%%%%%%%%%%%%%%%%%%%%%%%%%%%%%%%%%%%%%%%%%%%%%%%%%%%%%%%%%%%%%%%%%%%%%%%%%%
\gdef\Wphoton{
\ifcase\LINECONFIGURATION %\REG case
\setcoords(-2245,-1245,0)(-150,-400,0)[-2005,0]
\gdef\upperunitbox{\Frown}   \gdef\lowerunitbox{\Smile}
\or % \FLIPPED case
\setcoords(-2245,-1245,0)(-400,-150,0)[-2005,0]
\gdef\upperunitbox{\Smile}   \gdef\lowerunitbox{\Frown}
\else \UNIMPERROR
\fi
\particleadjustx=57 % Lengths are in centipoints.
\ifnum\numwigglespo=\number\numwiggles \particleadjustx=0  \fi
\numlowerunits=\numwigglespo   \numupperunits=\numwiggles
}
%%%%%%%%%%%%%%%%%%%%%%%%%%%%%%%%%%%%%%%%%%%%%%%%%%%%%%%%%%%%%%%%%%%%%%%%%%%
\gdef\NWphoton{  % NW   SLANTED PHOTONS:  RE-ORDERED IN PHOTONSETUP27
\ifcase\LINECONFIGURATION  %\REG case
\setcoords(-200,-1425,0)(625,625,0)[-1250,1250]   \SETUNITBOX(1250)[bl][tr]
\or % \FLIPPED case
\setcoords(-825,-825,0)(0,1250,0)[-1250,1250]     \SETUNITBOX(1250)[tr][bl]
\ifnum\numwigglespo > \number \numwiggles \particleadjusty=-15 \fi
\or % \CURLY case.
\setcoords(-200,-1600,0)(700,700,0)[-1400,1400]   \SETUNITBOX(1400)[bl][tr]
\or % \FLIPPEDCURLY case
\setcoords(-900,-900,0)(0,1400,0)[-1400,1400]     \SETUNITBOX(1400)[tr][bl]
\or % \FLAT case.
\setcoords(-200,-1400,0)(600,600,0)[-1200,1200]   \SETUNITBOX(1200)[bl][tr]  
\or % \FLIPPEDFLAT case
\setcoords(-800,-800,0)(0,1200,0)[-1200,1200]     \SETUNITBOX(1200)[tr][bl]  
\else \UNIMPERROR
\fi
}
  \fi   
\selectphoton
%CONTAINS PHOTON DEFINITIONS. 
}
%%%%%%%%%%% (4) GLUONS %%%%%%%%%%% 
\gdef\selectgluon{   % RECURSIVELY RE-DEFINED IN GLUONSETUP(25+).TEX.
% Input gluon-setup stuff ONLY IF HAVE NOT DONE SO YET.
% This avoids having to process a gluon if none are drawn.
\ifnum\gluoncount=0 %                            GLUONSETUP(31).TEX
% CALLED BY FEYNMAN(34).
% USED FOR GENERATING GLUON LINES IN FEYNMAN DIAGRAMS IN LATEX.
\global\newcount\gluonlength
\global\newcount\gluonlengthx
\global\newcount\gluonlengthy
\global\newcount\gluonfrontx  %}(x,y) co-ord of left of gluon
\global\newcount\gluonfronty  %}
\global\newcount\gluonbackx
\global\newcount\gluonbacky
%
%%%%%%%%%%%%%%%%%%%%%%%%%%%%%%%%%%%%%%%%%%%%%%%%%%%%%%%%%%%%%%%%%%%%%%%%%%%%
\gdef\setunitbox(#1)[#2][#3](#4)[#5]{
\gdef\upperunitbox{\oval(#1,#1)[#2]}
\gdef\lowerunitbox{\oval(401,401)[#3]}
\gdef\thirdunitbox{\oval(#4,#4)[#5]}
}
%%%%%%%%%%%%%%%%%%%%%%%%%%%%%%%%%%%%%%%%%%%%%%%%%%%%%%%%%%%%%%%%%%%%%%%%%%%%
\gdef\selectgluon{  % ORIGINALLY DEFINED IN FEYNMAN(29+).  RECURSIVELY RE-DEFINE
\global\advance\gluoncount by 1  % Counts number of gluons drawn. 
\global\gluonfrontx=\particlefrontx   % READ IN FROM FEYNMAN \selectgluon
\global\gluonfronty=\particlefronty   % READ IN FROM FEYNMAN \selectgluon
\global\particleadjustx=0     \global\particleadjusty=0
% Error messages for overly-long lines.  See FEYNMAN for negative-lengths.
\ifnum\unitboxnumber > 40
\message{   *** WARNING *** Gluon with 
\the\unitboxnumber\space loops requested ***   }
\ifnum\unitboxnumber > 85
\message{   *** Reducing gluon length to 6 loops (max 85) ***   }
\ifnum\unitboxnumber > 1000
\message{   *** Probable Cause:  Gluon selected instead of Fermion ***   }
\fi \global\unitboxnumber=6 \fi \fi  % end of length error
\global\unitboxnumberpo=\unitboxnumber  % DEFINED IN \drawsavedline (FEYNMAN).
\global\advance\unitboxnumberpo by 1 %\unitboxnumber if \unitboxnumber is odd.
% DEFAULTS:
\global\numlineparts = 3
\global\numupperunits=\unitboxnumber
\global\numlowerunits=\unitboxnumber
\global\numthirdunits=\unitboxnumber
% select gluon alignment:
\ifcase\LINEDIRECTION
\Ngluon    %\LINEDIRECTION=0 (NORTH) CASE
\or  \NEgluon  % 1 case
\or  \Egluon   % 2 case...horizontal gluon.
\or  \SEgluon
\or  \Sgluon
\or  \SWgluon 
\or  \Wgluon
\or  \NWgluon
\else\DIRECTERROR \fi
\setparticle  
\global\gluonlengthx=\particlelengthx  \global\gluonlengthy=\particlelengthy
\global\gluonbackx=\particlebackx      \global\gluonbacky=\particlebacky      
} 
%%%%%%%%%%%%%%%%%%%%%%%%%%%%%%%%%%%%%%%%%%%%%%%%%%%%%%%%%%%%%%%%%%%%%%%%%%%%
\gdef\Ngluon{   % VERTICAL gluons
\ifcase\LINECONFIGURATION   % \REG GLUON CONFIGURATION
\setcoords(600,540,600)(20,620,1220)[0,1050] 
\setunitbox(1600)[tl][r](1600)[bl]
\particleadjusty=195
\or % \FLIPPED
\setcoords(-990,-930,-990)(12,615,1215)[0,1050]
\setunitbox(1600)[tr][l](1600)[br]
\particleadjusty=195
\or % \CURLYGLUON
\setcoords(440,390,440)(-10,415,840)[0,850]  
\setunitbox(1250)[tl][r](1250)[bl]
\particleadjustx=0
\particleadjusty=-10
\or % \FLIPPED case
\setcoords(-820,-770,-820)(-25,400,825)[0,850]  % Matches change in \S\CURLY
\particleadjusty=-10  % goes with \CURLY \particleadjusty=-10
\setunitbox(1250)[tr][l](1250)[br]
\or \UNIMPERROR  % \LCONFIG=4 is \FLAT case
\or \UNIMPERROR  % \LCONFIG=5 is \FLIPPEDFLAT case
\or % \LCONFIG=6 is \CENTRALGLUON case
\numupperunits=\unitboxnumberpo
\numlowerunits=\unitboxnumber
\numthirdunits=\unitboxnumberpo
\setcoords(-200,-200,-200)(616,1041,616)[0,850]
\setunitbox(1250)[tl][r](1250)[bl]
\particleadjusty=1238
\particleadjusty=1233
\or % \FLIPPEDCENTRAL
\numupperunits=\unitboxnumberpo
\numlowerunits=\unitboxnumber
\numthirdunits=\unitboxnumberpo
\setcoords(-200,-200,-200)(620,1045,620)[0,850]
\setunitbox(1250)[tr][l](1250)[br]
\particleadjusty=1245
\else \UNIMPERROR % etc.
\fi
}
%%%%%%%%%%%%%%%%%%%%%%%%%%%%%%%%%%%%%%%%%%%%%%%%%%%%%%%%%%%%%%%%%%%%%%%%%%%%%%%%
\gdef\NEgluon{
\numupperunits=\unitboxnumberpo
\numlowerunits=\unitboxnumber
\numthirdunits=\unitboxnumber
\ifcase\LINECONFIGURATION
\setcoords(900,900,900)(0,900,900)[900,900]
\setunitbox(2200)[tl][tr](401)[b]
\particleadjustx=1100     \particleadjusty=1100
\or % \FLIPPED case
\setcoords(-180,720,720)(1090,1091,1091)[900,900]
\setunitbox(2200)[br][tr](401)[l]
\particleadjustx=1110     \particleadjusty=1050                      
\else \UNIMPERROR % etc.
\fi
}
%%%%%%%%%%%%%%%%%%%%%%%%%%%%%%%%%%%%%%%%%%%%%%%%%%%%%%%%%%%%%%%%%%%%%%%%%%%%%%%%
\gdef\Egluon{     % EASTWARD HORIZONTAL gluons
\ifcase\LINECONFIGURATION
\setcoords(-210,390,990)(-800,-745,-800)[1050,0]  % draws from outer line edge
\setunitbox(1600)[tr][b](1600)[tl]  
\particleadjustx=130  % draws from outer line edge
\or % \FLIPPED
\setcoords(-210,390,990)(800,745,800)[1050,0]  % draws from outer line edge
\setunitbox(1600)[br][t](1600)[bl]
\particleadjustx=130
\or % \CURLYGLUON
\setcoords(-200,225,650)(-625,-575,-625)[850,0]
\setunitbox(1250)[tr][b](1250)[tl]  
\or % \FLIPPED case
\setcoords(-200,225,650)(625,575,625)[850,0]
\setunitbox(1250)[br][t](1250)[bl]
\or % \LCONFIG=4 is \FLAT case
\setcoords(-200,430,1060)(-830,-780,-830)[1260,0]
\setunitbox(1660)[tr][b](1660)[tl]
\or % \LCONFIG=5 is \FLIPPEDFLAT case
\setcoords(-200,430,1060)(830,780,830)[1260,0]
\setunitbox(1660)[br][t](1660)[bl]
\or % \LCONFIG=6 is \CENTRALGLUON case
\numupperunits=\unitboxnumberpo
\numlowerunits=\unitboxnumber
\numthirdunits=\unitboxnumberpo
\setcoords(440,865,440)(0,50,0)[850,0]
\setunitbox(1250)[tr][b](1250)[tl]
\particleadjustx=1260
\or % \FLIPPEDCENTRALGLUON case 
\numupperunits=\unitboxnumberpo
\numlowerunits=\unitboxnumber
\numthirdunits=\unitboxnumberpo
\setcoords(430,855,430)(0,-50,0)[850,0]
\setunitbox(1250)[br][t](1250)[bl]
\particleadjustx=1250
\or % \LCONFIG=8 is \SQUASHEDGLUON case
\setcoords(-160,440,1040)(-600,-550,-600)[1200,0]
\gdef\upperunitbox{\oval(1600,1200)[tr]}
\gdef\thirdunitbox{\oval(1600,1200)[tl]}
\gdef\lowerunitbox{\oval(401,401)[b]}
\else \UNIMPERROR 
\fi
}
%%%%%%%%%%%%%%%%%%%%%%%%%%%%%%%%%%%%%%%%%%%%%%%%%%%%%%%%%%%%%%%%%%%%%%%%%%%%%%%%
\gdef\SEgluon{  
\numupperunits=\unitboxnumberpo
\numlowerunits=\unitboxnumber
\numthirdunits=\unitboxnumber
\ifcase\LINECONFIGURATION
\setcoords(-200,700,700)(-1100,-1100,-1100)[900,-900]
\setunitbox(2200)[tr][br](401)[l]
\particleadjustx=1100     \particleadjusty=-1100  
\or % \FLIPPED case
\setcoords(890,890,890)(0,-900,-900)[900,-900]
\setunitbox(2200)[bl][br](401)[t]
\particleadjustx=1050     \particleadjusty=-1100 
\else \UNIMPERROR % etc.
\fi
}
%%%%%%%%%%%%%%%%%%%%%%%%%%%%%%%%%%%%%%%%%%%%%%%%%%%%%%%%%%%%%%%%%%%%%%%%%%%%%%%%
\gdef\Sgluon{   % VERTICAL gluons
\ifcase\LINECONFIGURATION  % \REG STYLE GLUON
\setcoords(-1000,-940,-1000)(0,-595,-1195)[0,-1050]
\setunitbox(1600)[br][l](1600)[tr]
\particleadjusty=-150
\or % \FLIPPED
\setcoords(605,545,605)(-20,-615,-1215)[0,-1050]
\setunitbox(1600)[bl][r](1600)[tl]  
\particleadjusty=-150
\or % \CURLYGLUON
\setcoords(-820,-770,-820)(0,-425,-850)[0,-850]
\setunitbox(1250)[br][l](1250)[tr]  
\or % \FLIPPED case
\setcoords(440,390,440)(0,-425,-850)[0,-850]
\setunitbox(1250)[bl][r](1250)[tl]  
\or \UNIMPERROR % \LCONFIG=4 is \FLATGLUON case
\or \UNIMPERROR
\or % \LCONFIG=6 is \CENTRALGLUON case
\numupperunits=\unitboxnumberpo
\numlowerunits=\unitboxnumber
\numthirdunits=\unitboxnumberpo
\setcoords(-180,-180,-180)(-635,-1060,-635)[0,-850]
\setunitbox(1250)[br][l](1250)[tr]  
\particleadjusty=-1290
\or % \FLIPPEDCENTRAL case
\numupperunits=\unitboxnumberpo
\numlowerunits=\unitboxnumber
\numthirdunits=\unitboxnumberpo
\setcoords(-180,-180,-180)(-635,-1060,-635)[0,-850]
\setunitbox(1250)[bl][r](1250)[tl]
\particleadjusty=-1290
\else \UNIMPERROR % etc.
\fi
}
%%%%%%%%%%%%%%%%%%%%%%%%%%%%%%%%%%%%%%%%%%%%%%%%%%%%%%%%%%%%%%%%%%%%%%%%%%%%%%%%
\gdef\SWgluon{
\numupperunits=\unitboxnumberpo
\numlowerunits=\unitboxnumber
\numthirdunits=\unitboxnumber
\ifcase\LINECONFIGURATION
\setcoords(-1300,-1300,-1300)(0,-900,-900)[-900,-900]
\setunitbox(2200)[br][bl](401)[t]
\particleadjustx=-1100     \particleadjusty=-1100
\or % \FLIPPED case
\setcoords(-215,-1115,-1115)(-1107,-1107,-1107)[-900,-900]
\setunitbox(2200)[tl][bl](401)[r]
\particleadjustx=-1120     \particleadjusty=-1120
\else \UNIMPERROR % FLIPPED FLATGLUON etc.
\fi
}
%%%%%%%%%%%%%%%%%%%%%%%%%%%%%%%%%%%%%%%%%%%%%%%%%%%%%%%%%%%%%%%%%%%%%%%%%%%%%%%%
\gdef\Wgluon{   % HORIZONTAL gluons
\ifcase\LINECONFIGURATION
\setcoords(-190,-790,-1390)(800,745,800)[-1050,0]
\setunitbox(1600)[bl][t](1600)[br]  
\particleadjustx=-150  %302
\or % \UNIMPERROR % \FLIPPED
\setcoords(-190,-790,-1390)(-800,-745,-800)[-1050,0]
\setunitbox(1600)[tl][b](1600)[tr]  
\particleadjustx=-150  %302
\or % \CURLYGLUON
\setcoords(-200,-625,-1050)(625,575,625)[-850,0]
\setunitbox(1250)[bl][t](1250)[br]  
\or % \FLIPPED case
\setcoords(-200,-625,-1050)(-625,-575,-625)[-850,0] 
\setunitbox(1250)[tl][b](1250)[tr]  
\or % \LCONFIG=4 is \FLATGLUON case
\setcoords(-230,-860,-1490)(830,780,830)[-1260,0]
\setunitbox(1660)[bl][t](1660)[br]  
\or % \UNIMPERROR
\setcoords(-230,-860,-1490)(-830,-780,-830)[-1260,0]
\setunitbox(1660)[tl][b](1660)[tr]
\or % \LCONFIG=6 is \CENTRALGLUON case
\numupperunits=\unitboxnumberpo
\numlowerunits=\unitboxnumber
\numthirdunits=\unitboxnumberpo
\setcoords(-825,-1250,-825)(0,-50,0)[-850,0]
\setunitbox(1250)[bl][t](1250)[br]  
\particleadjustx=-1250
\or  % \FLIPPEDCENTRALGLUON
\numupperunits=\unitboxnumberpo
\numlowerunits=\unitboxnumber
\numthirdunits=\unitboxnumberpo
\setcoords(-825,-1250,-825)(0,50,0)[-850,0] 
\setunitbox(1250)[tl][b](1250)[tr]  
\particleadjustx=-1250
\else \UNIMPERROR % FLIPPED FLATGLUON etc.
\fi
}
%%%%%%%%%%%%%%%%%%%%%%%%%%%%%%%%%%%%%%%%%%%%%%%%%%%%%%%%%%%%%%%%%%%%%%%%%%%%%%%%
\gdef\NWgluon{
\numupperunits=\unitboxnumberpo
\numlowerunits=\unitboxnumber
\numthirdunits=\unitboxnumber
\ifcase\LINECONFIGURATION
\setcoords(-200,-1100,-1100)(1100,1100,1100)[-900,900]
\setunitbox(2200)[bl][tl](401)[r]
\particleadjustx=-1110   \particleadjusty=1100
\or  % \FLIPPED
\setcoords(-1309,-1309,-1309)(-15,885,885)[-900,900]
\setunitbox(2200)[tr][tl](401)[b]
\particleadjustx=-1120   \particleadjusty=1065
\else \UNIMPERROR % FLIPPED FLATGLUON etc.
\fi
}
%
%                         *** GLUON LINKS AND CAPS ***
%
% 
\gdef\gluonlink{    %  A `RECURSIVE' DEFINITION.
%                       GLUONLINKS.TEX (FEB. 5, 1989)
%  CALLED BY FEYNMAN(34).TEX.
%
%  Routines for drawing GLUON LINKS and GLUON CAPS used by FEYNMAN.TEX.
%  Formerly in separate files (GLUONLINKS & GLUONCAPS).
%  This file is read in when \gluonlink, \gluoncap, defined in GLUONSETUP(21+),
%  and related commands (in \drawvetex) are used.
%
%  GLUON LINKS are used to link \CENTRAL-type gluon lines together 
%  (eg: \REG & \FLIPPED gluons drawn diagonally).
%  Syntax:   \drawline\gluon[\W\CENTRAL](x,y)[number of loops]\gluonlink
%            \drawline\gluon[\W\CENTRAL](\pbackx,\pbacky)[number of loops]
%
%  GLUON CAPS are used on vertical and horizontal gluons to make the
%  terminus of the gluon line end along the gluon axis.  It then draws
%  a stem of length \stemlength (as though the command \BACKSTEMMED had
%  been issued).  Principal use is when \drawvertex is used.
%  Syntax:   \drawline\gluon[\W\FLAT](x,y)[number of loops]\gluoncap
%
%%%%%%%%%%%%%%%%%%%%%%%%%%%%%%%%%%%%%%%%%%%%%%%%%%%%%%%%%%%%%%%%%%%%%%%%%%%%%
%                                                                           %
%                                                                           %
%                                GLUON LINKS                                %
%                                                                           %
%                                                                           %
%%%%%%%%%%%%%%%%%%%%%%%%%%%%%%%%%%%%%%%%%%%%%%%%%%%%%%%%%%%%%%%%%%%%%%%%%%%%%
%
%        ROUTINE FOR LINKING \CENTRAL AND \FLIPPEDCENTRAL GLUONS
%
\gdef\gluonlink{
\global\stemlengthx=401 \SETDIR \setadjxy \divide\adjx by -2 \divide\adjy by -2
\ifcase\LDIR  
     \linksetupB \ifodd\LCONFIG\LINKPUT[l]  \else\LINKPUT[r] \fi %\Ngluon
\or  \ifodd\LCONFIG \linksetupBx \LINKPUT[tr]\LINKPUT[l]         %\NEgluon
     \else          \linksetupBy \LINKPUT[tr]\LINKPUT[b]     \fi
\or  \linksetupB \ifodd\LCONFIG\LINKPUT[t]  \else\LINKPUT[b] \fi %\Egluon
\or  \ifodd\LCONFIG \linksetupBy \LINKPUT[br]\LINKPUT[t]         %\SEgluon
     \else          \linksetupBx \LINKPUT[br]\LINKPUT[l]     \fi
\or  \linksetupB \ifodd\LCONFIG\LINKPUT[r]  \else\LINKPUT[l] \fi %\Sgluon
\or  \ifodd\LCONFIG \linksetupBx \LINKPUT[bl]\LINKPUT[r]         %\SWgluon
     \else          \linksetupBy \LINKPUT[bl]\LINKPUT[t]     \fi
\or  \linksetupB \ifodd\LCONFIG\LINKPUT[b]  \else\LINKPUT[t] \fi %\Wgluon
\or  \ifodd\LCONFIG \linksetupBy \LINKPUT[tl]\LINKPUT[b]         %\NWgluon
     \else          \linksetupBx \LINKPUT[tl]\LINKPUT[r]     \fi
\else \UNIMPERROR 
\fi
\ifodd\LDIR\relax 
\else \global\advance\pbackx by \adjx   \global\advance\pbacky by \adjy \fi
\linksetupC
}
%
%%%%%%%%%%%%%%%%%%%%%%%%%%%%%%%%%%%%%%%%%%%%%%%%%%%%%%%%%%%%%%%%%%%%%%%%%%%%%
%                                                                           %
%                                                                           %
%                                GLUONCAPS                                  %
%                                                                           %
%                                                                           %
%%%%%%%%%%%%%%%%%%%%%%%%%%%%%%%%%%%%%%%%%%%%%%%%%%%%%%%%%%%%%%%%%%%%%%%%%%%%%
%
%            ROUTINE FOR CENTRALIZING/CAPPING GLUONS
%
\gdef\gluoncap{  % For centralizing gluons
\global\stemlengthx=0  
\ifodd\LDIR\message{NOTE:  Diagonal Gluons are not Capped}\relax
\else
\ifcase\LCONFIG \global\stemlengthx=1000  %\REG case for NSEW
\or \global\stemlengthx=1000  %\FLIPPED case (NSEW)
\or \global\stemlengthx=825   %\CURLY case (NSEW)
\or \global\stemlengthx=825   %\FLIPPEDCURLY  case (NSEW)
\or \global\stemlengthx=1030  %\FLAT case (NSEW)
\or \global\stemlengthx=1030  %\FLIPPEDFLAT case (NSEW)
\or \global\stemlengthx=0     %\CENTRAL case (NSEW)
\or \global\stemlengthx=0     %\FLIPPEDCENTRAL case (NSEW)
\or \global\stemlengthx=800   %\SQUASHED case (NSEW)
\else\UNIMPERROR
\fi % end \ifcase\LCONFIG
\ifnum\stemlengthx>400
\global\advance\LDIR by 2 \moduloeight\LDIR   \SETDIR 
\global\advance\LDIR by 6 \moduloeight\LDIR   \setadjxy
\ifodd\LCONFIG \multiply \adjx by -1   \multiply \adjy by -1 \fi
\divide\adjx by 2  \divide\adjy by 2 \linksetupB
\ifcase\LDIR 
     \ifodd\LCONFIG\LINKPUT[tr]  \else\LINKPUT[tl]   \fi   %\NGLUON
\or  \relax % \NEGLUON
\or  \ifodd\LCONFIG\LINKPUT[br]  \else\LINKPUT[tr]   \fi   %\Egluon
\or  \relax % \SEgluon
\or  \ifodd\LCONFIG\LINKPUT[bl]  \else\LINKPUT[br]   \fi   %\Sgluon
\or  \relax % \SWgluon
\or  \ifodd\LCONFIG\LINKPUT[tl]  \else\LINKPUT[bl]   \fi   %\Wgluon
\or  \relax % \NWgluon
\else \UNIMPERROR 
\fi
\linksetupD
\else\message{* NOTE:  Attempt to use gluoncap of less that 401 centipoints *}
\fi % End of \ifnum\stemlengthx
\fi % end \ifodd
}
%
%%%%%%%%%%%%%%%%%%%%%%%%%%%%%%%%%%%%%%%%%%%%%%%%%%%%%%%%%%%%%%%%%%%%%%%%%%%%%
%                                                                           %
%                                                                           %
%       AUXILIARY ROUTINES USED BY BOTH \gluonlink and \gluoncap            %
%                                                                           %
%                                                                           %
%%%%%%%%%%%%%%%%%%%%%%%%%%%%%%%%%%%%%%%%%%%%%%%%%%%%%%%%%%%%%%%%%%%%%%%%%%%%%
%
\gdef\setadjxy{
\adjx=\stemlengthx   \adjy=\stemlengthx % Yes, it's `x', not `y'.
\multiply \adjx by \XDIR%  
\multiply \adjy by \YDIR%  
}
\gdef\advancegluonlength{
\global\advance\particlelengthx by \adjx \global\advance\particlelengthy by\adjy
\global\advance\gluonlengthx by \adjx  \global\advance\gluonlengthy by \adjy
% The new adjusted gluon lengths.
}
\gdef\LINKPUT[#1]{\ifnum\phantomswitch=0\put(\gluonbackx,\gluonbacky)
{\oval(\stemlengthx,\stemlengthx)[#1]}\fi}  
\gdef\linksetupBx{\global\advance\gluonbackx by \adjx}
\gdef\linksetupBy{\global\advance\gluonbacky by \adjy}
\gdef\linksetupB{\linksetupBx\linksetupBy}
\gdef\linksetupC{
\global\advance\pbackx by \adjx  \global\advance\pbacky by \adjy
\global\gluonbackx=\pbackx   \global\gluonbacky=\pbacky
% Take negative particle lengths into account.  \BOXLENGTHX,Y remain as before.
\ifnum\plengthx<0 \multiply\adjx by -1 \fi
\ifnum\plengthy<0 \multiply\adjy by -1 \fi
\advancegluonlength
}
\gdef\linksetupD{
\linksetupC
\SETDIR 
\setadjxy
\divide\adjx by 2  \divide\adjy by 2 \linksetupB
\global\pbackx=\gluonbackx   \global\pbacky=\gluonbacky
\advancegluonlength
\ifnum\phantomswitch=0\put(\pbackx,\pbacky){\line(\XDIR,\YDIR){\stemlength}}\fi
\stemlengthx=\stemlength  %TESTING
\setadjxy
\global\advance\pbackx by \adjx  \global\advance\pbacky by \adjy
\global\gluonbackx=\pbackx   \global\gluonbacky=\pbacky
\advancegluonlength
% The new adjusted gluon lengths and rear positions
}
   % \gluonlink is redefined here.
\gluonlink}  %  THIS RE-DEFINES \gluonlink FOR ALL FUTURE USES.
\gdef\gluoncap{    %  A `RECURSIVE' DEFINITION.
   % \gluoncap is redefined here.
\gluoncap}  %  THIS RE-DEFINES \gluoncap FOR ALL FUTURE USES.
  \fi
\selectgluon
%                  CONTAINS gluon DEFINITIONS.
}
%%%%%%%%%%% (5) SPECIAL - USER DEFINED %%%%%%%%%%% 
\gdef\selectespecial{\UNIMPERROR}
%
%%%%%%%%%%%%%%%%%%%%%%%%%%%%%%%%%%%%%%%%%%%%%%%%%%%%%%%%%%%%%%%%%%%%%%%%%%%%%%%
%                                                                             %
%                   ROUTINES FOR DRAWING VERTICES                             %
%                                                                             %
%%%%%%%%%%%%%%%%%%%%%%%%%%%%%%%%%%%%%%%%%%%%%%%%%%%%%%%%%%%%%%%%%%%%%%%%%%%%%%%
%
% Input vertex-setup stuff ONLY IF HAVE NOT DONE SO YET.                      %
% This avoids having to process a vertex if none are drawn. 
\gdef\checkvertex{ %immediately re-defines \drawvertex,\vertexcap,\linkvertex...
\ifnum\vertexcount=-1   %                          VERTEX(25)
% CALLED BY FEYNMAN(34).TEX.
% Vertex functions for FEYNMAN drawn via \drawvertex.
%
\global\advance\vertexcount by 1 % Needed and defined in FEYNMAN19&20.
\newsavebox{\vertexbox}   
\global\newcount\LDIRcount      % Counts the number of vertices drawn.
\global\newcount\VERTEXNUMBER   % 3 or 4; The number of particles in the vertex.
\global\newdimen\VERTEXLINKONE  % And now some switches.
\global\newdimen\VERTEXLINKTWO
\global\newdimen\VERTEXLINKTHREE
\global\newdimen\VERTEXLINKFOUR %\clearvertex  %sets them to zero
\global\newdimen\VERTEXCAPONE   % And now some GLUONCAP switches.
\global\newdimen\VERTEXCAPTWO
\global\newdimen\VERTEXCAPTHREE
\global\newdimen\VERTEXCAPFOUR  %\clearvertex  %sets them to zero
\global\newdimen\STEMVERTEXONE
\global\newdimen\STEMVERTEXTWO
\global\newdimen\STEMVERTEXTHREE
\global\newdimen\STEMVERTEXFOUR %\clearvertex  %sets them to zero
\global\newcount\stemlengthcopy
\global\newcount\vertexonex    \global\newcount\vertexoney
\global\newcount\vertextwox    \global\newcount\vertextwoy
\global\newcount\vertexthreex  \global\newcount\vertexthreey
\global\newcount\vertexfourx   \global\newcount\vertexfoury
\global\newcount\vertexmidx    \global\newcount\vertexmidy
\global\newcount\VERTEXLINE  
\global\newcount\FLIPVERTEX    \global\FLIPVERTEX=0
\gdef\flipvertex{\global\FLIPVERTEX=1}  %For drawing vertices with flipped lines
\newcount\vertadj              \newcount\negvertadj
\gdef\drawvertex#1[#2#3](#4,#5)[#6]{
\global\advance\vertexcount by 1 % Counts number of vertices draw.
\global\LINETYPE=#1           % either \gluon or \photon
\global\LINEDIRECTION=#2
\global\VERTEXNUMBER=#3  % either 3 or 4
\global\vertexonex=#4
\global\vertexoney=#5
\global\unitboxnumber=#6
\global\stemlengthcopy=\stemlength   % Record use-defined value.
%
% SELECT VERTEX:
%
\ifnum\LINETYPE<3 \LINEERROR \fi
\ifnum\LINETYPE=3   %  *** PHOTONS ***
% TEST WHETHER GLUONS HAVE BEEN PREVIOUSLY DRAWN:
\ifnum\gluoncount=0 \def\gluonlink{\relax} \def\gluoncap{\relax} \fi
  \ifnum\VERTEXNUMBER<3 \UNIMPERROR \fi
  \ifnum\VERTEXNUMBER=3 \THREEPHOTON\fi % 3-PHOTON VERTEX
  \ifnum\VERTEXNUMBER=4 \FOURPHOTON \fi % 4-PHOTON VERTEX
  \ifnum\VERTEXNUMBER>4 \UNIMPERROR \fi
\fi
\ifnum\LINETYPE=4   %  *** GLUONS ***
  \ifnum\VERTEXNUMBER<3 \UNIMPERROR \fi
  \ifnum\VERTEXNUMBER=3 \THREEGLUON \fi % 3-GLUON VERTEX
  \ifnum\VERTEXNUMBER=4 \FOURGLUON  \fi % 4-GLUON VERTEX
  \ifnum\VERTEXNUMBER>4 \UNIMPERROR \fi
\fi
\ifnum\LINETYPE>4 \LINEERROR \fi
\clearvertex   %  Resets VERTEXLINKs, VERTEXCAPS etc.
}  % END \drawvertex
%%%%%%%%%%%%%%%%%%%%%%%%%%%%%%%%%%%%%%%%%%%%%%%%%%%%%%%%%%%%%%%%%%%%%%%%%%%%%%
%                                                                            %
%                            PHOTON VERTICES                                 %
%                                                                            %
%%%%%%%%%%%%%%%%%%%%%%%%%%%%%%%%%%%%%%%%%%%%%%%%%%%%%%%%%%%%%%%%%%%%%%%%%%%%%%
%
\gdef\advtwomodeight#1{
\global\advance\LDIRcount by2\moduloeight\LDIRcount \diagFOURVERT#1[\LDIRcount]}
%
%                        %%%%%%%%%%%%%%%%%%%%%%%
%                        % Three-Photon Vertex %
%                        %%%%%%%%%%%%%%%%%%%%%%%
\gdef\THREEPHOTON{
\ifcase\LDIR  % N vertex
\setvertexA[\S\REG]
\setvertexB[\NW\CURLY](0,0)[2]  \setvertexB[\NE\FLIPPEDCURLY](0,0)[3]
\or  % NE vertex
\setvertexA[\SW\CURLY]
\setvertexB[\N\REG](70,-100)[2] \setvertexB[\E\FLIPPED](0,0)[3]
\or  % E vertex
\setvertexA[\W0]
\setvertexB[\NE\CURLY](20,0)[2] \setvertexB[\SE\FLIPPEDCURLY](20,0)[3]
\or  % SE VERTEX
\setvertexA[\NW\CURLY]
\setvertexB[\E\REG](0,0)[2] \setvertexB[\S1](20,100)[3]
\or  % S VERTEX
\setvertexA[\N\REG]
\setvertexB[\SE\CURLY](0,0)[2]  \setvertexB[\SW\FLIPPEDCURLY](0,0)[3]
\or  % SW vertex
\setvertexA[\NE\CURLY]
\setvertexB[\S\REG](-20,100)[2] 
\setvertexB[\W\FLIPPED](0,0)[3]
\or  % W vertex
\setvertexA[\E\REG]
\setvertexB[\SW\CURLY](0,0)[2]  \setvertexB[\NW\FLIPPEDCURLY](0,0)[3]
\or  % NW vertex
\setvertexA[\SE\CURLY]
\setvertexB[\W\REG](0,0)[2]  
\setvertexB[\N\FLIPPED](-40,-100)[3] 
\else \DIRECTERROR
\fi
} %end \THREEPHOTON
%
%%%%%%%%%%%%%%%%%%%%%%%%%%%%%%%%%%%%%%%%%%%%%%%%%%%%%%%%%%%%%%%%%%%%%%%%%%%%%
%
%                        %%%%%%%%%%%%%%%%%%%%%%
%                        % Four-Photon Vertex %
%                        %%%%%%%%%%%%%%%%%%%%%%
\gdef\FOURPHOTON{
\ifnum\LDIR>-1
\global\LDIRcount=\LDIR  \global\advance\LDIRcount by 4 \moduloeight\LDIRcount
\setvertexA[\LDIRcount\REG] \advtwomodeight2 \advtwomodeight3 \advtwomodeight4
\else \UNIMPERROR \fi
\global\FLIPVERTEX=0
}
%
%%%%%%%%%%%%%%%%%%%%%%%%%%%%%%%%%%%%%%%%%%%%%%%%%%%%%%%%%%%%%%%%%%%%%%%%%%%%%%
%                                                                            %
%                            GLUON VERTICES                                  %
%                                                                            %
%%%%%%%%%%%%%%%%%%%%%%%%%%%%%%%%%%%%%%%%%%%%%%%%%%%%%%%%%%%%%%%%%%%%%%%%%%%%%%
%
%                        %%%%%%%%%%%%%%%%%%%%%%
%                        % Three-Gluon Vertex %
%                        %%%%%%%%%%%%%%%%%%%%%%
\gdef\THREEGLUON{
\vertadj=0   \adjvert  % No special adjustments required for this case.
\ifcase\LDIR  % N vertex
\setvertexA[\S\CENTRAL]
\setvertexB[\NW\REG](0,0)[2]  \setvertexB[\NE\FLIPPED](0,0)[3]
\or  % NE vertex
\setvertexA[\SW\REG]
\setvertexB[\N\CURLY](-170,442)[2] \setvertexB[\E 3](420,-183)[3]
\setvertexC(442,442)[bl]
\or  % E vertex
\setvertexA[\W6]
\setvertexB[\NE\REG](0,0)[2] \setvertexB[\SE\FLIPPED](0,0)[3]
\or  % SE VERTEX
\setvertexA[\NW\REG]
\setvertexB[\E\CURLY](420,183)[2]  \setvertexB[\S3](-183,-442)[3]
\setvertexC(442,-442)[tl]
\or  % S VERTEX
\setvertexA[\N\CENTRAL]
\setvertexB[\SE\REG](0,0)[2]  \setvertexB[\SW\FLIPPED](0,0)[3]
\or  % SW vertex
\setvertexA[\NE\REG]
\setvertexB[\S\CURLY](170,-442)[2]
\setvertexB[\W\FLIPPEDCURLY](-420,183)[3]
\setvertexC(-442,-442)[tr]
\or  % W vertex
\setvertexA[\E\CENTRAL]
\setvertexB[\SW\REG](0,0)[2]  \setvertexB[\NW\FLIPPED](0,0)[3]
\or  % NW vertex
\setvertexA[\SE\REG]
\setvertexB[\W\CURLY](-420,-183)[2]
\setvertexB[\N\FLIPPEDCURLY](170,442)[3]
\setvertexC(-442,442)[br]
\else \DIRECTERROR
\fi
} %end \THREEGLUON
%
%%%%%%%%%%%%%%%%%%%%%%%%%%%%%%%%%%%%%%%%%%%%%%%%%%%%%%%%%%%%%%%%%%%%%%%%%%%%%%
%
%                         %%%%%%%%%%%%%%%%%%%%%
%                         % Four-Gluon Vertex %
%                         %%%%%%%%%%%%%%%%%%%%%
%
%
\gdef\FOURGLUON{
\ifodd\LDIR\vertadj=0 \else  \vertadj=412 \fi    \adjvert
\ifcase\LDIR  % N vertex
\setvertexA[\S\CURLY]         \MIDADJUST(\negvertadj,\vertadj)
\WFOURVERT2    \NFOURVERT3    \EFOURVERT4
\or \FOURPHOTON  % NE vertex
\or  % E vertex
\setvertexA[\W\CURLY]         \MIDADJUST(\vertadj,\vertadj)
\NFOURVERT2  \EFOURVERT3  \SFOURVERT4
\or \FOURPHOTON    % SE VERTEX
\or  % S VERTEX
\setvertexA[\N\CURLY]         \MIDADJUST(\vertadj,\negvertadj)
\EFOURVERT2   \SFOURVERT3   \WFOURVERT4
\or \FOURPHOTON    % SW vertex
\or  % W vertex
\setvertexA[\E\CURLY]         \MIDADJUST(\negvertadj,\negvertadj)
\SFOURVERT2  \WFOURVERT3  \NFOURVERT4
\or \FOURPHOTON    % NW vertex
\else \DIRECTERROR
\fi
\global\FLIPVERTEX=0   % reset \FLIPVERTEX to unflipped case.
} % End of \FOURGLUON
%
%%%%%%%%%%%%%%%%%%%%%%%%%%%%%%%%%%%%%%%%%%%%%%%%%%%%%%%%%%%%%%%%%%%%%%%%%%%%
%                %                                    %                    %
%                %    FOUR GLUON UTILITY ROUTINES:    %                    %
%                %                                    %                    %
%%%%%%%%%%%%%%%%%%%%%%%%%%%%%%%%%%%%%%%%%%%%%%%%%%%%%%%%%%%%%%%%%%%%%%%%%%%%
%
% \MIDADJUST Adjusts the position of the vertex midpoint and draws the
%            central four-gluon pattern.
\gdef\MIDADJUST(#1,#2){  
\ifnum\FLIPVERTEX=0
\global\advance\vertexmidx by #1   \global\advance\vertexmidy by #2
\setvertexC(0,\vertadj)[bl]        \setvertexC(0,\negvertadj)[tr]
\setvertexC(\vertadj,0)[tl]        \setvertexC(\negvertadj,0)[br]
\else  % \FLIPPED case
\global\particleadjustx=#1  \global\particleadjusty=#2  
\ifnum\LDIR=0 \global\multiply\particleadjustx by -1 \fi
\ifnum\LDIR=2 \global\multiply\particleadjusty by -1 \fi
\ifnum\LDIR=4 \global\multiply\particleadjustx by -1 \fi
\ifnum\LDIR=6 \global\multiply\particleadjusty by -1 \fi
\global\advance\vertexmidx by \particleadjustx  
\global\advance\vertexmidy by \particleadjusty
\setvertexC(0,\negvertadj)[tl]        \setvertexC(0,\vertadj)[br]
\setvertexC(\negvertadj,0)[tr]        \setvertexC(\vertadj,0)[bl]
\fi
}
\gdef\NFOURVERT#1{
  \ifnum\FLIPVERTEX=0 \setvertexB[\N\CURLY](\negvertadj,\vertadj)[#1]
  \else \setvertexB[\N\FLIPPEDCURLY](\vertadj,\vertadj)[#1] \fi}
\gdef\SFOURVERT#1{
  \ifnum\FLIPVERTEX=0 \setvertexB[\S\CURLY](\vertadj,\negvertadj)[#1]
  \else \setvertexB[\S\FLIPPEDCURLY](\negvertadj,\negvertadj)[#1] \fi}
\gdef\EFOURVERT#1{
  \ifnum\FLIPVERTEX=0 \setvertexB[\E\CURLY](\vertadj,\vertadj)[#1]
  \else \setvertexB[\E\FLIPPEDCURLY](\vertadj,\negvertadj)[#1] \fi}
\gdef\WFOURVERT#1{
  \ifnum\FLIPVERTEX=0 \setvertexB[\W\CURLY](\negvertadj,\negvertadj)[#1] 
  \else \setvertexB[\W\FLIPPEDCURLY](\negvertadj,\vertadj)[#1] \fi}
\gdef\diagFOURVERT#1[#2]{\setvertexB[#2\FLIPVERTEX](0,0)[#1]} %for\NE,\SE,\SW\NW
%
%%%%%%%%%%%%%%%%%%%%%%%%%%%%%%%%%%%%%%%%%%%%%%%%%%%%%%%%%%%%%%%%%%%%%%%%%%%%%%
%                                                                            %
%          Utility Routines Used in Setting Up General Vertices              %
%                                                                            %
%%%%%%%%%%%%%%%%%%%%%%%%%%%%%%%%%%%%%%%%%%%%%%%%%%%%%%%%%%%%%%%%%%%%%%%%%%%%%%
%
%
\gdef\setvertexA[#1#2]{
\global\savebox\vertexbox(0,0){
\begin{picture}(0,0)
%eg: \drawline\LINETYPE[\S\CENTRAL](0,0)[\unitboxnumber]
\global\adjx=#2
\ifnum\FLIPVERTEX=1 \global\advance\adjx by 1  
      \ifnum\VERTEXNUMBER=3 \global\FLIPVERTEX=0 \fi  
  \fi  %  Checks for flipped vertex and then Resets \FLIPVERTEX in 3-glue case.
\ifdim\STEMVERTEXONE=1pt\backstemmed \fi
\drawline\LINETYPE[#1\adjx](0,0)[\unitboxnumber]
\global\stemlength=\stemlengthcopy  % Re-set to use-defined value.
\ifdim\VERTEXLINKONE=1pt\gluonlink \fi % For linking external gluons to vertex
\ifdim\VERTEXCAPONE=1pt\gluoncap \fi % For capping external gluons to vertex
\global\vertexmidx=\particlebackx  \global\vertexmidy=\particlebacky  
\end{picture}
}%end of \savebox
\global\multiply\vertexmidx by -1  \global\multiply\vertexmidy by -1  
\global\advance\vertexmidx by \vertexonex  
\global\advance\vertexmidy by \vertexoney
%eg: \drawline\LINETYPE[\S\CENTRAL](\vertexmidx,\vertexmidy)[\unitboxnumber]
\ifdim\STEMVERTEXONE=1pt\backstemmed \fi
\drawline\LINETYPE[#1\adjx](\vertexmidx,\vertexmidy)[\unitboxnumber]
\global\stemlength=\stemlengthcopy  % Re-set to use-defined value.
\ifdim\VERTEXLINKONE=1pt\gluonlink \fi % For linking external gluons to vertex
\ifdim\VERTEXCAPONE=1pt\gluoncap \fi % For capping external gluons to vertex
} %End of \setvertexA
\gdef\setvertexB[#1#2](#3,#4)[#5]{
\global\adjx=\vertexmidx   \global\adjy=\vertexmidy
\global\advance\adjx by #3   \global\advance\adjy by #4
\VERTEXLINE=#5
\ifcase\VERTEXLINE\UNIMPERROR  %=0
\or \UNIMPERROR  %line 1                      %%%%%%%%%%%%%%%%%%%%%
\or \ifdim\STEMVERTEXTWO=1pt\backstemmed \fi  %  For stemming     %
\or \ifdim\STEMVERTEXTHREE=1pt\backstemmed\fi %  external gluons  %
\or \ifdim\STEMVERTEXFOUR=1pt\backstemmed\fi  %  to vertex        %
\else \UNIMPERROR                             %%%%%%%%%%%%%%%%%%%%%
\fi
\drawline\LINETYPE[#1#2](\adjx,\adjy)[\unitboxnumber]
\global\stemlength=\stemlengthcopy  % Re-set to use-defined value.
\ifcase\VERTEXLINE\UNIMPERROR  %=0
\or \UNIMPERROR  %line 1
\or \ifdim\VERTEXLINKTWO=1pt\gluonlink \fi % For linking external gluons to vert
    \ifdim\VERTEXCAPTWO=1pt\gluoncap \fi % For capping external gluons to vertex
    \global\vertextwox=\particlebackx  \global\vertextwoy=\particlebacky  
\or \ifdim\VERTEXLINKTHREE=1pt\gluonlink\fi %For linking external gluons to vert
    \ifdim\VERTEXCAPTHREE=1pt\gluoncap\fi %For capping external gluons to vertex
    \global\vertexthreex=\particlebackx  \global\vertexthreey=\particlebacky  
\or \ifdim\VERTEXLINKFOUR=1pt\gluonlink\fi % For linking external gluons to vert
    \ifdim\VERTEXCAPFOUR=1pt\gluoncap\fi % For capping external gluons to vertex
    \global\vertexfourx=\particlebackx  \global\vertexfoury=\particlebacky  
\else \UNIMPERROR  
\fi
}
\gdef\setvertexC(#1,#2)[#3#4]{
\global\adjx=\vertexmidx   \global\adjy=\vertexmidy
\global\advance\adjx by #1   \global\advance\adjy by #2
\absstemlength=1250  % Default for 3-gluon vertex; \absstemlength unused var.
\ifnum \VERTEXNUMBER=4 \absstemlength=\vertadj\double\absstemlength \fi
\ifnum\phantomswitch=0 
   \put(\adjx,\adjy) {\oval(\absstemlength,\absstemlength)[#3#4]}\fi
}
\gdef\adjvert{\negvertadj=\vertadj  \multiply\negvertadj by -1}
%
%%%%%%%%%%%%%%%%%%%%%%%%%%%%%%%%%%%%%%%%%%%%%%%%%%%%%%%%%%%%%%%%%%%%%%%%%%%%%
%                                                                           %
%                        LINKING GLUONS TO VERTICES                         %
%                                                                           %
%%%%%%%%%%%%%%%%%%%%%%%%%%%%%%%%%%%%%%%%%%%%%%%%%%%%%%%%%%%%%%%%%%%%%%%%%%%%%
%
\gdef\vertexlink#1{
\global\VERTEXLINE=#1
\ifcase\VERTEXLINE\UNIMPERROR
\or \global\VERTEXLINKONE=1pt    \or \global\VERTEXLINKTWO=1pt
\or \global\VERTEXLINKTHREE=1pt  \or \global\VERTEXLINKFOUR=1pt  
\else\UNIMPERROR\fi}
\gdef\vertexlinks{
\global\VERTEXLINKONE=1pt     \global\VERTEXLINKTWO=1pt
\global\VERTEXLINKTHREE=1pt  \global\VERTEXLINKFOUR=1pt  }
% counters cleared by \clearvertex now defined in STEM section below.
%
%
%%%%%%%%%%%%%%%%%%%%%%%%%%%%%%%%%%%%%%%%%%%%%%%%%%%%%%%%%%%%%%%%%%%%%%%%%%%%%
%                                                                           %
%                        CAPPING GLUONS ON VERTICES                         %
%                                                                           %
%%%%%%%%%%%%%%%%%%%%%%%%%%%%%%%%%%%%%%%%%%%%%%%%%%%%%%%%%%%%%%%%%%%%%%%%%%%%%
%
\gdef\vertexcap#1{
\global\VERTEXLINE=#1
\ifcase\VERTEXLINE\UNIMPERROR
\or \global\VERTEXCAPONE=1pt    \or \global\VERTEXCAPTWO=1pt
\or \global\VERTEXCAPTHREE=1pt  \or \global\VERTEXCAPFOUR=1pt  
\else\UNIMPERROR\fi}
\gdef\vertexcaps{
\global\VERTEXCAPONE=1pt     \global\VERTEXCAPTWO=1pt
\global\VERTEXCAPTHREE=1pt  \global\VERTEXCAPFOUR=1pt  }
% counters cleared by \clearvertex now defined in STEM section below.
%
%%%%%%%%%%%%%%%%%%%%%%%%%%%%%%%%%%%%%%%%%%%%%%%%%%%%%%%%%%%%%%%%%%%%%%%%%%%%%
%                                                                           %
%                         LINKING STEMS TO VERTICES                         %
%                                                                           %
%%%%%%%%%%%%%%%%%%%%%%%%%%%%%%%%%%%%%%%%%%%%%%%%%%%%%%%%%%%%%%%%%%%%%%%%%%%%%
%
%
\gdef\stemvertex#1{
\global\VERTEXLINE=#1
\ifcase\VERTEXLINE\UNIMPERROR
\or \global\STEMVERTEXONE=1pt    \or \global\STEMVERTEXTWO=1pt  
\or \global\STEMVERTEXTHREE=1pt  \or \global\STEMVERTEXFOUR=1pt  
\else\UNIMPERROR\fi}
\gdef\stemvertices{
\global\STEMVERTEXONE=1pt     \global\STEMVERTEXTWO=1pt  
\global\STEMVERTEXTHREE=1pt  \global\STEMVERTEXFOUR=1pt  }
\gdef\clearvertex{
\global\STEMVERTEXONE=0pt    \global\STEMVERTEXTWO=0pt  
\global\STEMVERTEXTHREE=0pt  \global\STEMVERTEXFOUR=0pt 
\global\VERTEXLINKONE=0pt    \global\VERTEXLINKTWO=0pt  
\global\VERTEXLINKTHREE=0pt  \global\VERTEXLINKFOUR=0pt  
\global\VERTEXCAPONE=0pt    \global\VERTEXCAPTWO=0pt  
\global\VERTEXCAPTHREE=0pt  \global\VERTEXCAPFOUR=0pt  
\global\stemlength=275}  % Re-set default.
\global\stemlengthcopy=\stemlength
\clearvertex  %set them to zero initially.
\global\stemlength=\stemlengthcopy
  \fi}
% RECURSIVE DEFINITIONS:
\gdef\drawvertex#1[#2#3](#4,#5)[#6]{\checkvertex\drawvertex#1[#2#3](#4,#5)[#6]}
\gdef\vertexcap#1{\checkvertex\vertexcap#1}
\gdef\vertexcaps{\checkvertex\vertexcaps}
\gdef\vertexlink#1{\checkvertex\vertexlink#1}
\gdef\vertexlinks{\checkvertex\vertexlinks}
\gdef\stemvertex#1{\checkvertex\stemvertex#1}
\gdef\stemvertices{\checkvertex\stemvertices}
\gdef\flipvertex{\checkvertex\flipvertex}
%
%%%%%%%%%%%%%%%%%%%%%%%%%%%%%%%%%%%%%%%%%%%%%%%%%%%%%%%%%%%%%%%%%%%%%%%%%%%%%%%
%                                                                             %
%                   ROUTINES FOR DRAWING ARROWS                               %
%                                                                             %
%%%%%%%%%%%%%%%%%%%%%%%%%%%%%%%%%%%%%%%%%%%%%%%%%%%%%%%%%%%%%%%%%%%%%%%%%%%%%%%
%
% SYNTAX:  \drawarrow[\NW\ATBASE](\pmidx,\pmidy)  etc.
\global\arrowlength=349  % Length of arrow
\gdef\drawarrow[#1#2](#3,#4){
\global\LDIR=#1
\SETDIR
\global\boxlengthx=#3  % Just a convenient variable name.  No significance.
\global\boxlengthy=#4  % The Arrow co-ordinates.
\ifdim#2=1pt  % CASE \ATBASE WHERE THE CO-ORDS ARE AT THE ARROWS BASE.
   %   #2 IS either \ATTIP or \ATBASE...Depending where it is to be positioned.
\adjx=\arrowlength      \adjy=\arrowlength
\multiply\adjx by \XDIR \multiply\adjy by \YDIR  % Set in \SETDIR
\slanttest(\adjx,\adjy)
\global\advance\boxlengthx by \adjx    \global\advance\boxlengthy by \adjy
\fi
\ifnum\phantomswitch=0\put(\boxlengthx,\boxlengthy){\vector(\XDIR,\YDIR){0}}\fi
}  % END OF \drawarrow.
%
%%%%%%%%%%%%%%%%%%%%%%%%%%%%%%%%%%%%%%%%%%%%%%%%%%%%%%%%%%%%%%%%%%%%%%%%%%%%%%%
%                                                                             %
%                     ROUTINES FOR DRAWING STEMS                              %
%                                                                             %
%%%%%%%%%%%%%%%%%%%%%%%%%%%%%%%%%%%%%%%%%%%%%%%%%%%%%%%%%%%%%%%%%%%%%%%%%%%%%%%
%
\gdef\SETFRONTSTEM{
\EITHERSTEM=\FRONTSTEM   \advance\EITHERSTEM by \BACKSTEM
\ifdim\EITHERSTEM>0pt
\global\stemlengthx=\stemlength   \global\stemlengthy=\stemlength   
\global\absstemlength=\stemlength   
\SETDIR
\gslanttest(\stemlengthx,\stemlengthy)
\gslanttest(\absstemlength,\REG)  % the \REG is to use up the parameter space.
\ifnum\XDIR=0 \stemlengthx=0 \fi
\ifnum\YDIR=0 \stemlengthy=0 \fi
\global\multiply\stemlengthx by \XDIR
\global\multiply\stemlengthy by \YDIR
\ifdim\FRONTSTEM=1pt 
\ifnum\phantomswitch=0
          \put(\pfrontx,\pfronty){\line(\XDIR,\YDIR){\absstemlength}}\fi
\global\advance\plengthx by \stemlengthx
\global\advance\plengthy by \stemlengthy
\global\advance\PFRONTx by \stemlengthx   
\global\advance\PFRONTy by \stemlengthy
\global\advance\pmidx by \stemlengthx
\global\advance\pmidy by \stemlengthy
\global\advance\pbackx by \stemlengthx
\global\advance\pbacky by \stemlengthy
% FOR STEMMED PHOTONS AND GLUONS, \photonfront,back(x,y) are for the
% photon proper (no stem) while \pbackx,y include the stems.
\ifnum\LTYPE=3
\global\photonfrontx=\PFRONTx  \global\photonfronty=\PFRONTy
\global\photonbackx=\pbackx    \global\photonbacky=\pbacky
\fi  % END LTYPE
\ifnum\LTYPE=4
\global\gluonfrontx=\PFRONTx  \global\gluonfronty=\PFRONTy
\global\gluonbackx=\pbackx    \global\gluonbacky=\pbacky
\fi  % END LTYPE
\fi  % END FRONTSTEM
\fi  % END EITHERSTEM
}    % end of \SETFRONTSTEM
\gdef\SETBACKSTEM{
\ifdim\BACKSTEM=1pt 
\ifnum\phantomswitch=0
       \put(\pbackx,\pbacky){\line(\XDIR,\YDIR){\absstemlength}}\fi
\global\advance\plengthx by \stemlengthx
\global\advance\plengthy by \stemlengthy
\global\advance\pbackx by \stemlengthx
\global\advance\pbacky by \stemlengthy
\fi  % END BACKSTEM
\global\stemlength=275  \FRONTSTEM=0pt  \BACKSTEM=0pt % Reset default switches.
}    % END OF \SETBACKSTEM 
%%%%%%%%%%%%%%%%%%%%%%%%%%%%%%%%%%%%%%%%%%%%%%%%%%%%%%%%%%%%%%%%%%%%%%%%%%%%%
%                              LOOPS                                        %
%%%%%%%%%%%%%%%%%%%%%%%%%%%%%%%%%%%%%%%%%%%%%%%%%%%%%%%%%%%%%%%%%%%%%%%%%%%%%
\gdef\drawloop#1[#2#3](#4,#5){  %RECURSIVE.  
%                           LOOPS(1).TEX
% DRAWS LOOPS FOR FEYNMAN(32+).TEX:  GLUON LOOPS ONLY
\global\newcount\loopfrontx    \global\newcount\loopfronty
\global\newcount\loopbackx    \global\newcount\loopbacky
\global\newcount\loopmidx    \global\newcount\loopmidy
\global\newdimen\CENTRALLOOP
\gdef\drawloop#1[#2#3](#4,#5){
\global\CENTRALLOOP=0pt  % non-central is default
\global\LINETYPE=#1  
\ifnum\LTYPE=\gluon\relax\else\UNIMPERROR\LTYPE=1\message{Reverting to Gluons}
\fi
\global\LINEDIRECTION=#2  %initial loop direction
\global\fourthlineadjx=#3 %number of eighths of loop
\ifnum\fourthlineadjx=0 % (x,y) now midpoint.
  \global\CENTRALLOOP=1pt  % non-central is default
  \global\fourthlineadjx=8
  \global\LDIR=0
\fi   
\global\fourthlineadjy=\fourthlineadjx  % a conveniently unused variable.
\global\advance\fourthlineadjy by -4
\global\loopfrontx=#4   \global\loopfronty=#5
\ifdim\CENTRALLOOP=1pt
  \global\advance\loopfrontx by -2413  \global\advance\loopfronty by -425
\fi                          % diameter of gluonloop is 4825 to 4830 cpt.
\global\unitboxnumber=1  % default; \gluoncase
\ifnum\LINETYPE=\photon \unitboxnumber=2 \fi
\checkdir
\drawline\LINETYPE[\LDIR\LCONFIG](\loopfrontx,\loopfronty)[\unitboxnumber]
\DRAWLOOP
\ifnum\fourthlineadjy>-1 % at least 1/2 a loop
\global\loopmidx=\loopfrontx   \global\loopmidy=\loopfronty
\global\advance\loopmidx by \loopbackx  \global\advance\loopmidy by \loopbacky
\divide\loopmidx  by 2 \divide\loopmidy by 2  % midpoints of loop
\ifdim\CENTRALLOOP=1pt
  \global\advance\loopfrontx by 200    \global\advance\loopfronty by 425
  \global\advance\loopbackx by -200    \global\advance\loopbacky by -425
\fi
\fi % end of \ifnum\fourthlineadjy>-1 
}
\gdef\DRAWLOOP{
\global\advance\fourthlineadjx by -1
\ifnum\fourthlineadjx=0\relax  % finished!
\else
\ifnum\fourthlineadjx=\fourthlineadjy % opposite side of loop
   \global\loopbackx=\pbackx   \global\loopbacky=\pbacky
\fi
\global\advance\LDIR by 1
\moduloeight\LDIR
\checkdir
\drawline\LINETYPE[\LDIR\LCONFIG](\pbackx,\pbacky)[\unitboxnumber]
\fi % end \ifnum\fourthlineadjx
\ifnum\fourthlineadjx>1 \DRAWLOOP  \fi  % recursive
}
\gdef\checkdir{
\ifnum\LTYPE=\gluon
\ifodd\LDIR \global\LCONFIG=0 \else \global\LCONFIG=2 \fi
\fi %end of \gluon
}
  % contains loops definitions
\drawloop#1[#2#3](#4,#5)}
%%%%%%%%%%%%%%%%%%%%%%%%%%%%%%%%%%%%%%%%%%%%%%%%%%%%%%%%%%%%%%%%%%%%%%%%%%%%%
\Feynmanlength  % Set length scale to centipoints.

\newcommand{\ecm}{\mathrm{\sqrt{s}}}
\newcommand{\eplus}{\mathrm{e}^+}
\newcommand{\eminus}{\mathrm{e}^-}
\newcommand{\epem}{\eplus\eminus}
\newcommand{\lplm}{l^+l^-}
\newcommand{\mpmm}{\mu^+\mu^-}
\newcommand{\tptm}{\tau^+\tau^-}
\newcommand{\eeX}{\epem X}
\newcommand{\mmX}{\mpmm X}
\newcommand{\ZH}{\mathrm{ZH}}
\newcommand{\Zzero}{\mathrm{Z}}
\newcommand{\Zo}{\mathrm{Z^0}}
\newcommand{\Wboson}{\mathrm{W}}
\newcommand{\WpWm}{\Wboson^+\Wboson^-}
\newcommand{\Higgs}{\mathrm{H}}
\newcommand{\qq}{\mathrm{q}\overline{\mathrm{q}}}
\newcommand{\uubar}{\mathrm{u}\overline{\mathrm{u}}}
\newcommand{\ddbar}{\mathrm{d}\overline{\mathrm{d}}}
\newcommand{\ssbar}{\mathrm{s}\overline{\mathrm{s}}}
\newcommand{\GammaZ}{\Gamma_\Zzero}
\newcommand{\GammaW}{\Gamma_\Wboson}
\newcommand{\mZ}{\mathrm{M_\Zzero}}
\newcommand{\mW}{\mathrm{M_\Wboson}}
\newcommand{\mH}{\mathrm{M_\Higgs}}
\newcommand{\Mdl}{\mathrm{M_{dl}}}
\newcommand{\Mrecoil}{\mathrm{M_{recoil}}}
\newcommand{\Ptdl}{\mathrm{P_{Tdl}}}
\newcommand{\Pdl}{\mathbf{P_{dl}}}
\newcommand{\roots}{\mathrm{\sqrt{s}}}
\newcommand{\invfb}{\mathrm{fb^{-1}}}
\newcommand{\GeV}{\mathrm{GeV}}
\newcommand{\MeV}{\mathrm{MeV}}
\newcommand{\polRL}{\mathrm{e^-_R e^+_L}}
\newcommand{\polLR}{\mathrm{e^-_L e^+_R}}
\newcommand{\rmsn}{\mathrm{rms}_{90}}
\newcommand{\cthadl}{\mathrm{cos\theta_{dl}}}
\newcommand{\Ptbal}{\mathrm{\Delta P_{Tbal.}}}
\newcommand{\Ptgamma}{\mathrm{P_{T\gamma}}}
\newcommand{\Pt}{\mathrm{P_{T}}}
\newcommand{\Naddtks}{\mathrm{N_{add.TK}}}
\newcommand{\dthattk}{\mathrm{|\Delta \theta_{2tk}|}}
\newcommand{\dthamin}{\mathrm{|\Delta \theta_{min}|}}
\newcommand{\cthamiss}{\mathrm{|cos\theta_{missing}|}}

\hyphenation{brems-strah-lung}

%\begin{samepage}
%\begin{center}
%\baselineskip 9pt plus 2pt
%Universit\'e de Paris-Sud 
%\hfill LAL 00-00
%\end{center}
%\fancypagestyle{plain}{My fancy style}
%\markboth{the left head}{right head} 
%\markright{the right head}
%\thispagestyle{myheadings}
\title{
%%%%   Paper title goes here  %%%%%%%%%%%%%%
$\Higgs\Zzero$ Recoil Mass and Cross Section Analysis in ILD} %% 
%***********************************************************************
% AUTHORS INFORMATION AREA
%***********************************************************************
\author{ 
H. Li$^{1,4}$, K. Ito$^2$, R. P\"oschl$^1$, F. Richard$^1$, \\
M. Ruan$^{1,3}$, Y. Takubo$^2$, H. Yamamoto$^2$\\
and the ILD Design Study Group
%M. Ruan$^{2,3}$, Y. Takubo$^1$, H. Yamamoto$^1$, Z. Zhang$^2$
% Optional short acknowledgment: remove next line if non-needed
%\thanks{This is an optional funding source acknowledgment.}
% DO NOT MODIFY THE FOLLOWING '\vspace' ARGUMENT
%\begin{footnotesize}
\vspace{.3cm}\\
% Addresses and institutions (remove "1- " in case of a single institution)
%\begin{footnotesize} 
{\footnotesize {\it 1- Laboratoire de l'Acc\'el\'erateur Lin\'eaire (LAL) -CNRS/IN2P3;
B.P. 34, 91898 Orsay, France}}
%\end{footnotesize}
%% Remove the next three lines in case of a single institution
\vspace{.1cm}\\
{\footnotesize {\it 2- Tohoku University, Department of Physics;
Aoba District, Sendai, Miyagi 980-8578, Japan}}
\vspace{.1cm}\\
{\footnotesize {\it 3- Laboratoire Leprince-Ringuet  (LLR), \'Ecole Polytechnique - CNRS/IN2P3}}\\
{\footnotesize {\it Route de Saclay, 91128 Palaiseau Cedex, France}}\\
{\footnotesize {\it 4- Laboratoire de Physique Subatomique et de Cosmologie - Universit\'{e} Joseph Fourier Grenoble 1}}\\
{\footnotesize {\it CNRS/IN2P3 - Institut Polytechnique de Grenoble}}
{\footnotesize {\it 53, rue des Martyrs, 38026 Grenoble CEDEX, France}}
%\end{footnotesize}
}
%\author{K. Ito, H. Li, R. P\"oschl, F. Richard, \\
%M. Ruan, Y. Takubo, H. Yamamoto, Z. Zhang}
%%***********************************************************************
% END OF AUTHORS INFORMATION AREA
%***********************************************************************
\date{}
\maketitle
%\markboth{both right head} 
%\markright{another  right head}
%\pagestyle{myheadings}
\thispagestyle{fancy}
%\rightline{[TODAY'S DATE]}
%\end{samepage}
\begin{abstract}
This note describes the details of a simulation study of the Higgs boson production for processes 
in which the Higgs is produced together with a well measurable di-lepton system using the proposal of the ILD detector for its Letter of Intent~\cite{ild09}.  The analysis is optimised for the measurement of the Higgs-strahlung process, i.e. $\epem\rightarrow\Higgs\Zzero$. The cross section can be determined with a 
precision of 2-3\% and by combining the decay channels a precision of $\sim$30\,MeV is obtained for the mass of the Higgs boson.  The background can be largely reduced and the analysis exhibits a sensitivity to the configuration of the accelerator. 
\end{abstract}

\section{Introduction}
The understanding of electro-weak symmetry breaking is intimately coupled to the study of the Higgs boson.
It arises as a consequence of the observation of massive gauge bosons which can be generated 
by the spontaneous breaking of the $\mathrm{SU(2)\times U(1)}$ symmetry of the electroweak Lagrangian.

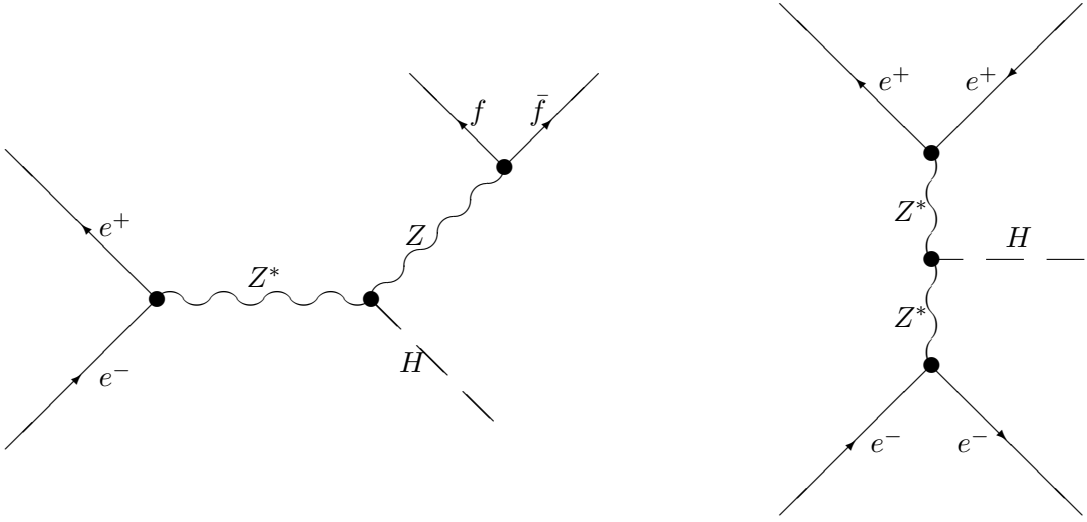
\begin{figure}
\begin{picture}(10000,20000)(0,0)
%Draw the intermediate Vektor Boson
\drawline\photon[\E\REG](6000,10000)[8]  % Even number of half-wiggles.
\global\advance\pmidx by -650
\global\advance\pmidy by 450
\put(\pmidx,\pmidy){$Z^{\ast}$}
\global\advance\pmidx by -1600
\global\advance\pmidy by -4500
%\put(\pmidx,\pmidy){$s$-channel}
\put(\photonbackx,\photonbacky){\circle*{600}}
\put(\photonfrontx,\photonbacky){\circle*{600}}
%Draw the initial state leptons
% Make the fermions the same length as the photon.
\drawline\fermion[\NW\REG](\photonfrontx,\photonfronty)[\photonlengthx]
\drawarrow[\NW\ATTIP](\pmidx, \pmidy)
\global\advance\pmidx by 650
\global\advance\pmidy by -450
\put(\pmidx,\pmidy){$e^+$}
\drawline\fermion[\SW\REG](\photonfrontx,\photonfronty)[\photonlengthx]
\drawarrow[\NE\ATTIP](\pmidx, \pmidy)
\global\advance\pmidx by 650
\global\advance\pmidy by -450
\put(\pmidx,\pmidy){$e^-$}
%Draw the outgoing particles
\global\divide\photonlengthx by 5
\global\seglength=\photonlengthx
\drawline\scalar[\SE\REG](\photonbackx,\photonbacky)[3]
\global\advance\pmidx by  -1250
\global\advance\pmidy by -450
\put(\pmidx,\pmidy){$H$}
\drawline\photon[\NE\REG](\photonbackx,\photonbacky)[8]
\put(\photonbackx,\photonbacky){\circle*{600}}
\global\advance\pmidx by  -1250
\global\advance\pmidy by -450
\put(\pmidx,\pmidy){$Z$}
\global\divide\fermionlength by 4 
\drawline\fermion[\NW\REG](\photonbackx,\photonbacky)[\photonlengthx]
\drawarrow[\NW\ATTIP](\pmidx, \pmidy)
\global\advance\pmidx by  450
\put(\pmidx,\pmidy){$f$}
\drawline\fermion[\NE\REG](\photonbackx,\photonbacky)[\photonlengthx]
\drawarrow[\NE\ATTIP](\pmidx, \pmidy)
\global\advance\pmidx by  -850
\put(\pmidx,\pmidy){$\bar{f}$}
%The ZZ fusion 
\drawline\photon[\N\REG](35000,7500)[8]  % Even number of half-wiggles.
\put(\pmidx,\pmidy){\circle*{600}}
\global\advance\pmidx by -1650
%\global\advance\pmidy by -1650
%\put(\pmidx,\pmidy){$Z^{\star}$}
%\global\advance\pmidy by -10500
\put(\photonbackx,\photonbacky){\circle*{600}}
\put(\photonfrontx,\photonfronty){\circle*{600}}
%Draw the outgoing higgs
%\global\divide\photonlengthx by 5
%\global\seglength=\photonlengthx
\global\advance\pmidx by 1450
\drawline\scalar[\E\REG](\pmidx,\pmidy)[3]
%\global\advance\pmidx by  -1250
\global\advance\pmidy by 450
\put(\pmidx,\pmidy){$H$}
\global\advance\pmidx by  -4200
\global\advance\pmidy by   1000
\put(\pmidx,\pmidy){$Z^{\ast}$}
\global\advance\pmidy by  -4000
\put(\pmidx,\pmidy){$Z^{\ast}$}
%\put(\pmidx,\pmidy){$t$-channel}
%Draw the initial state leptons
% Make the fermions the same length as the photon.
\drawline\fermion[\NW\REG](\photonbackx,\photonbacky)[\photonlengthy]
\drawarrow[\NW\ATTIP](\pmidx, \pmidy)
\global\advance\pmidx by 850
\global\advance\pmidy by -450
\put(\pmidx,\pmidy){$e^+$}
\drawline\fermion[\NE\REG](\photonbackx,\photonbacky)[\photonlengthy]
\drawarrow[\SW\ATTIP](\pmidx, \pmidy)
\global\advance\pmidx by -1550
\global\advance\pmidy by -450
\put(\pmidx,\pmidy){$e^+$}
\drawline\fermion[\SW\REG](\photonfrontx,\photonfronty)[\photonlengthy]
\drawarrow[\NE\ATTIP](\pmidx, \pmidy)
\global\advance\pmidx by 550
\global\advance\pmidy by -450
\put(\pmidx,\pmidy){$e^-$}
\drawline\fermion[\SE\REG](\photonfrontx,\photonfronty)[\photonlengthy]
\drawarrow[\SE\ATTIP](\pmidx, \pmidy)
\global\advance\pmidx by -1850
\global\advance\pmidy by -450
\put(\pmidx,\pmidy){$e^-$}
\end{picture}
\caption{Higgs boson production via the Higgs-strahlung process (left) and $\Zzero\Zzero$ fusion (left) and associated final state fermions with opposite charge at $\epem$-colliders.}
\label{fig:higgs-ff}
\end{figure}

If existing, a Higgs boson with a mass $\mH$ of 120\,GeV as favoured by recent analyses of electro-weak data~\cite{lepwww} will be discovered at the LHC or even at the TEVATRON. The ILC will allow for the detailed investigation of the nature of the Higgs boson as has been demonstrated in~\cite{Higgswg, martin, manqi} and references therein. The relevant processes for the present study are the recoil reaction $\epem\rightarrow\Higgs\Zzero\rightarrow\Higgs f\bar{f}$ (where $f$=leptons and quarks), also called \emph{Higgs-strahlung}, or  $\epem\rightarrow\Higgs\epem$, also called $\Zzero\Zzero$ fusion. The Feynman diagrams are shown in Figure~\ref{fig:higgs-ff}. Please note that the cross section of the Higgs-strahlung dominates largely over that of the $\Zzero\Zzero$ fusion. Hence, the analysis will be optimised for the measurement of the Higgs-strahlung process. 

By detecting the decay products of the $\Zzero$ boson, the introduced processes and in particular the Higgs-strahlung process allow for the search of Higgs signals without any further assumption on its decay modes.  In contrast to the LHC, the initial state is very well known at the ILC. These two items together allow for an unbiased
search for the Higgs boson also called {\it Model Independent Analysis} which is only possible at a 
Lepton Collider such as the ILC. The presumably cleanest way to study the Higgs is given by 
the process $\epem\rightarrow\Higgs\Zzero$ and the subsequent decay $\Zzero\rightarrow\mpmm$ or $\Zzero\rightarrow\epem$, i.e. searching for the decay leptons of the well known $\Zzero$ boson. 
These channels, also named $\mu\mu X$-channel  and  $ee X$-channel hereafter, will be examined in detail in this note for a centre-of-mass energy of $\sqrt{s}={\rm 250\,GeV}$ as proposed in the definition of the benchmark scenarios for the {\it Letter of Intent Studies for ILC detectors}~\cite{ilcbench}.

\section{ILD Detector}
A detailed description of the current model of the ILD detector can be found elsewhere~\cite{ild09}. The $z$-axis of the right handed co-ordinate system is given by the direction of the incoming electron beam. Polar angles given in this note are defined with respect to this axis.
The most important sub-detectors for this study  are described in the following.
\begin{itemize}
\item The vertex detector consists of three double layers of silicon extending between 16\,mm and 60\,mm in radius and between 62.5\,mm and 125\,mm in $z$ direction. It is designed for an impact parameter resolution of $\sigma_{r\phi}=\sigma_{rz}=5\oplus 10/(p {\rm sin}^{\frac{3}{2}}\theta)\,\mathrm{\mu m}$.
\item The measurement of charged tracks is supported by an inner Silicon Tracker (SIT) in the central region and by a set of silicon disks in forward 
direction, i.e. towards large absolute values of $cos\theta$.
\item The ILD detector contains a large Time Projection Chamber (TPC) with an inner sensitive radius of 395\,mm and an outer sensitive radius of 1743\,mm.  The half length in $z$ is 2250\,mm. Recent simulation studies confirm that the momentum of charged particle tracks can be measured to a precision of $\delta(1/P_T)\sim 2\times10^{-5}\,\GeV^{-1}$. Here $P_T$ denotes the transverse component of the three  momentum $P$ of the particles.
\item The electromagnetic calorimeter is a SiW sampling calorimeter. Its longitudinal depths of 24\,${\rm X_{0}}$ allows
for the complete absorption of photons with energies of up to 50 GeV as relevant for the studies here. The simulated energy
resolution of the electromagnetic calorimeter is  $\frac{\Delta E}{E}=15\%/\sqrt{E\,\mathrm{[GeV]}}$
\item The hadronic calorimeter surrounds the electromagnetic calorimeter and comprises 4.5 interaction length $\lambda_I$.  Two proposals exist for the hadronic calorimeter. A digital variant consisting of steel absorbers and gas RPC chambers with a pixel size of $1 \times 1$\,${\rm cm^2}$ as active material. The second one features scintillating tiles with size of $3 \times 3$\,${\rm cm^2}$ as active material. The latter option is employed in the present work.
%\item The BeamCal calorimeter is mainly used for beam diagnostics and covers polar angles between 5\,mrad and 40\,mrad. It features
%tungsten as absorber and radiation hard material such as diamond as sensitive medium. In the present note it is employed to
%veto events created by $\gamma\gamma$ scattering, see APPENDIX B.  

\end{itemize}

In the current design of the ILC the initial beams enter with a crossing angle of 14\,mrad. This crossing angle is not taken into account in the present study.

\section{Event Generation, Detector Simulation and Event Reconstruction}\label{sec:gensimrec}

All data analysed for this note have been centrally produced by the ILD Group in autumn/winter 2008/09 based on generator files known as \emph{SLAC samples}. For the event generation the version $1.40$ of the event generator WHIZARD~\cite{whizard} has been used. The incoming beams have been simulated with the GUINEA-PIG package~\cite{guinea}. The energy of the incoming beams is smeared with an energy spread of 0.28\% for the electron beam and with 0.18\% for the positron beam. In addition the energy is modulated by beamstrahlung. The impact on the precision of the physics result of this uncertainly will be discussed below. The generated signal and background samples are given in the Table~\ref{tab:reaction_lr} for the beam polarisation mode\\
\begin{center} $\eminus_L\eplus_R$: $P_{\eminus}= -80\%$ and $P_{\eplus}=+30\%$\\ \end{center} 
and in Table~\ref{tab:reaction_rl} for the beam polarisation mode\\
\begin{center} $\eminus_R\eplus_L$: $P_{\eminus}= +80\%$ and $P_{\eplus}= -30\%$.\\\end{center}
 The initially generated samples of the signal are combined such that they yield $\mathcal{L} = 10\,{\rm ab^{-1}}$ in each of the polarisation modes. For background samples the integrated luminosity is mostly larger than 250\,${\rm fb^{-1}}$. Where it is smaller, it is still provided that the samples contain considerable statistics.  Note, that in Tables~\ref{tab:reaction_lr} and~\ref{tab:reaction_rl} the background samples have been grouped by the resulting final state\footnote{Please note, that due to a bug in the luminosity spectrum in the initial generation, the background has been re-weighted to the correct spectrum according to~\cite{mikael}.}. 
%The tables with a breakdown of the contributing sub-processes are given in Tables~A-1 and~A-2 in APPENDIX A.

\begin{table}[htb]
\centering
\begin{minipage}[b]{0.5\textwidth}
%muon
\centering
\begin{tabular}{|c|c|}

\hline
Process &  Cross-Section \\

\hline
\boldmath{$\mu\mu X$} & \bf{11.67 fb} \\
 \hline
$\mu\mu$ & 10.44 pb (84.86 fb) \\
 \hline
$\tau\tau$ & 6213.22 fb \\
 \hline
$\mu\mu\nu\nu$ & 481.68 fb \\
 \hline
$\mu\mu ff$ & 1196.79 fb \\
 \hline

\end{tabular}
\end{minipage}%
\begin{minipage}[b]{0.5\textwidth}
%electron
\centering
\begin{tabular}{|c|c|}
\hline
Process & Cross-Section \\
\hline
\boldmath{$ee X$} & \bf{12.55 fb} \\
 \hline
$ee$ & 17.30 nb (357.14 fb) \\
 \hline
$\tau\tau$ & 6213.22 fb \\
 \hline
$ee\nu\nu$ & 648.51 fb \\
 \hline
$ee ff$ & 4250.58 fb \\
 \hline
\end{tabular}

\end{minipage}
\caption{Processes and cross sections for polarisation mode $\polLR$. The signal is indicated by bold face letters; the cross-section in the parentheses of $\epem$ and $\mpmm$ 
are that after Pre-Cuts, see Table \ref{tab:pre-cuts} for the Pre-Cuts definition.}
\label{tab:reaction_lr}
\end{table}

\begin{table}[htb]
\centering
\begin{minipage}[b]{0.5\textwidth}
%muon
\centering
\begin{tabular}{|c|c|}
\hline
Process & Cross-Section \\
\hline
\boldmath{$\mu\mu X$} & \bf{7.87 fb} \\
 \hline
$\mu\mu$ & 8.12 pb (58.26 fb) \\
 \hline
$\tau\tau$ & 4850.05 fb \\
 \hline
$\mu\mu\nu\nu$ & 52.37 fb \\
 \hline
$\mu\mu ff$ & 1130.01 fb \\
 \hline

\end{tabular}
\end{minipage}%
\begin{minipage}[b]{0.5\textwidth}
%electron
\centering
\begin{tabular}{|c|c|}
\hline
Process & Cross-Section \\
\hline
\boldmath{$ee X$} & \bf{8.43 fb} \\
 \hline
$ee$ & 17.30 nb (335.47 fb) \\
 \hline
$\tau\tau$ & 4814.46 fb \\
 \hline
$ee\nu\nu$ & 107.88 fb \\
 \hline
$ee ff$ & 4135.97 rb \\
 \hline
\end{tabular}

\end{minipage}
\caption{Processes and cross sections for polarisation mode $\polRL$. The signal is indicated by bold face letters; the cross-section in the parentheses of $\epem$ and $\mpmm$ 
are that after Pre-Cuts, see Table \ref{tab:pre-cuts} for the Pre-Cuts definition.}
\label{tab:reaction_rl}
\end{table}

Due to the large cross section of the Bhabha Scattering, i.e. $\epem\rightarrow\epem$ and muon pair production, i.e.
$\epem\rightarrow\mpmm$, pre-cuts have been applied in order to reduce the simulation time. These cuts are given
in Table~\ref{tab:pre-cuts} and will be later on referred to as \emph{Pre-cuts}.

\begin{table}[!tbp]
\centering
\begin{tabular}{c|c}
$\epem\rightarrow\epem$ & $\epem\rightarrow\mpmm$ \\
\hline
$|cos\theta_{e^+/e^-}|<0.95$ & \\
$M_{\epem} \in (71.18, 111.18)\ \GeV$ & $M_{\mpmm} \in (71.18, 111.18)\ \GeV$ \\
$P_{T\epem} > 10\ \GeV$ & $P_{T\mpmm} > 10\ \GeV$\\
$\Mrecoil \in (105, 165)\ GeV$ & $\Mrecoil \in (105, 165)\ GeV$ \\
\end{tabular}
\caption{List of Pre-cuts applied to Bhabha scattering and muon pair production in order to reduce the  simulation time.}
\label{tab:pre-cuts}
\end{table}

Here, $M_{\epem}$ and $M_{\mpmm}$, respectively, are the invariant mass of the {\it di-lepton} system for signal events, while $P_{T\epem}$ and $P_{T\mpmm}$ denote the transverse momentum calculated from the vectorial sum of the two leptons.

The generated events are subject to a detailed detector simulation. The simulation is performed with the MOKKA~\cite{ilcsoft} software package which provides the geometry interface to the GEANT4~\cite{geant4} simulation toolkit. The event reconstruction is performed using the MarlinReco framework. For this study the versions as contained in the Software Package {\it ILCSoft v01-06}~\cite{ilcsoft} are employed.
The main output of this framework for the present study are the so-called {\it LDC Tracks} which is a combination of track segments measured in the vertex detector, the Silicon Inner Tracker and the TPC or Forward Tracking Disks. Their momenta are compared with the energy of calorimeter clusters composed from signals in the electromagnetic and hadronic calorimeter for the particle identification.

\section{Signal Selection and Background Rejection}\label{sec:sigbak}
 
The signal is selected by identifying two well measured leptons in the final state which yield the
mass of the $\Zzero$ boson. The mass $M_{recoil}$ of the system recoiling to the di-lepton system is computed using the expression: 
\begin{displaymath}
M^2_{recoil}=s+M^2_{\Zzero}-2E_{\Zzero}\roots
\end{displaymath}
Here $M_{Z}$ denote the mass of the $\Zzero$ boson as reconstructed from the di-lepton system and $E_Z$ its corresponding energy.
A number of background processes have  to be suppressed. Techniques of background suppression similar to those presented in this note were already introduced in~\cite{jcb}. 
This section firstly defines the criteria of lepton identification and then addresses the means to suppress the background. This will be done under two assumptions: 1) model independent 2) model dependent, i.e. assuming a Higgs boson as predicted by the Standard Model. The latter excludes decay modes in which the Higgs boson decays invisibly.

\subsection{Lepton Identification} 

\begin{figure}[ht]
\begin{center}
\includegraphics[width=0.9\textwidth]{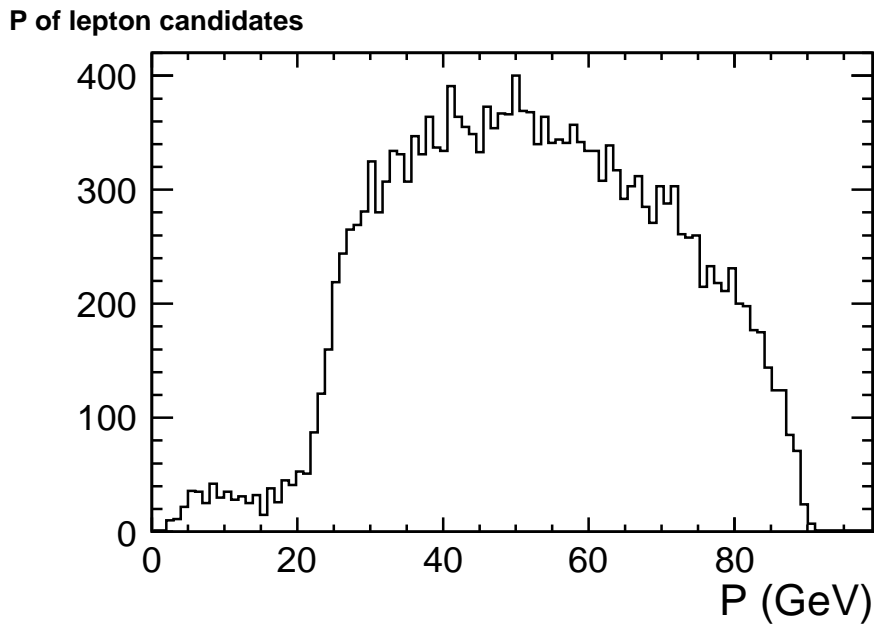}
\caption{Momentum range of the final state leptons as produced in $\Zzero\rightarrow \mpmm$ decays from $\epem\rightarrow\Higgs\Zzero$ events at $\roots =250\,\GeV$.}
\label{fig:mom}
\end{center}
\end{figure}

\begin{figure}[ht]
\begin{center}
\includegraphics[width=0.9\textwidth]{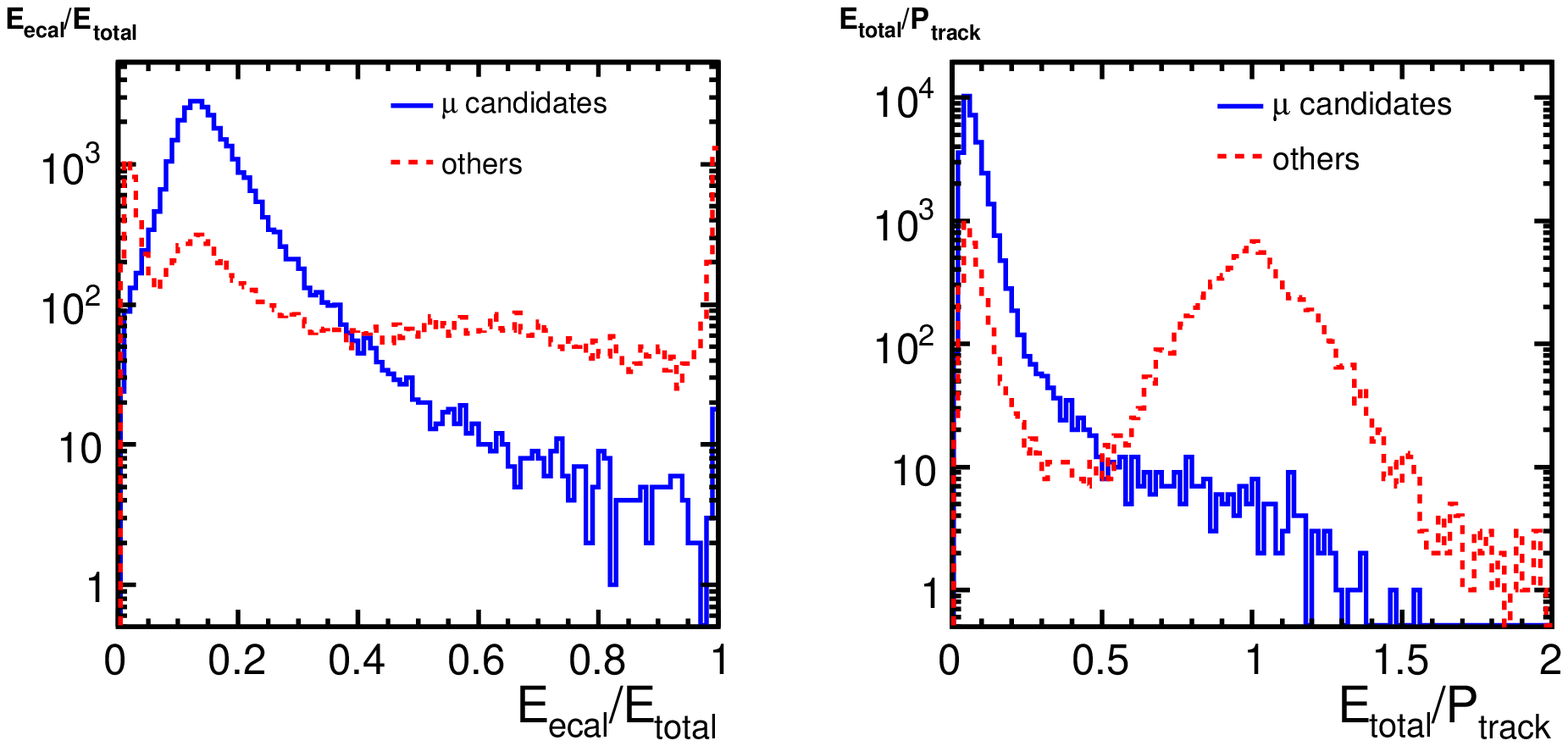}
\includegraphics[width=0.9\textwidth]{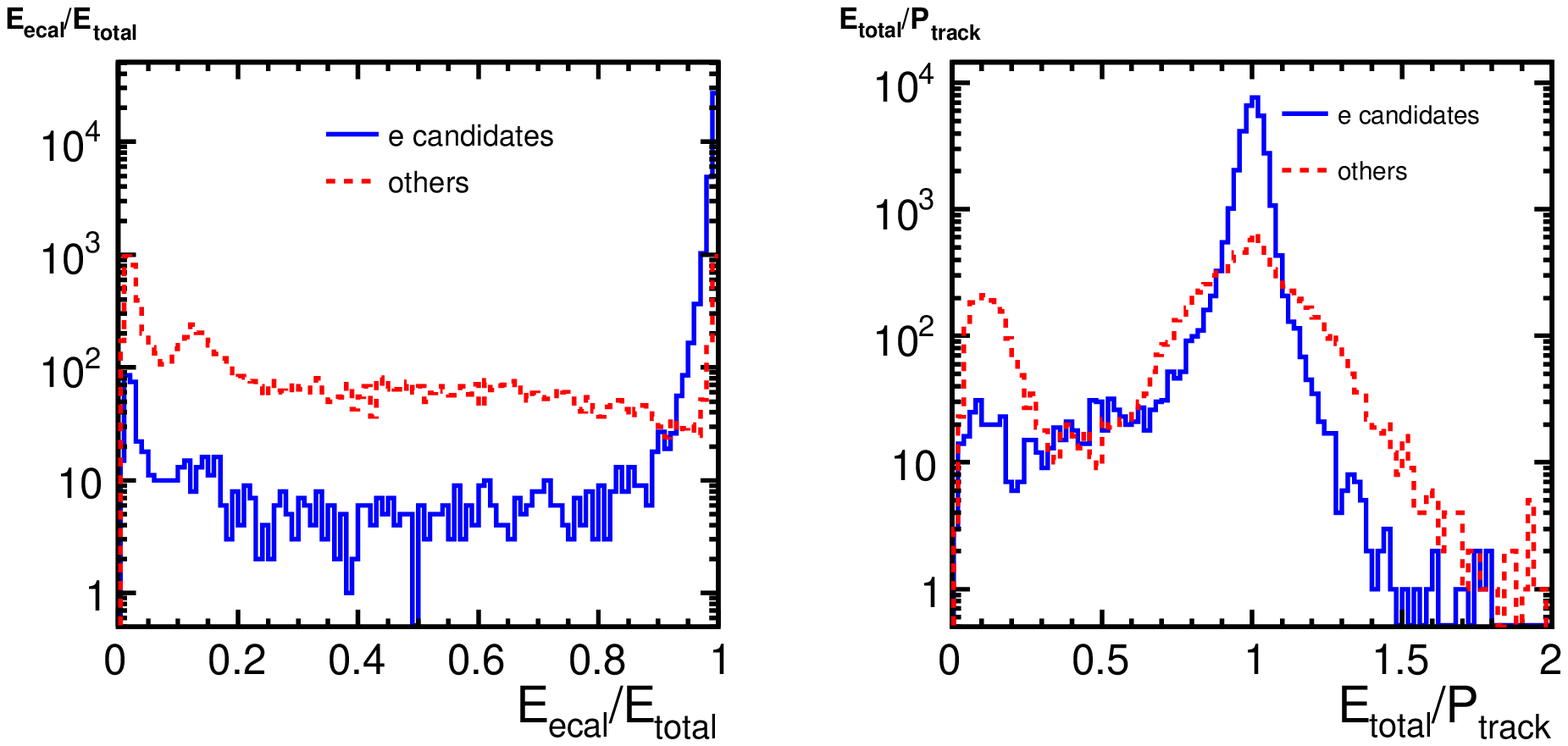}
\caption{Distributions of the variables for lepton identification of lepton candidates and other particles with $P > 15\,\GeV$.}
\label{fig:lid}
\end{center}
\end{figure}

The task is to identify the muons and electrons produced in the decay of the $\Zzero$ boson. In a first step, the energy deposition in the electromagnetic calorimeter ($E_{ECAL}$), the total calorimetric energy $E_{total}$ and the measured track momentum $P_{track}$ are compared accordingly for each final state particle. The lepton identification is mainly based on the 
assumption that an electron deposits all its energy in the electromagnetic calorimeter while a muon in the considered energy
range, see Figure~\ref{fig:mom},  passes both the electromagnetic and the hadronic calorimeter as a minimal ionising particle.
The observables and cut values are summarised in Table~\ref{tab:lepton-id}. The motivation of the cut values can be inferred from Figure~\ref{fig:lid} where the spectra for the corresponding lepton type in the relevant momentum range $P > 15\,\GeV$ are compared with those from other particles are displayed.

\begin{table}[htb]
\centering
\begin{tabular}{|c|c|c|}
\hline
&$\mu$-Identification & $e$-Identification \\
$E_{ECAL}/E_{total}$ & $< 0.5$ & $> 0.6$\\
$E_{total}/P_{track}$ & $< 0.3$ & $> 0.9$\\
\hline
\end{tabular}\label{tab:lepton-id}
\end{table}

The criteria to estimate the quality of the lepton identification and hence the signal selection are the \emph{Efficiency} and \emph{Purity}. These are defined as follows:
\begin{displaymath}
{\rm Efficiency=\frac{N_{true \cap iden}}{N_{true}}}
\end{displaymath}
\begin{displaymath}
{\rm Purity=\frac{N_{true \cap iden}}{N_{iden}}}
\end{displaymath}
Here $N_{true}$ defines the generated number of the corresponding lepton type and $N_{iden}$ defines the reconstructed number of   the corresponding lepton type according to the selection criteria. For electrons and muons with $P>15\,\GeV$ in the signal samples the obtained values are listed in Table~\ref{tab:lidsig}. 

\begin{table}[!h]
\centering 

\begin{tabular}{|l|l|l|}
\hline
& $\mu\ ID\ in\ \mu\mu X$ & $e\ ID\ in\ eeX$  \\
\hline
$N_{true}$   & 31833 & 34301  \\
\hline
$N_{true\cap iden}$   & 31063 & 33017  \\
\hline
$N_{iden}$ & 33986 & 34346  \\
\hline
Efficiency & 97.6\% & 96.3\%  \\
\hline
Purity & 91.4\% & 96.1\% \\
\hline
\end{tabular}
\caption{Lepton ID Efficiency and Purity for reconstructed particles with $P>15\,\GeV$.}
\label{tab:lidsig}

\end{table}

The efficiencies and purities are well above 95\% except for the purity of the muon identification. This is caused by final state
charged pions which pass the detector as minimal ionising particles and which are  indistinguishable from muons with the applied selection criteria. This deficiency is partially balanced by the fact that two leptons of the same type with opposite charge are required for the 
reconstruction of the $\Zzero$ boson and that they should yield the mass of the $\Zzero$ boson. Indeed, using the above selection cuts, the efficiency to identify both leptons from the $\Zzero$ boson decay is 95.4\% for the case $\Zzero\rightarrow\mu\mu$ and 98.8\% for the case $\Zzero\rightarrow ee$. Note, that the cut on $P>15\,\GeV$ has been omitted in this case.

\subsection{Track Selection}

As the invariant mass of the $\Zzero$ boson and thus the recoil mass will be calculated from the four momenta of the LDC Tracks, badly measured LDC Tracks need to be discarded from the analysis. The track quality can be estimated by the ratio $\Delta P/P^2$ where the uncertainty $\Delta P$ is derived from the error matrix of the given track by error propagation.

\begin{figure}[htb]
\centering
\includegraphics[width=0.4\textwidth]{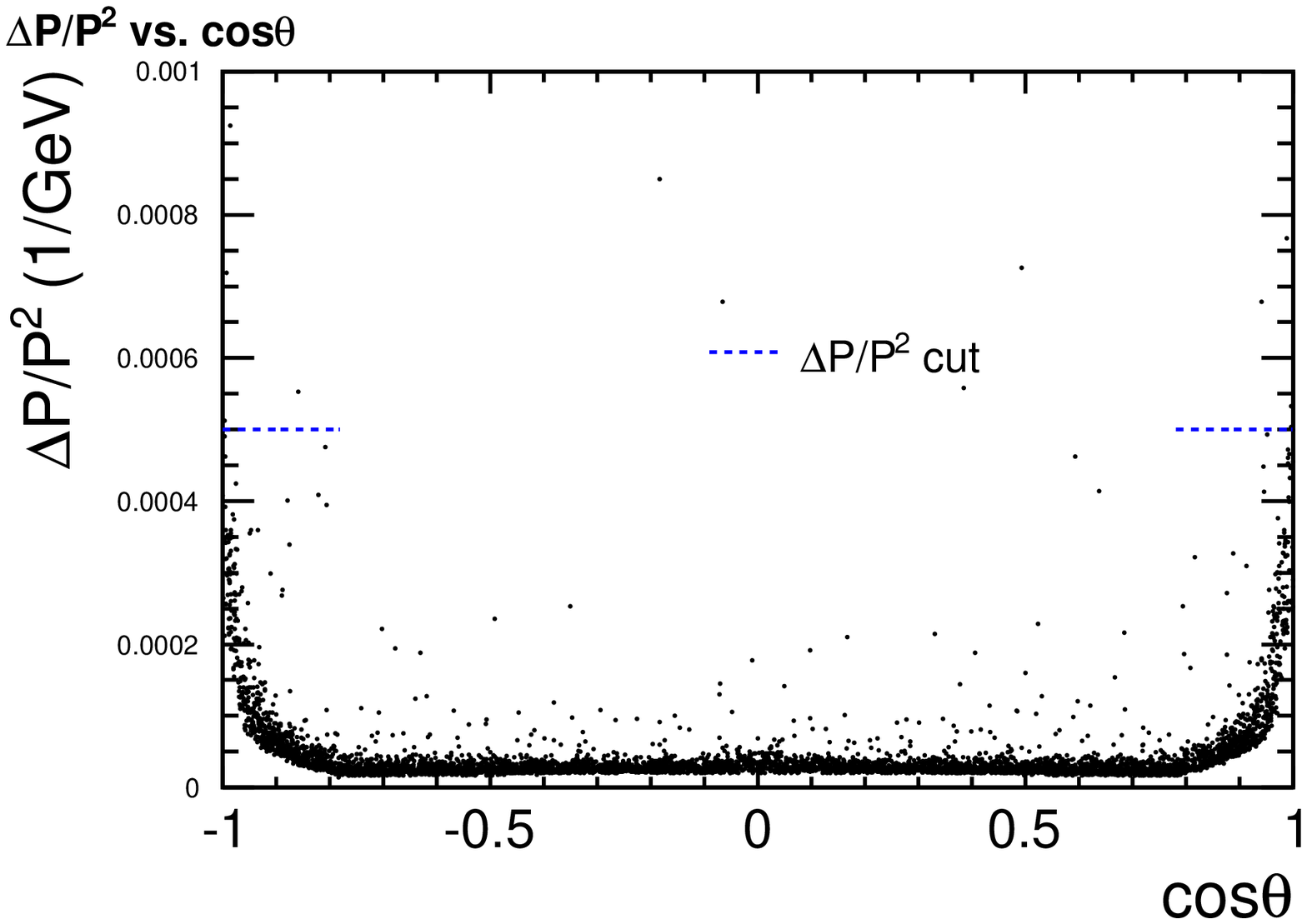}%
\includegraphics[width=0.4\textwidth]{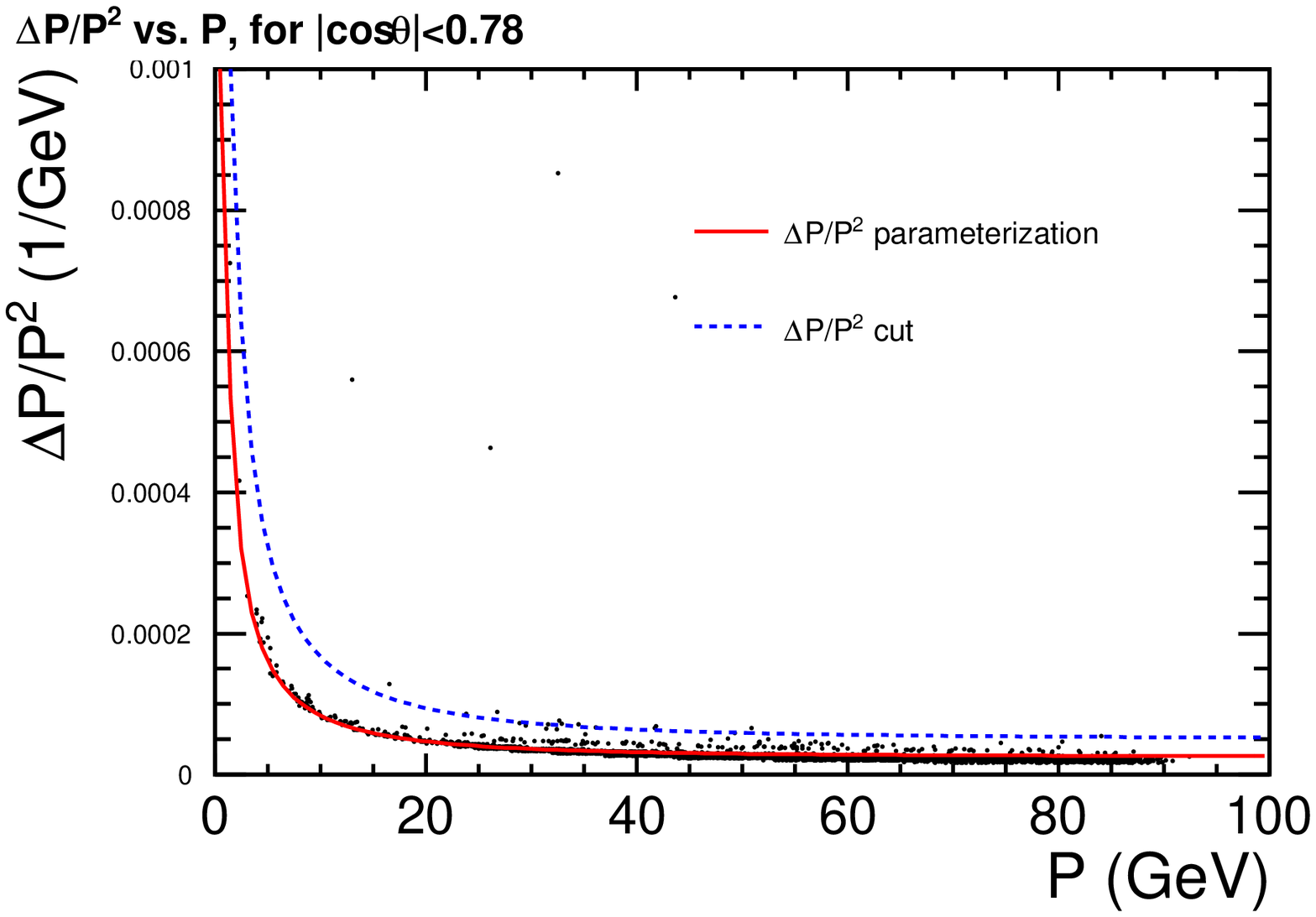}
\caption{2D $\Delta P/P^2$ distribution vs. $cos\theta$ (left) and $\Delta P/P^2$ distribution vs. track momentum (right) of muon candidates}
\label{fig:2ddpop2_mmx}
\end{figure}

\begin{figure}[htb]
\centering
\includegraphics[width=0.4\textwidth]{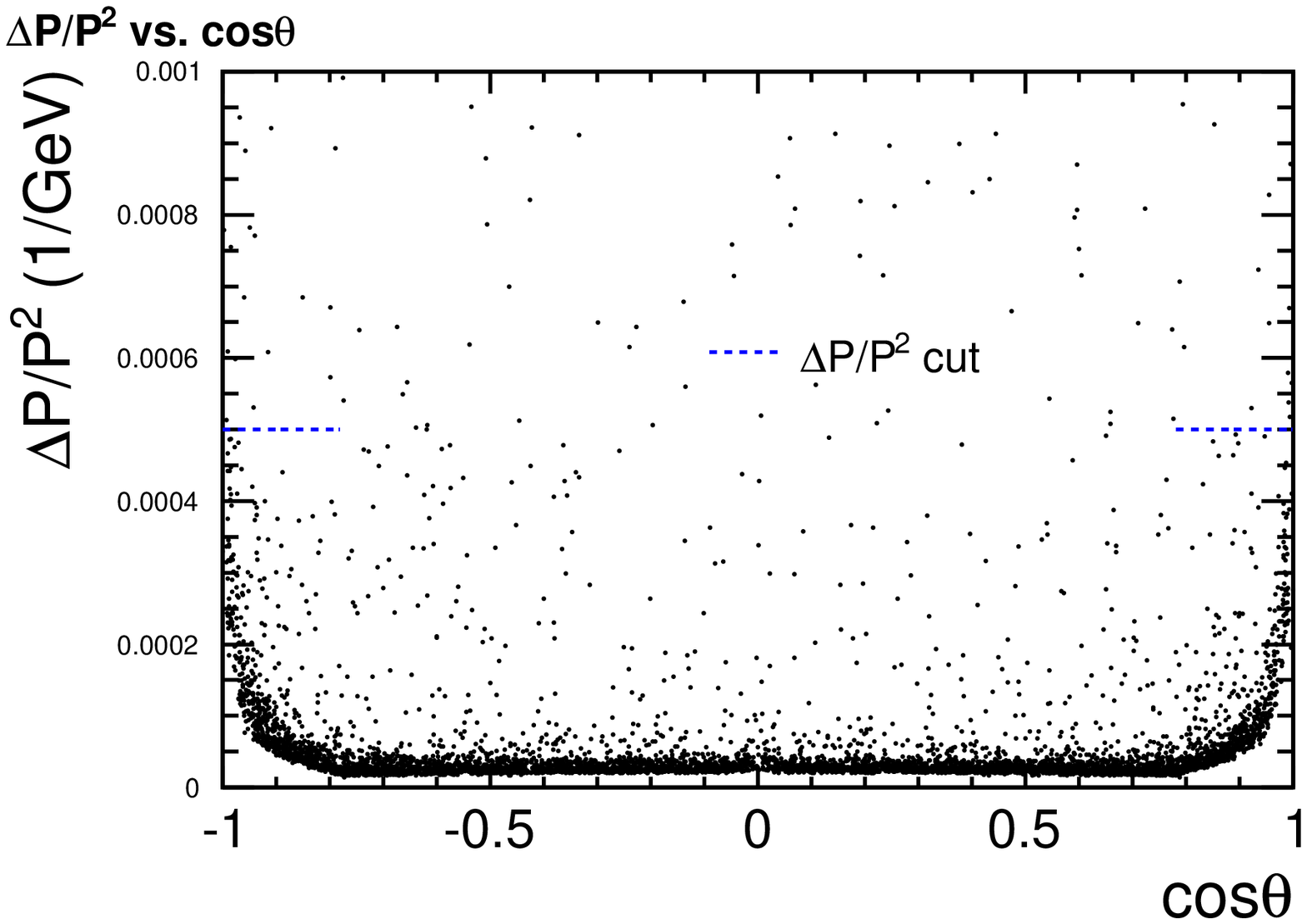}%
\includegraphics[width=0.4\textwidth]{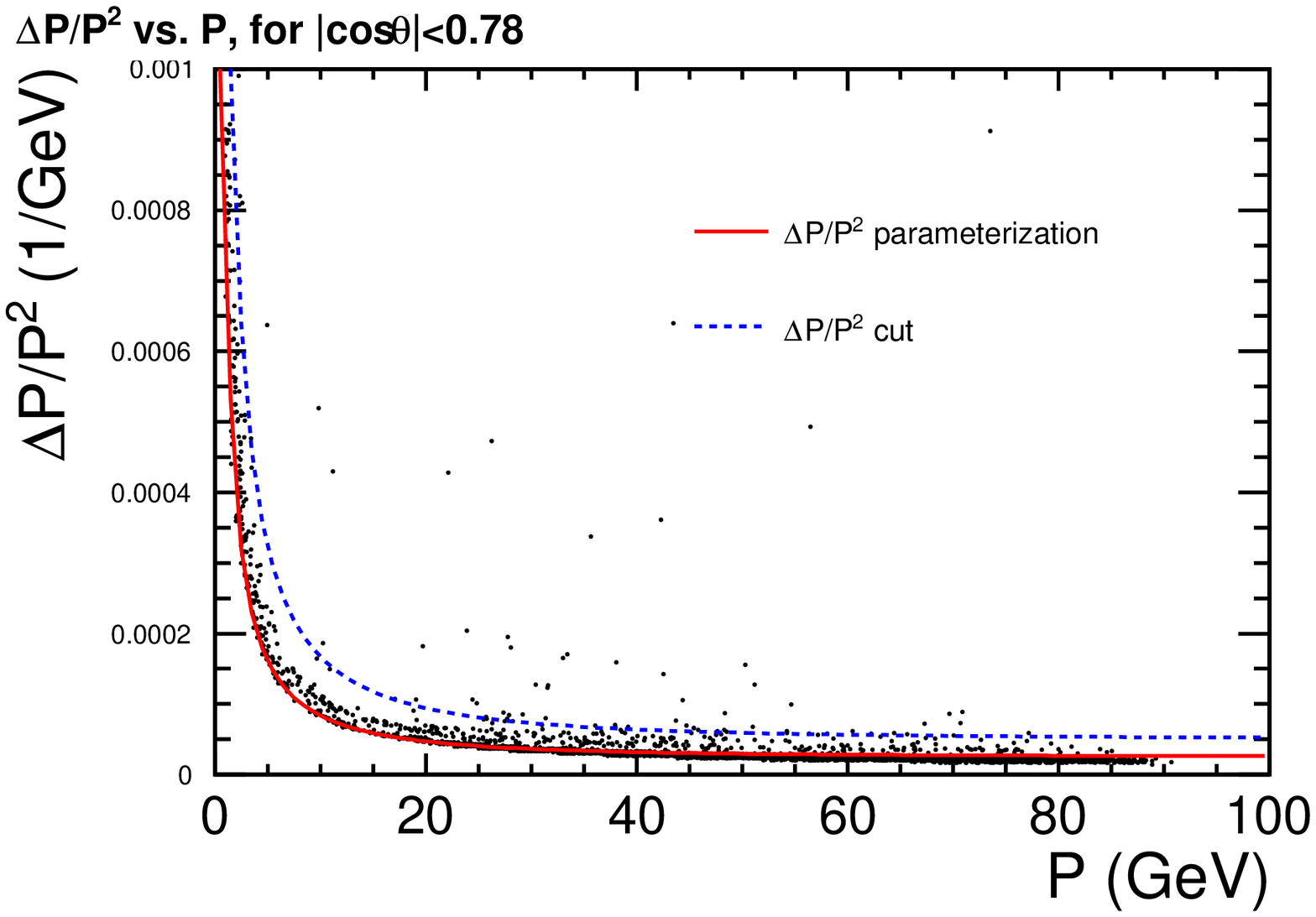}
\caption{2D $\Delta P/P^2$ distribution vs. $cos\theta$ (left) and $\Delta P/P^2$ distribution vs. track momentum (right) of electron candidates}
\label{fig:2ddpop2_eex}
\end{figure}

The Figures~\ref{fig:2ddpop2_mmx} and~\ref{fig:2ddpop2_eex} show, for muons and electrons separately, the dependency of $\Delta P/P^2$ on the polar angle $cos\theta$ and on the track momentum $P$. 
For reasons discussed in the following the latter has been restricted to $|cos\theta|<0.78$, i.e. the central region. 
For both variables the distributions exhibit for muon tracks a narrow band with well measured momenta equivalent to small $\Delta P/P^2$. The track quality decreases as expected towards large $|cos\theta|$, i.e. towards the acceptance limits of the TPC which motivates the restriction to the central region when displaying $\Delta P/P^2$ versus $P$. These distributions show a decrease in track quality towards small particle momenta as expected from multiple scattering effects. Beyond that, it is clearly visible that for electrons the situation is much more diluted and the number of badly measured tracks is significantly higher than that for muons. This can be explained by the Bremsstrahlung of the electrons in the detector material. 
%The conversion of photons or the radiation of high energetic photons lead to an accumulation of tracks with small momenta, as visible in Figure~\ref{fig:2ddpop2_eex}. 
%{\bf Why not visible, I think we have discussed this}.

The procedure for track rejection is developed for the better measured muon induced tracks:

\begin{itemize}
\item For $|cos\theta|<0.78$ the shape of $\Delta P/P^2$ versus $P$ is approximated by:
\begin{equation}
\delta(1/P) = \Delta P/P^2 = a\oplus b/P=c(P);\ {\rm with}\ a=2.5 \times 10^{-5}\,\GeV^{-1}\ {\rm and}\ b = 8 \times 10^{-4} .
\end{equation}
Tracks are rejected if $\delta(1/P)>2c(P)$.
\item For $|cos\theta|>0.78$ tracks are rejected  if $\Delta P/P^2 > 5 \times 10^{-4}\,\GeV^{-1}$.
\end{itemize}

The cuts are indicated in Figure~\ref{fig:2ddpop2_mmx} and~\ref{fig:2ddpop2_eex} and underline that tracks created by electrons are rejected considerably more often which will reduce the number of reconstructed $\Zzero$ bosons in the corresponding channel. 

%\section{Model Independent Analysis}
 
\subsection{Background Rejection}

The recoil analysis is based on the identification of the di-lepton system as produced by the decay of the $\Zzero$ boson.  It is thus necessary to distinguish the processes which lead to two leptons in the final state as given in Table~\ref{tab:reaction_lr} and Table~\ref{tab:reaction_rl} from the ones produced in the Higgs-strahlung process. 

\begin{figure}[htbp]
\centering
\includegraphics[width=0.9\textwidth]{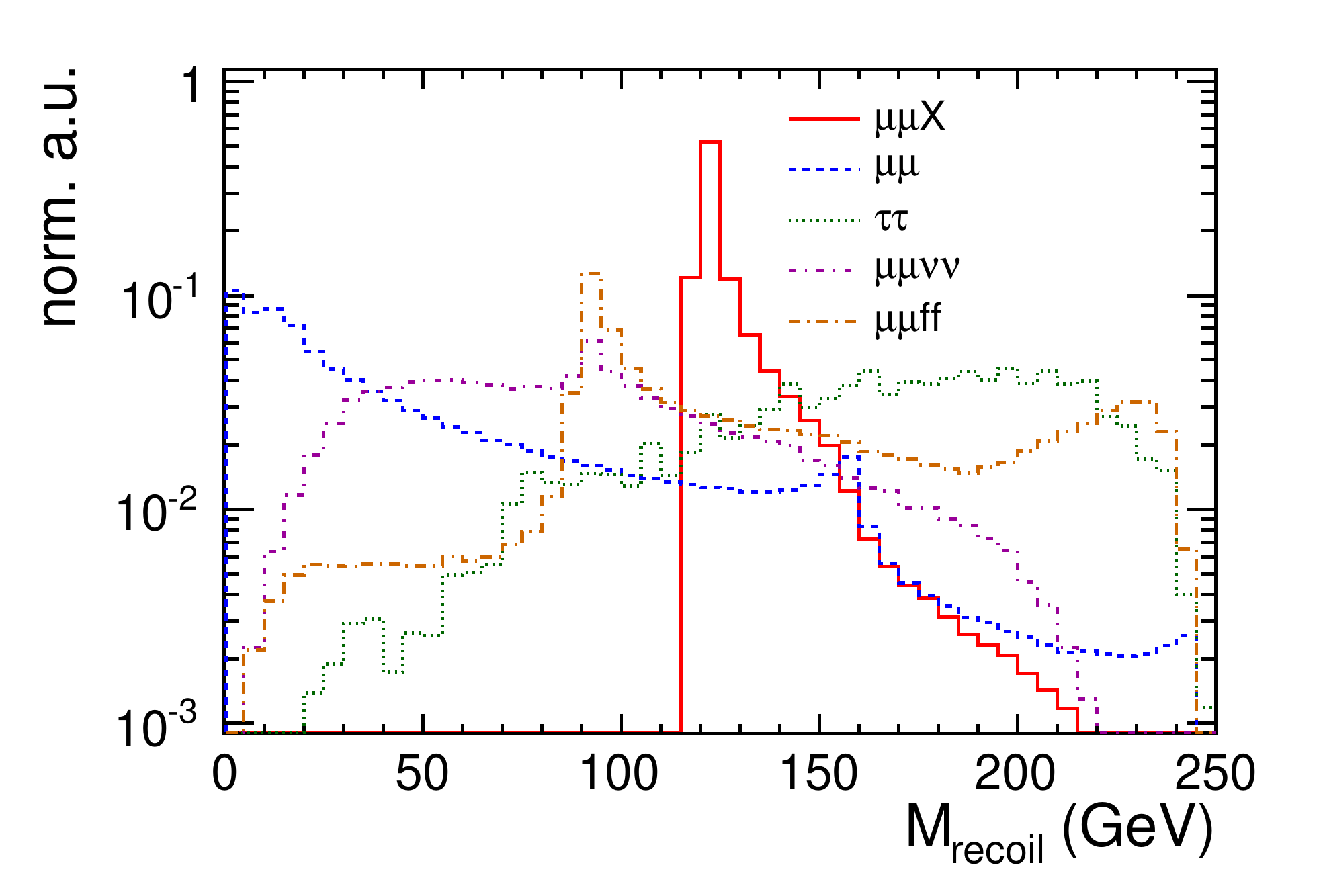}
\includegraphics[width=0.9\textwidth]{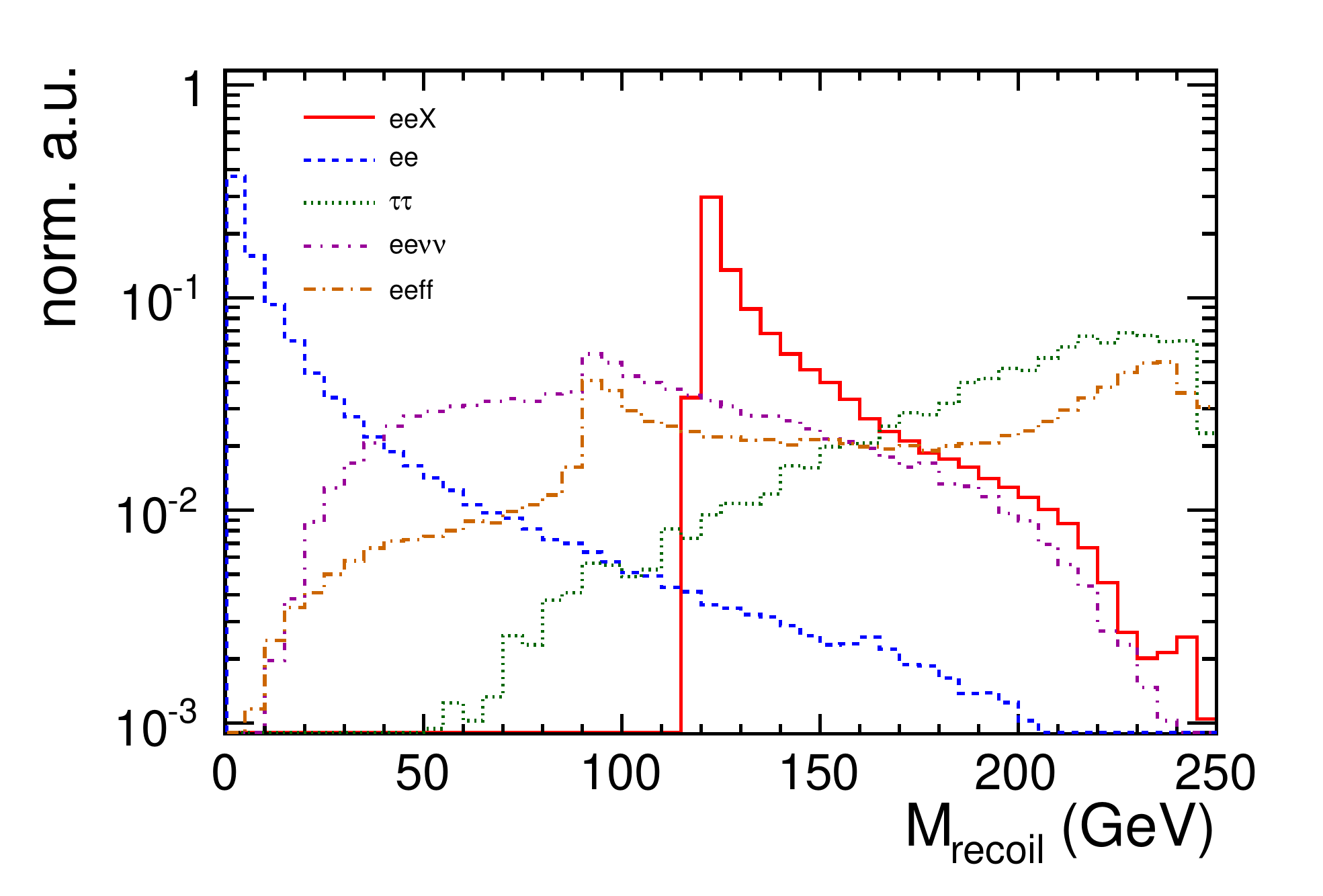}
\caption{Normalised signal and background distributions of the recoil mass $\Mrecoil$ distributions for the $\mmX$ (top) and the $\eeX$ Channel (bottom). Here, $\tau\tau$ refers to the $\mu\mu$ or $ee$ created in the decay of $\tau\tau$. Note that the Pre-cuts defined in Section~\ref{sec:gensimrec} have been applied to the $\mu\mu$ background sample.}
\label{fig:var_mrecoil}
\end{figure}

For the Higgs-strahlung process the invariant mass of the di-lepton system $M_{dl}$ should be equal to the $\Zzero$ boson mass while the invariant mass of the recoiling system, $M_{recoil}$ is expected to yield the introduced mass of the Higgs boson of 120\,GeV. It is unlikely that combinations of background processes fulfil both conditions at once. This argumentation is supported by Figures~\ref{fig:var_mdl} and~\ref{fig:var_mrecoil} which show the invariant mass distributions for the di-lepton system and the recoil mass for both, the di-lepton system consisting of muons and the di-lepton system consisting of electrons. These distributions suggest  to restrict the analysis to the following mass ranges:

\begin{figure}[htbp]
\centering
\includegraphics[width=0.9\textwidth]{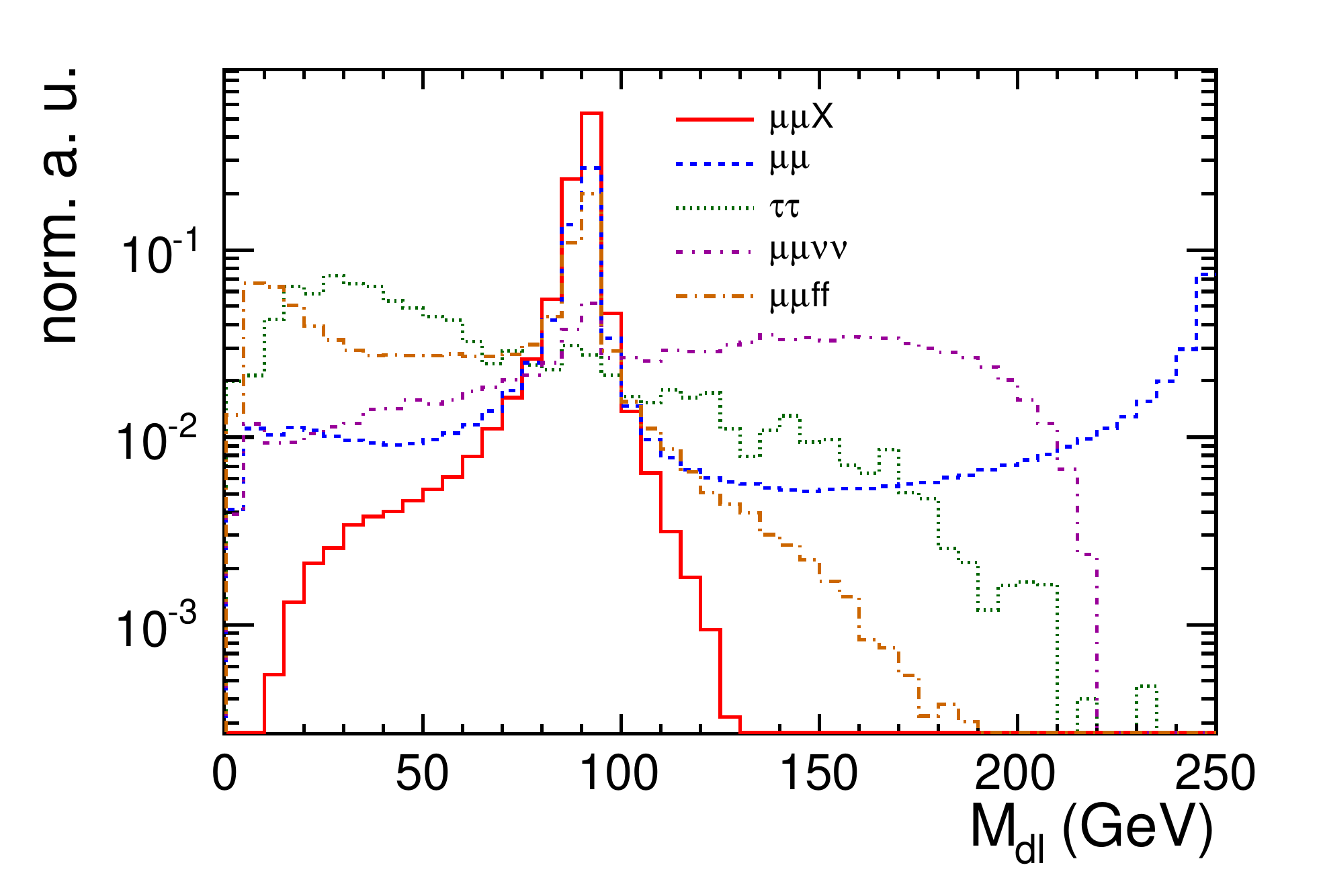}
\includegraphics[width=0.9\textwidth]{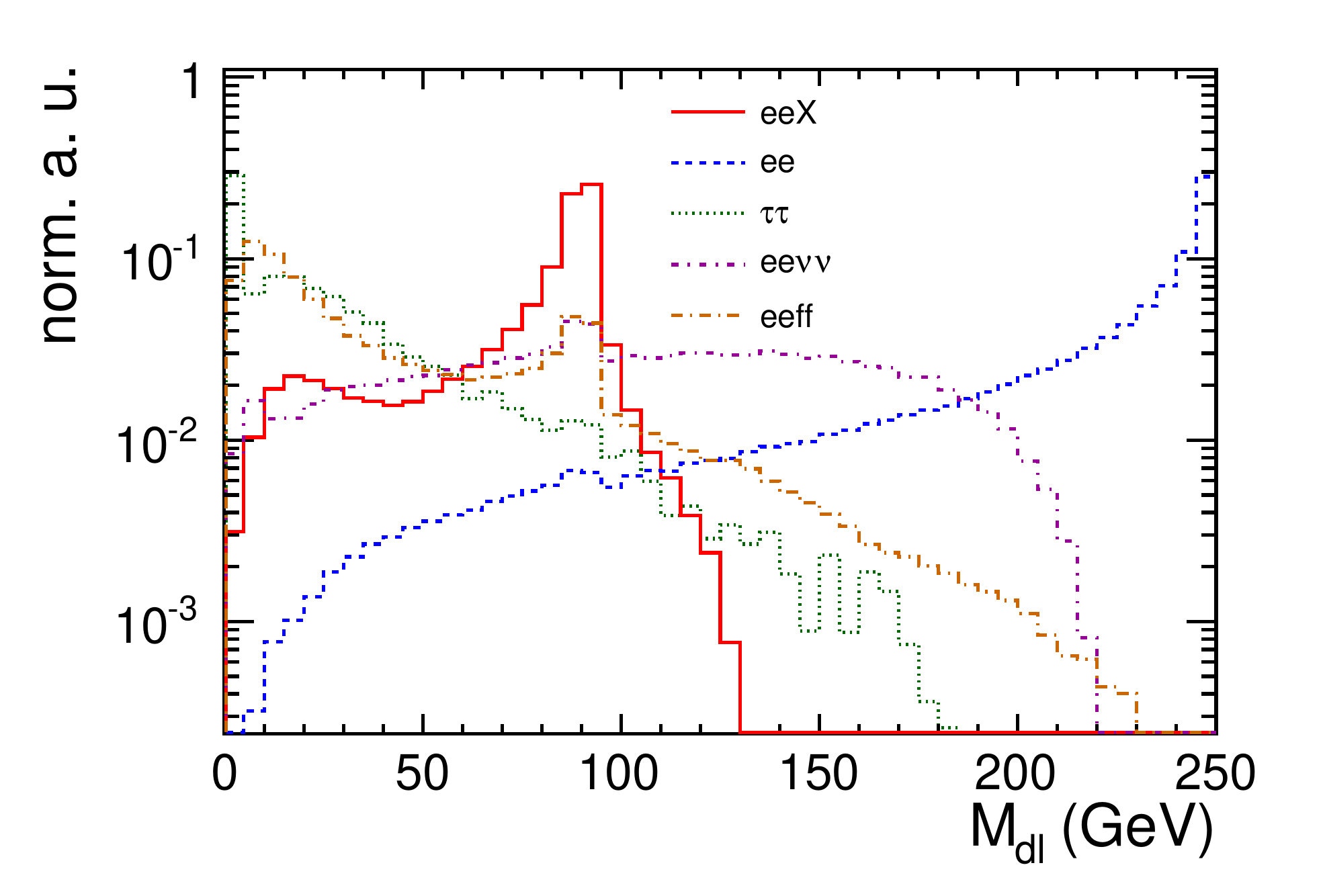}
\caption{Normalised signal and background distributions of the invariant mass of the di-lepton system $\Mdl$ for the $\mmX$ (top) and $\eeX$ Channel (bottom). Here, $\tau\tau$ refers to the $\mu\mu$ or $ee$ created in the decay of $\tau\tau$. Note that the Pre-cuts defined in Section~\ref{sec:gensimrec} have been applied to the $\mu\mu$ background sample.}
\label{fig:var_mdl}
\end{figure} 

\begin{itemize} 
\item $80 < M_{dl} < 100\,\GeV$ 
\item $115 < M_{recoil} < 150\,\GeV$ 
\end{itemize}

\begin{figure}[htbp]
\centering
\includegraphics[width=0.99\textwidth]{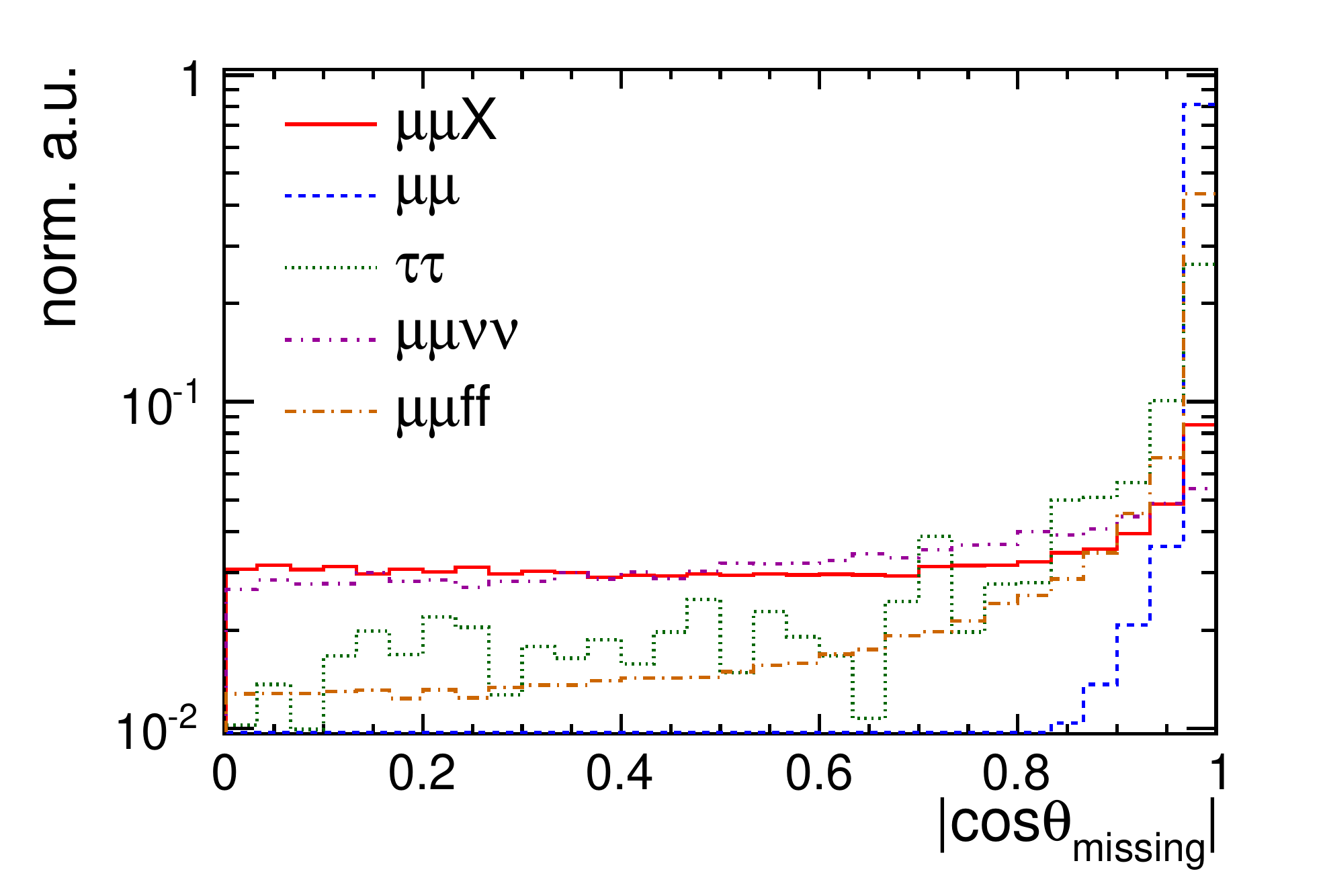}
\includegraphics[width=0.99\textwidth]{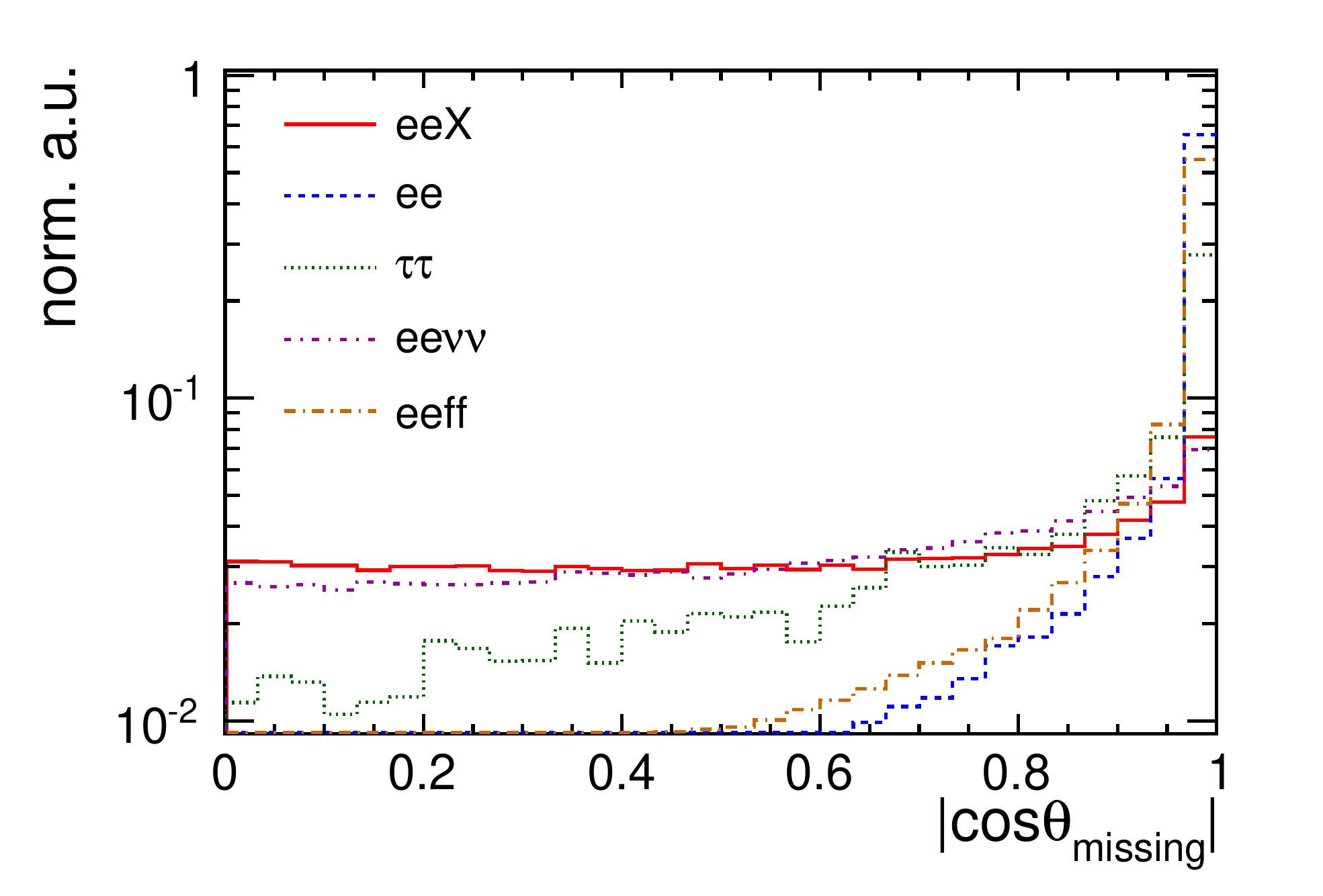}
\caption{Normalised signal and background distributions of the  $\cthamiss$ of the system of undetected particles for the  $\mmX$-Channel (top) and $\eeX$-Channel (bottom). }
\label{fig:var_cthamiss}
\end{figure}

In a next step the selection is to be made by means of the different kinematic properties.  In the following the variables used to distinguish signal events from background events will be introduced. 
\begin{itemize}
\item Acoplanarity {\it acop}, see Figure~\ref{fig:var_acop}: As for $\epem$ collisions with beams of equal energy the centre-of-mass system is at rest, it is expected that in processes in which the leptons are produced at the $Z^{\ast}$ vertex these two leptons are back-to-back in azimuth angle. The distance in azimuth angle is  expressed by the acoplanarity $acop$, defined as $acop=|\phi_{\ell^+}-\phi_{\ell^-}|$, where $\phi_{\ell^{\pm}}$ is the azimuth angle of the an individual lepton of the di-lepton system. If the particles are produced from a intermediate particle with a given transverse momentum, the exact back-to-back configuration is modulated.  Therefore a cut on $0.2 < acop < 3$ is applied.

\begin{figure}[htbp]
\centering
\includegraphics[width=0.9\textwidth]{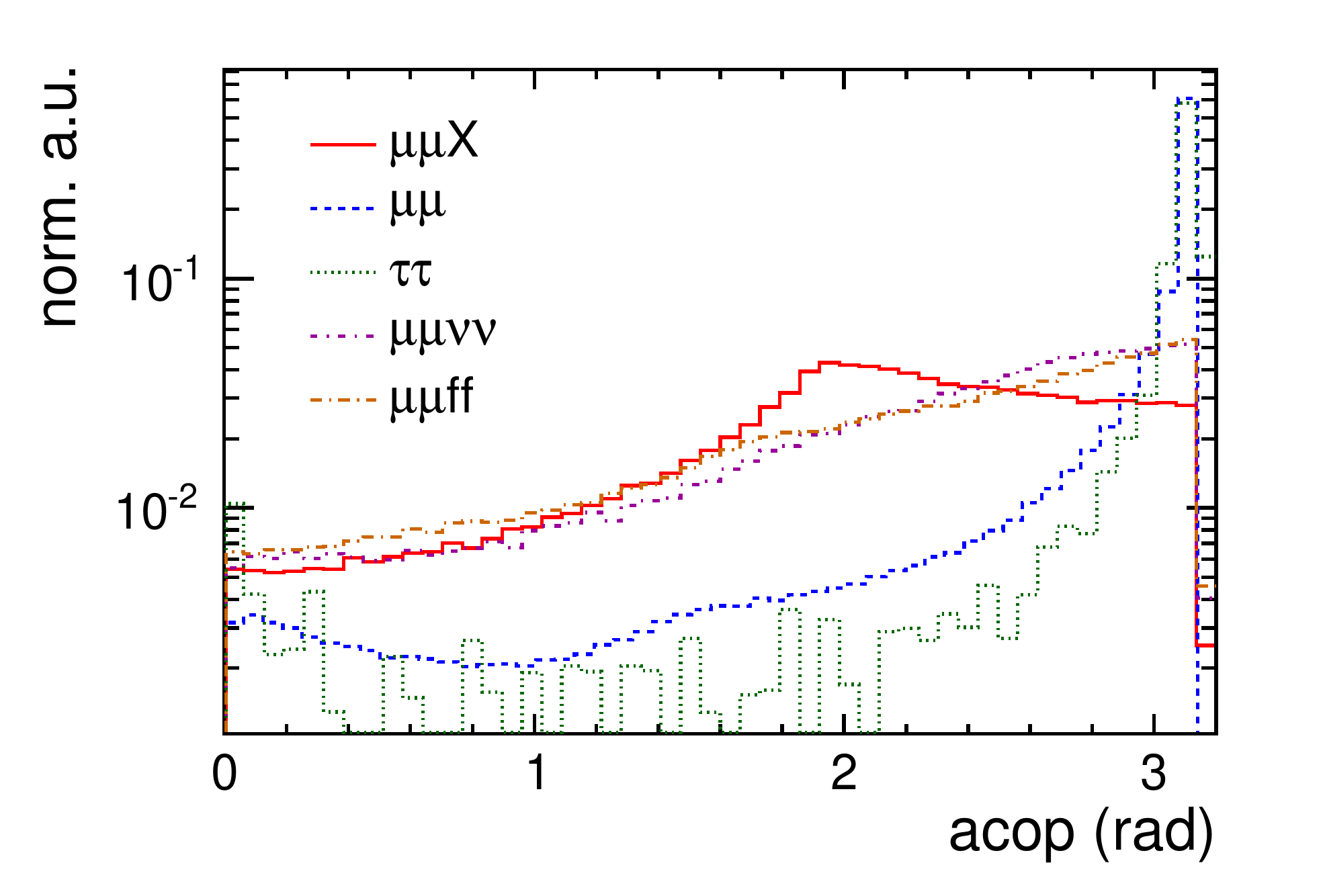}
\includegraphics[width=0.9\textwidth]{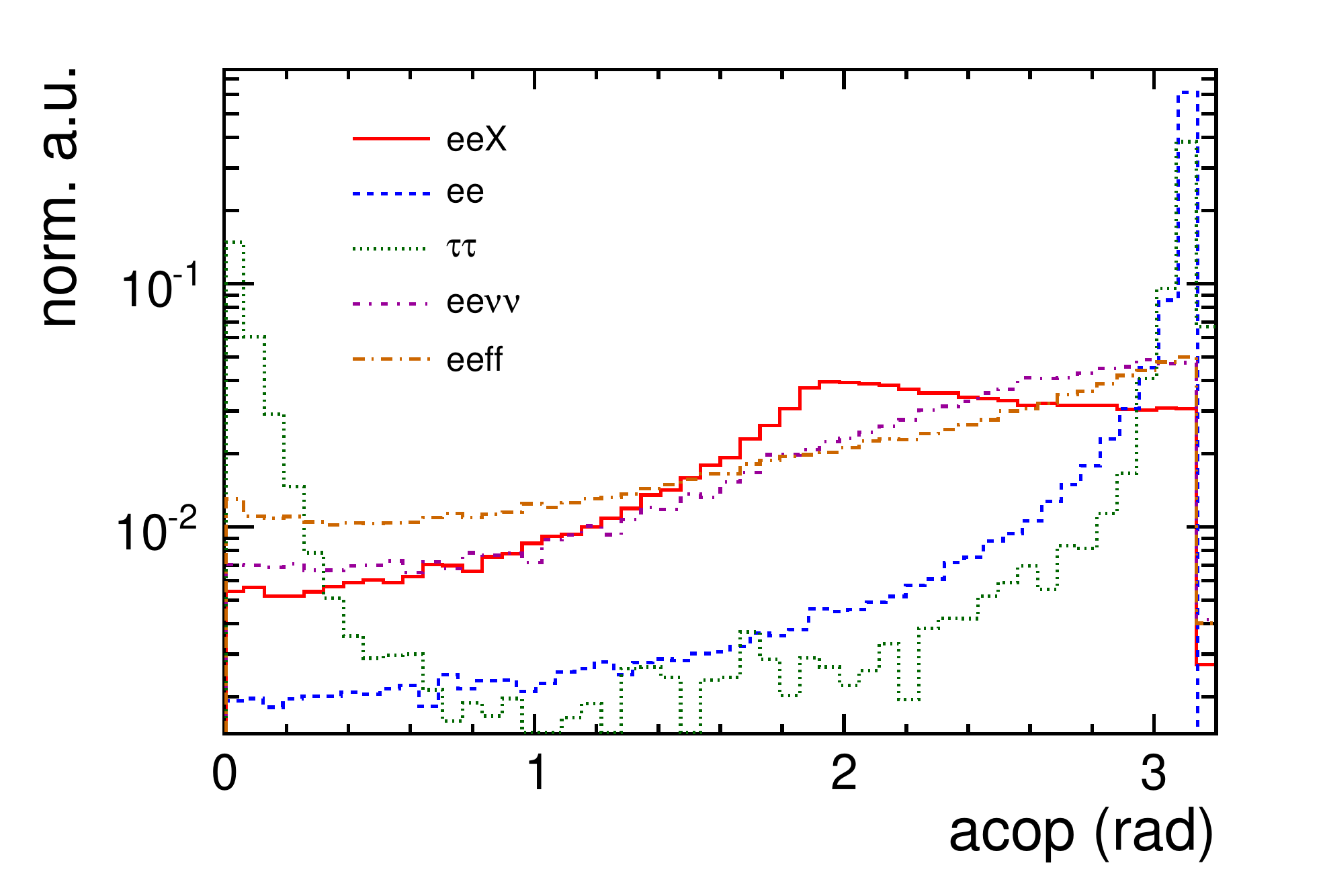}
\caption{Normalised signal and background distributions of the acoplanarity $acop$ for the $\mmX$-Channel (top) and $\eeX$-Channel (bottom). Here, $\tau\tau$ refers to the $\mu\mu$ or $ee$ created in the decay of $\tau\tau$. Note that the Pre-cuts defined in Section~\ref{sec:gensimrec} have been applied to the $\mu\mu$ background sample.}
\label{fig:var_acop}
\end{figure}

\item Transverse Momentum $P_{Tdl}$ of the di-lepton system, see Figure~\ref{fig:var_ptdl}: As the Higgs-strahlung process can be interpreted as a two body decay, both bosons gain equal transverse momentum which is conserved  by their decay products. The total final state for muon pair production or Bhabha Scattering has in first approximation no transverse momentum. In order to suppress this background, a cut $P_{Tdl}>20\,\GeV$  of the  di-lepton system is applied. This cut cannot suppress events in which initial state radiation of the incoming beams leads to a transverse momentum of the colliding system. This case will be discussed separately.

\begin{figure}[htbp]
\centering
\includegraphics[width=0.9\textwidth]{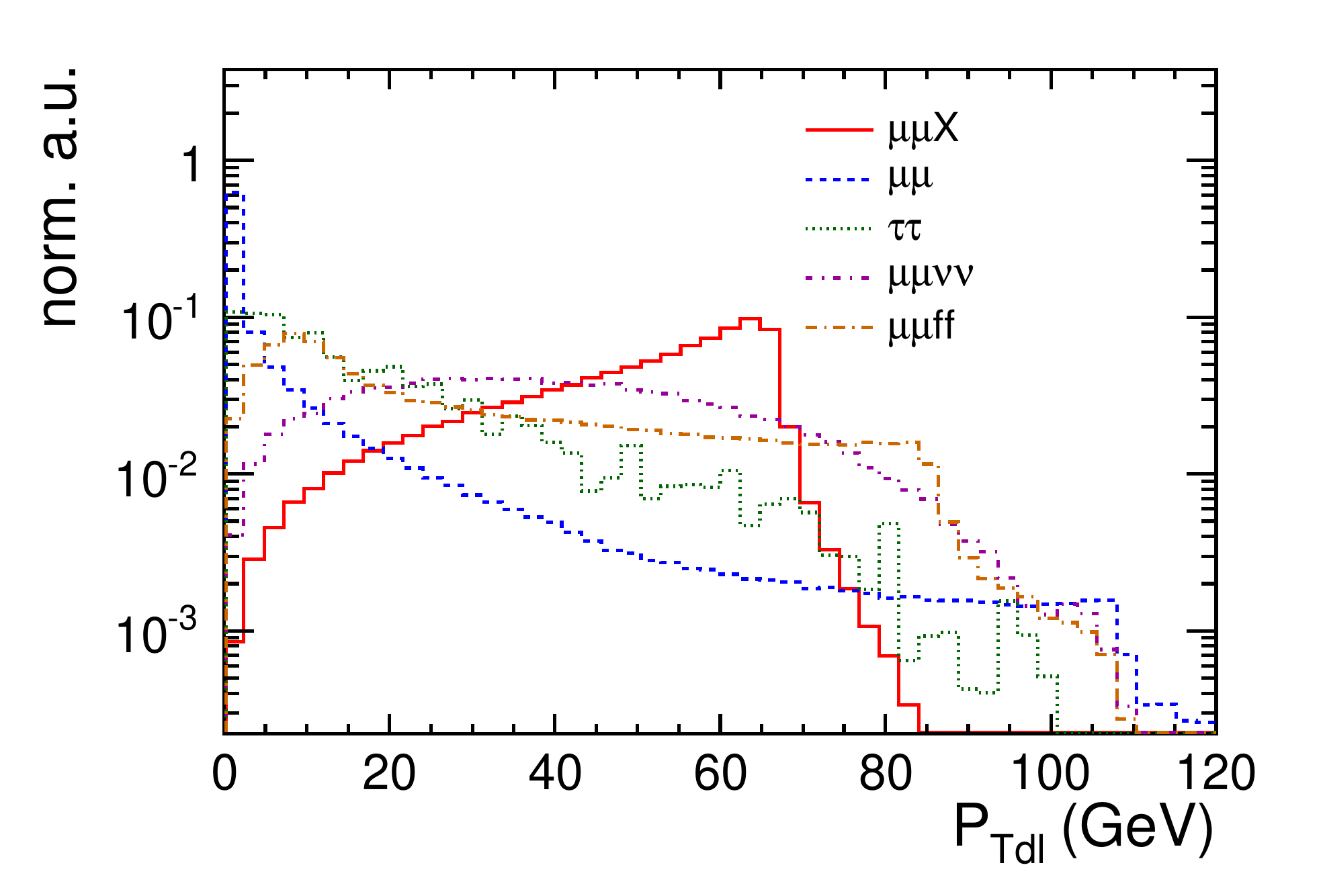}
\includegraphics[width=0.9\textwidth]{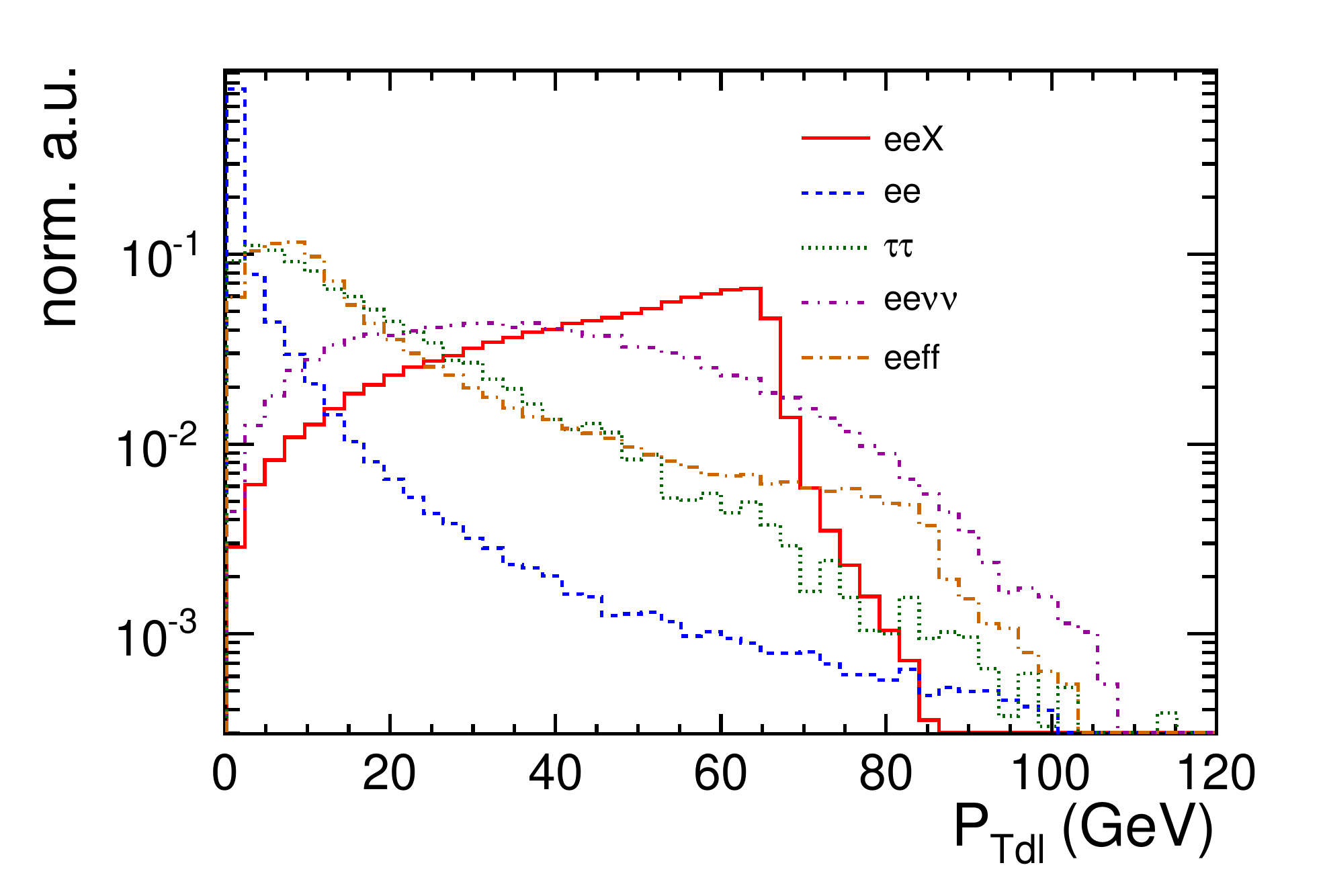}
\caption{Normalised signal and background distribution of the transverse momentum $\Ptdl$ of the dilepton system for the $\mmX$ Channel (top) and the $\eeX$-Channel (bottom). Here $\tau\tau$ refers to the $\mu\mu$ or $ee$ created in the decay of $\tau\tau$. Note that the Pre-cuts defined in Section~\ref{sec:gensimrec} have been applied to the $\mu\mu$ background sample. }
\label{fig:var_ptdl}
\end{figure} 

\item $cos\theta_{missing}$: this cut discriminates events which are unbalanced in longitudinal momentum, essentially, those of the type $\epem \rightarrow \lplm \gamma$. The distributions in Figure~\ref{fig:var_cthamiss} motivate a cut
on $|cos\theta_{missing}|<0.99$.
\end{itemize}

\begin{figure}[!h]
\centering
\includegraphics[width=0.5\textwidth]{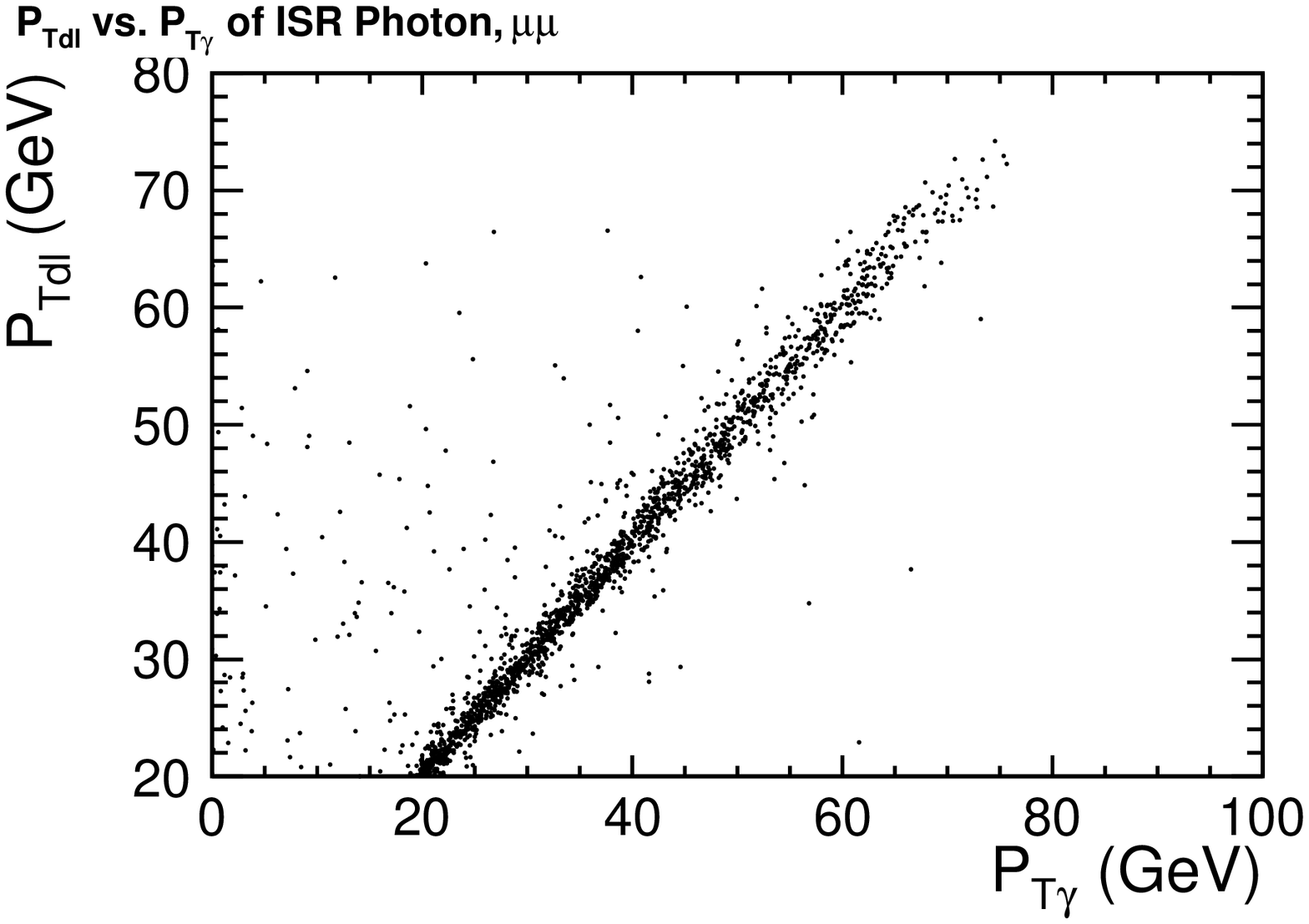}%
\includegraphics[width=0.5\textwidth]{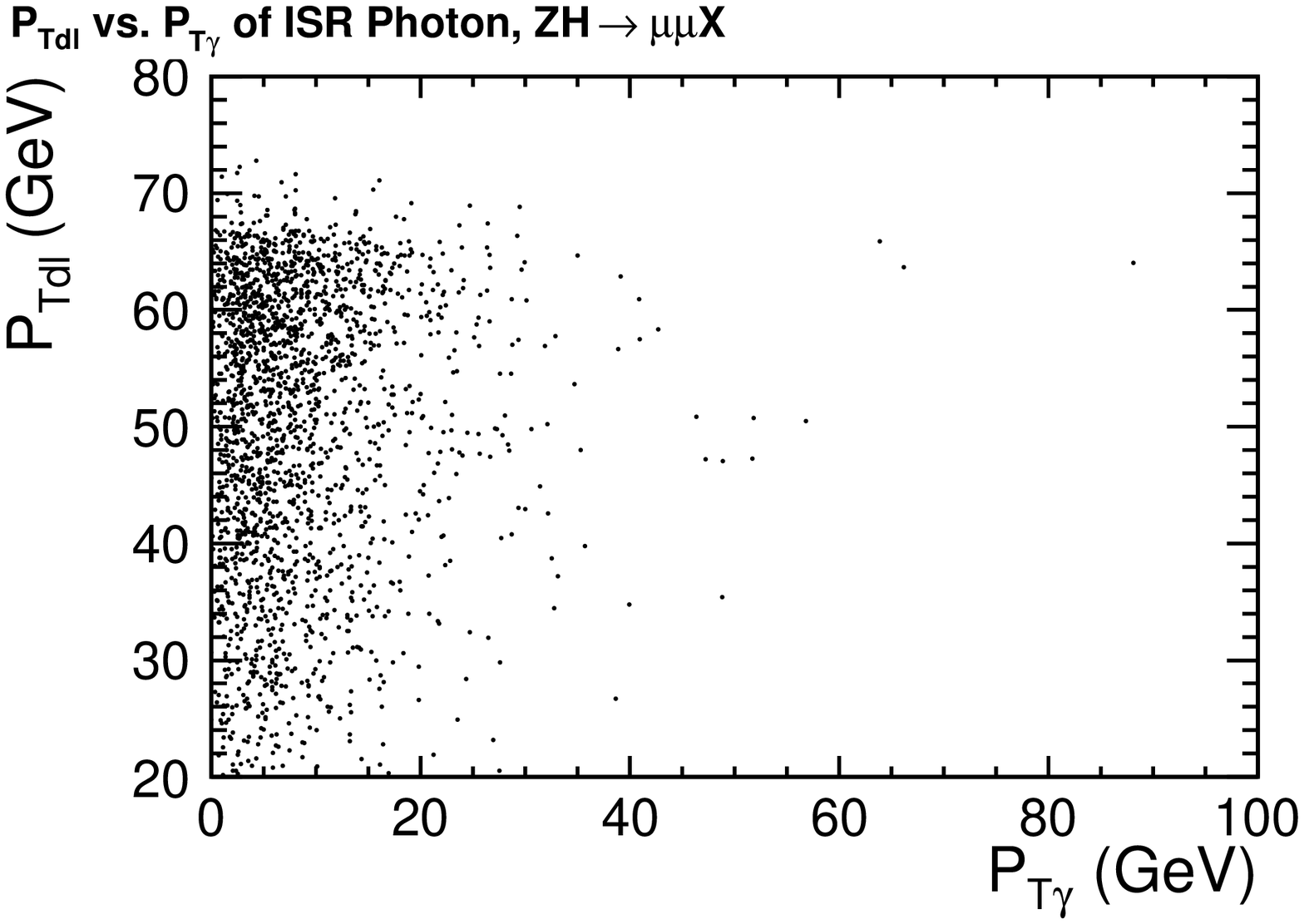}
\caption{$P_{Tdl}$ vs. $P_{T\gamma}$ for background by $\epem \rightarrow \mpmm$ (left) and for signal (right) in the $\mu\mu X$-channel. } 
\label{fig:pt_balance}
\end{figure}

\begin{figure}[!h]
\centering
\includegraphics[width=0.5\textwidth]{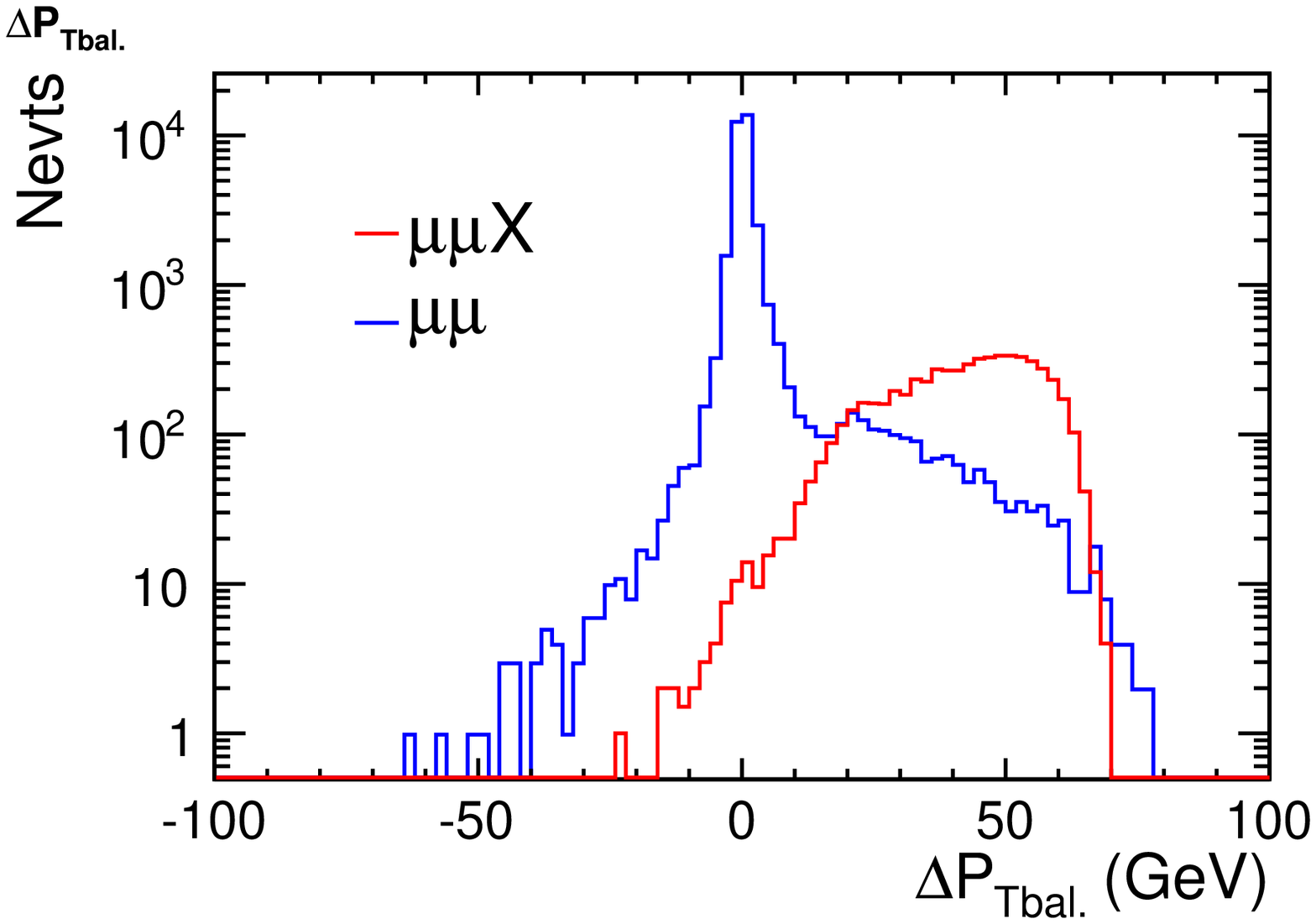}
\caption{$\Delta P_{Tbal.}$ distributions for background by $\epem \rightarrow \mpmm$ and signal in the $\mu\mu X$-channel. } 
\label{fig:dpbal_mu}
\end{figure}

The last introduced cut also suppresses events with initial state radiation happening approximately collinear with the incoming beams. The final state in $\epem\rightarrow \mpmm(\epem)$ can, however, gain sizeable transverse momentum by initial state radiation of a high energetic photon. Figure~\ref{fig:pt_balance} shows the correlation between the transverse momentum $P_{T\gamma}$ of a detected high energetic photon, assumed to be created by initial state radiation,  and the transverse momentum $P_{Tdl}$  of the di-lepton system for both, events in which only a muon pair is created at the $Z^{\ast}$- Boson vertex and signal events. The first type shows a clear correlation in transverse momentum. In order to suppress this background the variable  $\Delta P_{Tbal.} = P_{Tdl} - P_{T\gamma}$ is introduced which is shown in Figure~\ref{fig:dpbal_mu} for signal events and background events  superimposed with each other. By selecting events with  $\Delta P_{Tbal.} > 10\,\GeV$, a considerable fraction of background can be suppressed. It should finally be noted that background events of type $\epem\rightarrow \mpmm(\epem)$ which are undergoing final state radiation are suppressed by the requirement that the lepton should yield the $\Zzero$ boson mass.

The number of events remaining after each of these cuts for signal and backgrounds are given in Tabs.~\ref{tab:rej_mu_mi_lr} through~\ref{tab:rej_el_mi_rl} for two beam polarisations and the different compositions of the di-lepton system. The combination of cuts will be later referred as \emph{MI Cuts}. Please note that the cut variables $f_L$ and  $|\Delta\theta_{2tk}|$ will be introduced later.

From the tables the following conclusions can be drawn

\begin{itemize}
\item The requirement to have two well measured leptons retains always more than 95\% of the signal while it suppresses
in most of the cases the largest part of background events.
\item The requirement of a minimum $P_{Tdl}$ of the di-lepton system is very efficient for events in which the di-lepton system is produced directly at the $\Zzero^{\ast}$ vertex, see Figure~\ref{fig:higgs-ff}. This type of background is further reduced by comparing the transverse momentum of the di-lepton system with the transverse momentum of a radiative photon. The cut is particulary efficient to suppress background events
generated by Bhabha Scattering. 
\item Although largely suppressed, the number of events generated by Bhabha background still exceeds the number of signal events. This remains an irreducible background. 
\item The acoplanarity $acop$ is particularly efficient against background in which the di-lepton system is composed by $\tau$- Leptons. The larger mass of this particle reduces the phase space for radiative processes. Hence this lepton type is more often produced in a back-to-back configuration than the lighter lepton types.
\end{itemize}

The tables demonstrate that mostly events in which the di-lepton system is produced at the $Z^{\ast}$ vertex can be efficiently rejected by the defined cuts. The background by events in which two bosons are produced, i.e.  $\epem\rightarrow\Zzero\Zzero/\gamma\gamma$ or $\epem\rightarrow\WpWm$, is less well distinguishable from the  signal events. As these events however have slightly different spectra. Further rejection can be achieved by a multi-variate analysis of a set of suited discriminative variables. These variables are introduced in the following.

\begin{itemize}
\item The $\gamma$-pair production leads to a flat distribution in the di-lepton mass spectrum 
in the $\Zzero$-mass region The shape of the invariant mass $M_{dl}$ of the di-lepton system and hence also that of the transverse momentum $P_{Tdl}$ of the di-lepton system can be employed to
suppress background from $\gamma$-pair production.
\item The production of $\Zzero$\,boson and $\Wboson$\,boson pairs happens predominantly via exchange reactions which lead to a strong increase of the differential cross section towards large absolute values of the cosine of the polar angle. On the contrary, the Higgs-strahlung process is expected to decrease towards the forward and backward direction. Therefore the polar angle spectrum as shown in Figure~\ref{fig:var_cthadl} of the di-lepton system is expected to discriminate between signal events and background from $\Zzero$ and $\Wboson$ pair production. 
\begin{figure}[htbp]
\centering
\includegraphics[width=0.9\textwidth]{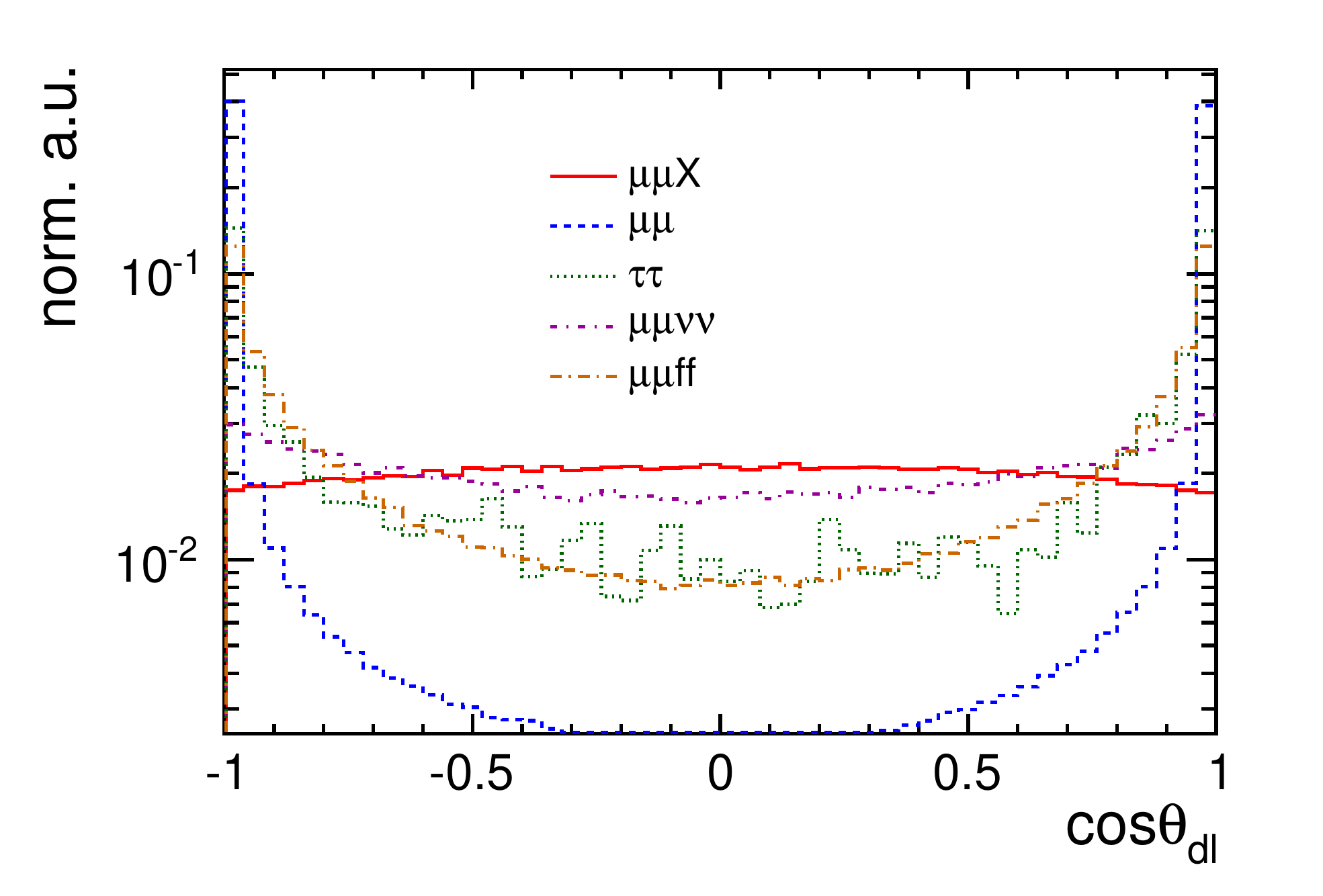}
\includegraphics[width=0.9\textwidth]{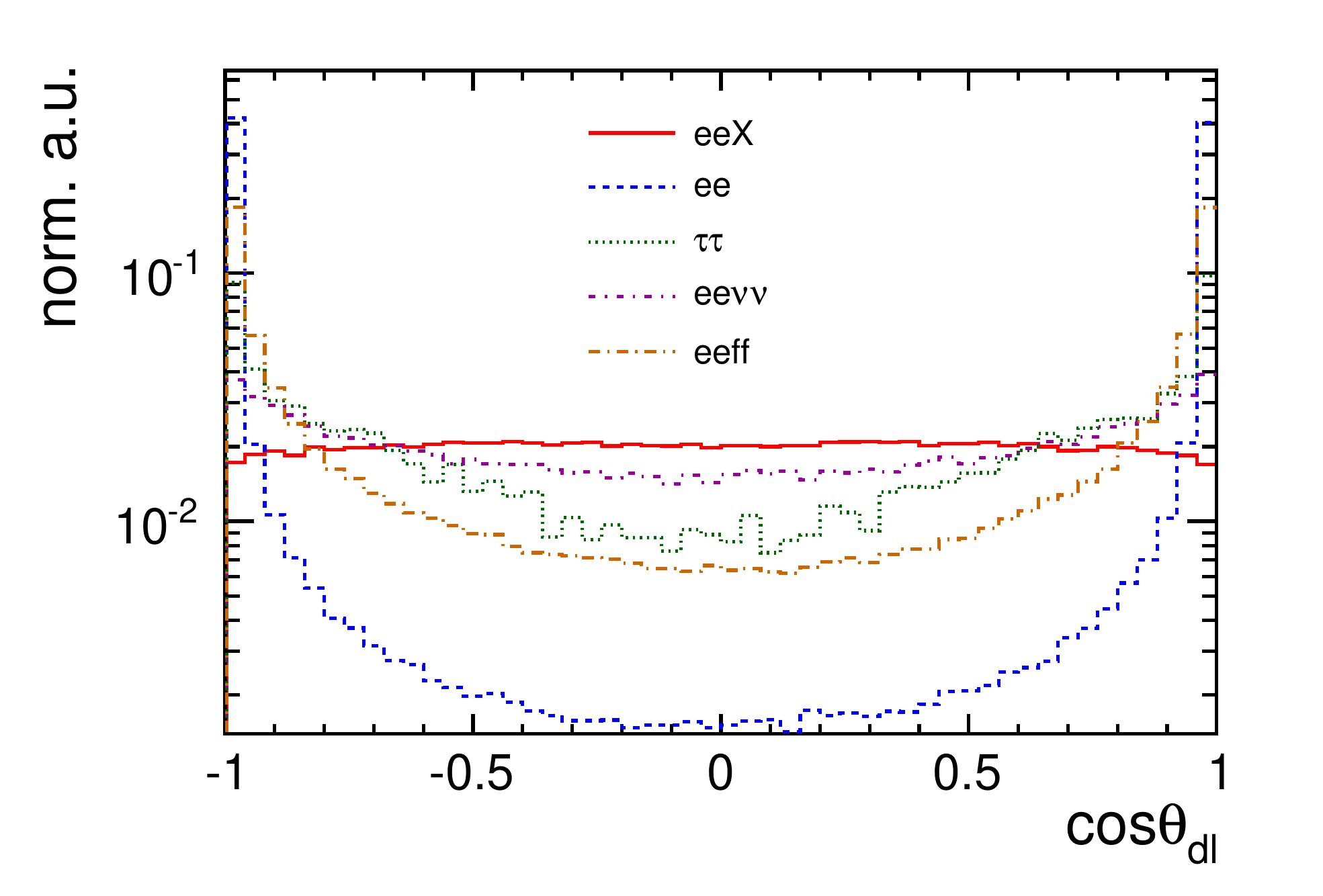}
\caption{Normalised signal and background distribution of the cosine of the polar angle $\cthadl$ of  the di-lepton system for the $\mmX$ Channel (top) and the $\eeX$-Channel (bottom). Here $\tau\tau$ refers to the $\mu\mu$ or $ee$ created in the decay of $\tau\tau$. Note that the Pre-cuts defined in Section~\ref{sec:gensimrec} have been applied to the $\mu\mu$ background sample.}
\label{fig:var_cthadl}
\end{figure} 
\item The acollinearity, defined as $acol=acos({\bf P_{\ell^+}}{\bf P_{\ell^-}}/|{\bf P_{\ell^+}}| |{\bf P_{\ell^-}}|)$, is sensitive to the boost of the di-lepton system. In case of $\Zzero$ pair production the decay products are expected to be boosted more strongly than in the case of Higgs-strahlung. This results in a different position of the Jacobian Peak in the $d\sigma/d(acol)$ differential cross section as demonstrated in Figure~\ref{fig:var_acol}.
\begin{figure}[htbp]
\centering
\includegraphics[width=0.9\textwidth]{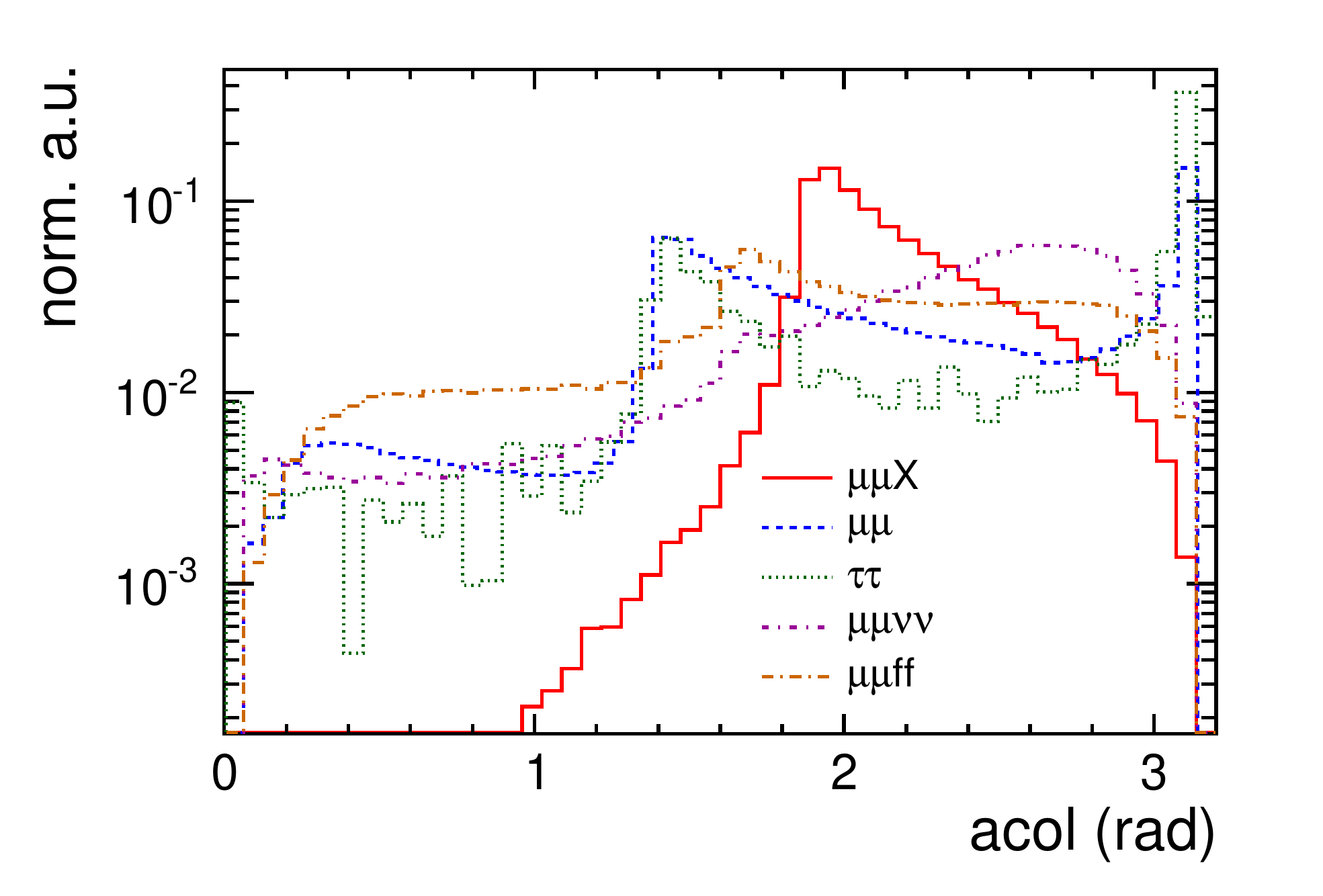}
\includegraphics[width=0.9\textwidth]{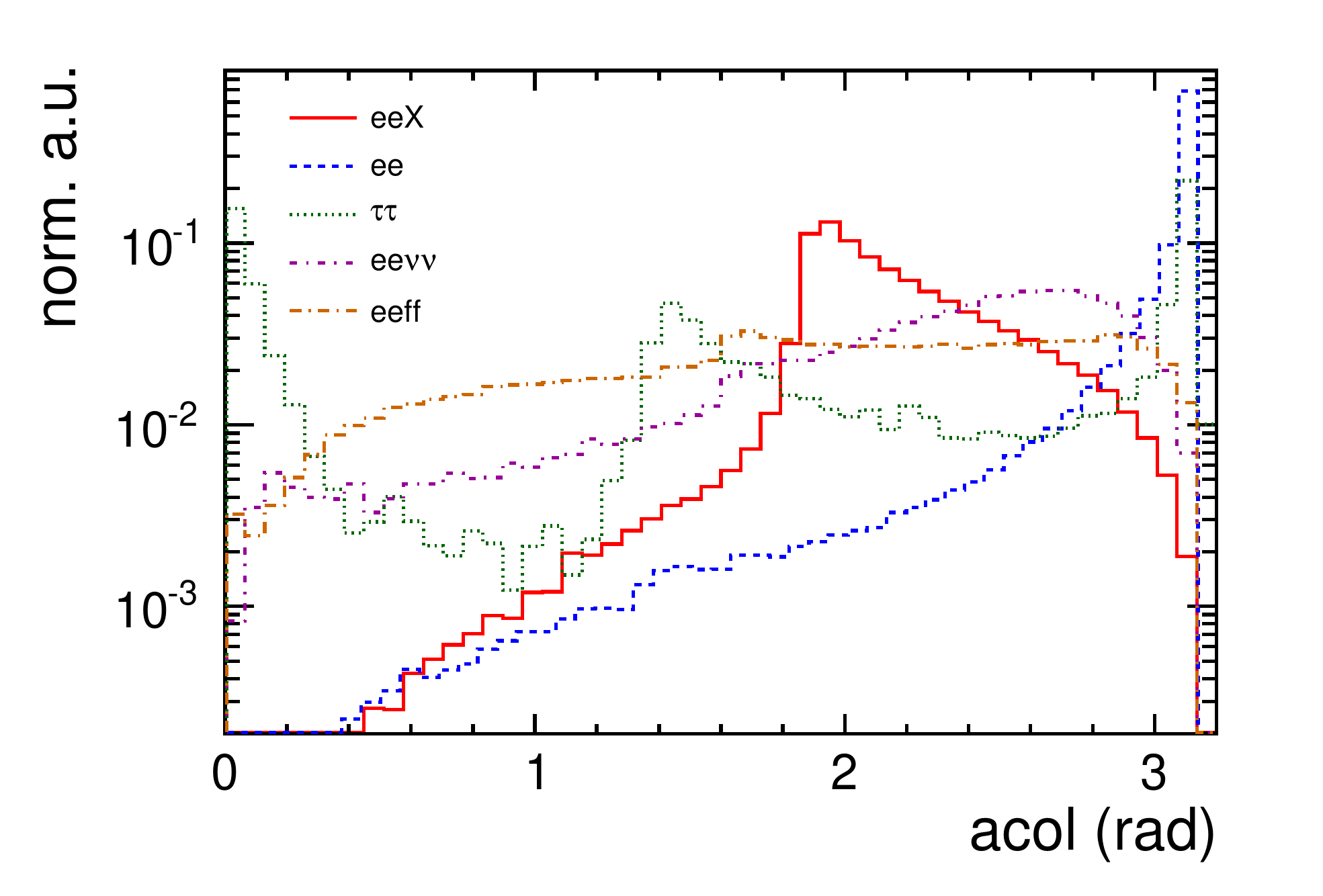}
\caption{Normalised signal and background distribution of angle $acol$ between the partners of  the di-lepton system for the $\mmX$ Channel (top) and the $\eeX$-Channel (bottom). Here $\tau\tau$ refers to the $\mu\mu$ or $ee$ created in the decay of $\tau\tau$. Note that the Pre-cuts defined in Section~\ref{sec:gensimrec} have been applied to the $\mu\mu$ background sample.}
\label{fig:var_acol}
\end{figure} 

\end{itemize}

\begin{figure}[!h]
\centering
\includegraphics[width=0.8\textwidth]{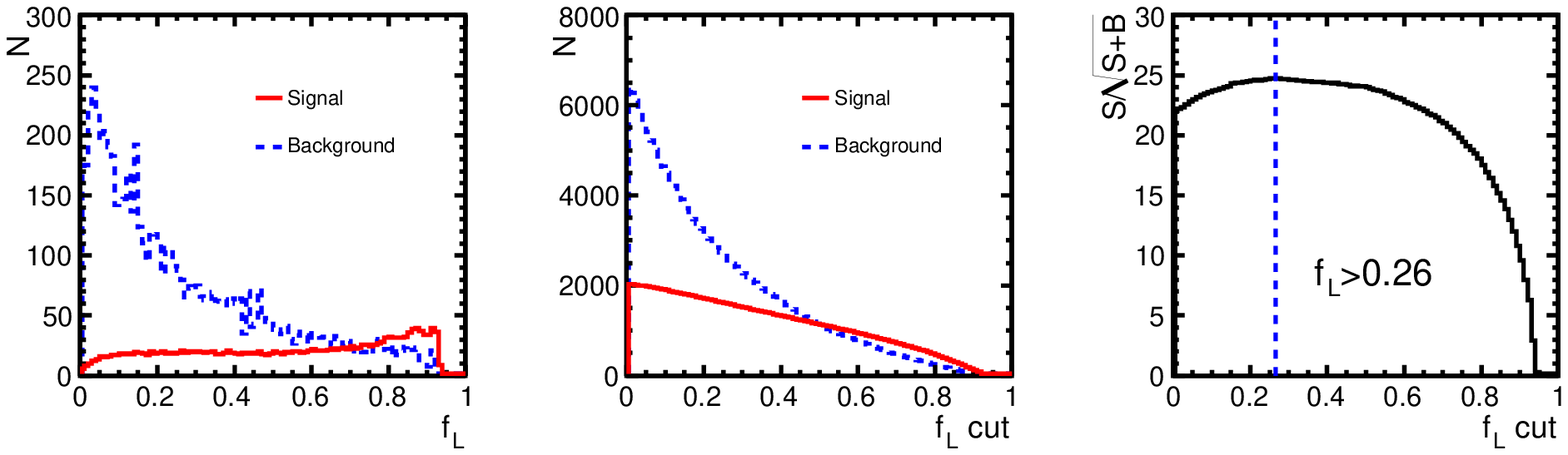}
\caption{The distributions of the Likelihood Fraction $f_L$ (left), the number of remaining  events versus the cut on $f_L$ (middle), and the significance versus $f_L$ cuts (right). The distributions are shown for the $\mu\mu X$-channel in the Model Independent Analysis and for the  polarisation mode $\eminus_L\eplus_R$.} 
\label{fig:lh_lr_mu}
\end{figure}

\begin{figure}[!h]
\centering
\includegraphics[width=0.8\textwidth]{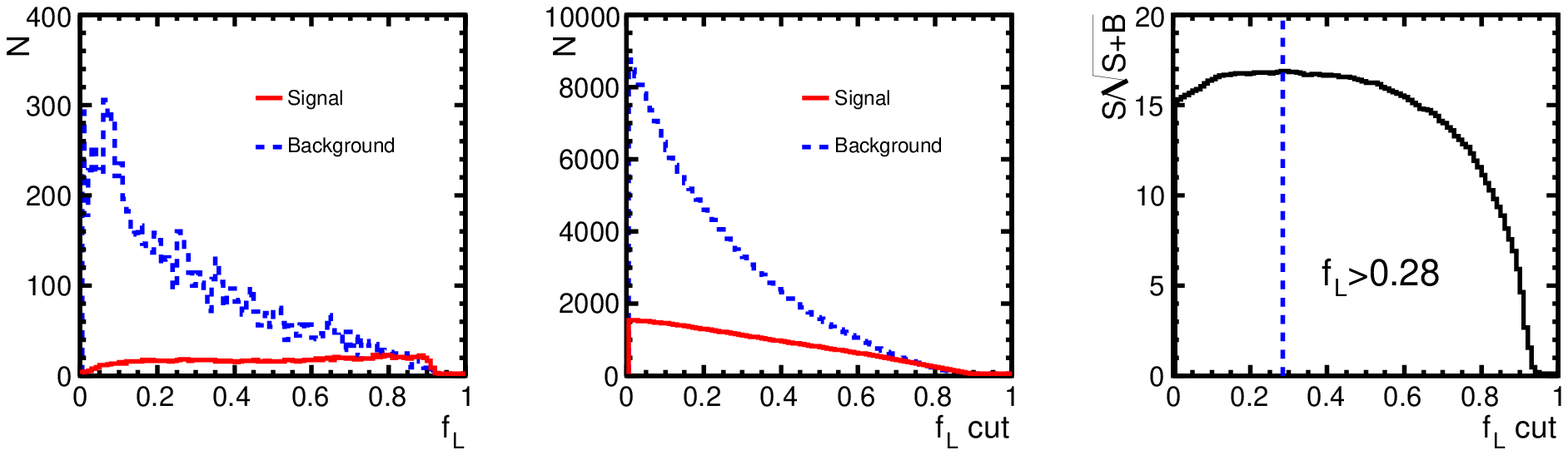}
\caption{The distributions of the Likelihood Fraction $f_L$ (left), the number of remaining  events versus the cut on $f_L$ (middle), and the significance versus $f_L$ cuts (right). The distributions are shown for the $ee X$-channel in the Model Independent Analysis and for the polarisation mode $\eminus_L\eplus_R$.}  
\label{fig:lh_lr_el}
\end{figure}

\begin{figure}[!h]
\centering
\includegraphics[width=0.8\textwidth]{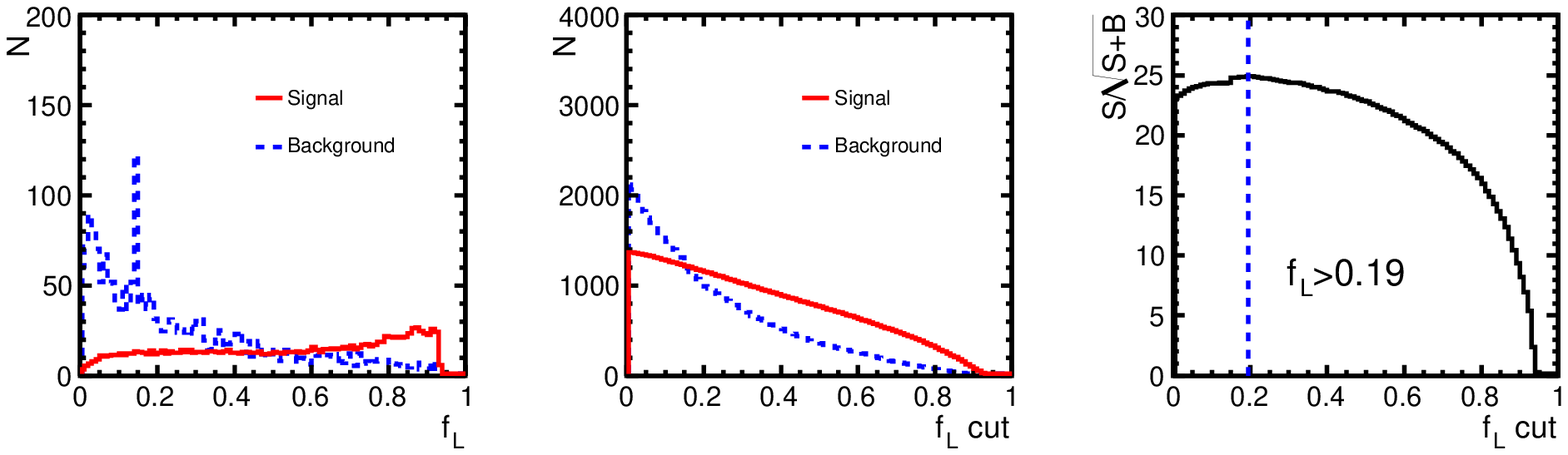}
\caption{The distributions of the Likelihood Fraction $f_L$ (left), the number of remaining  events versus the cut on $f_L$ (middle), and the significance versus $f_L$ cuts (right). The distributions are shown for the $\mu\mu X$-channel in the Model Independent Analysis and for the  polarisation mode $\eminus_R\eplus_L$.} 
\label{fig:lh_rl_mu}
\end{figure}

\begin{figure}[!h]
\centering
\includegraphics[width=0.8\textwidth]{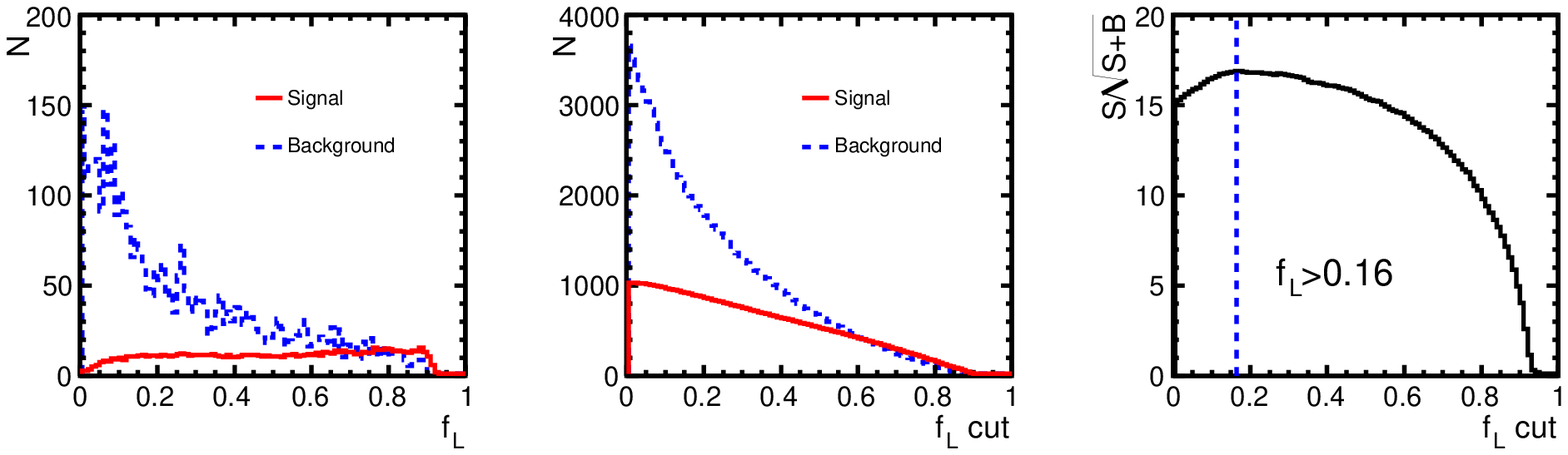}
\caption{The distributions of the Likelihood Fraction $f_L$ (left), the number of remaining  events versus the cut on $f_L$ (middle), and the significance versus $f_L$ cuts (right). The distributions are shown for the $ee X$-channel in the Model Independent Analysis and for the polarisation mode $\eminus_R\eplus_L$.}  
\label{fig:lh_rl_el}
\end{figure}

The \emph{likelihood} of an event to be the signal is defined as $L_S=\prod{P^S_i}$, where the $P^S_i$ is the probability of the event to be the signal according to the PDF of the signal of the $i$th selection variable. Similarly, the likelihood of an event to be the background is defined as $L_B=\prod{P^B_i}$. Hereafter, the \emph{Likelihood Fraction} is defined as $f_{L}=L_S/(L_S+L_B)$, which is within $(0,1)$. The Figures~\ref{fig:lh_lr_mu} through~\ref{fig:lh_rl_el} outline the optimisation procedure  in the likelihood analysis separately for the two analysis channels and polarisation modes using the four variables introduced above, for details see~\cite{phd_hengne}. It is clearly visible that the separation between signal and background improves towards small values of $f_L$. The cut on $f_L$ is optmised according to the maximum in the significance $S/\sqrt{S+B}$ where $S$ and $B$ are the number of remaining signal and background events, respectively. The cut on $f_L$ is adjusted for each polarisation mode of the incoming beams and the type of the di-lepton system under study. 
 
The final number of events also included in Tables~\ref{tab:rej_mu_mi_lr} through~\ref{tab:rej_el_mi_rl} shows that with the multi-variate analysis the number of background events are further reduced by roughly 50\% while the major part of the signal events is kept. 
 
\section{Model Dependent Analysis}\label{sec:smana}
 
\begin{figure}[hbt]
\centering
\includegraphics[width=0.8\textwidth]{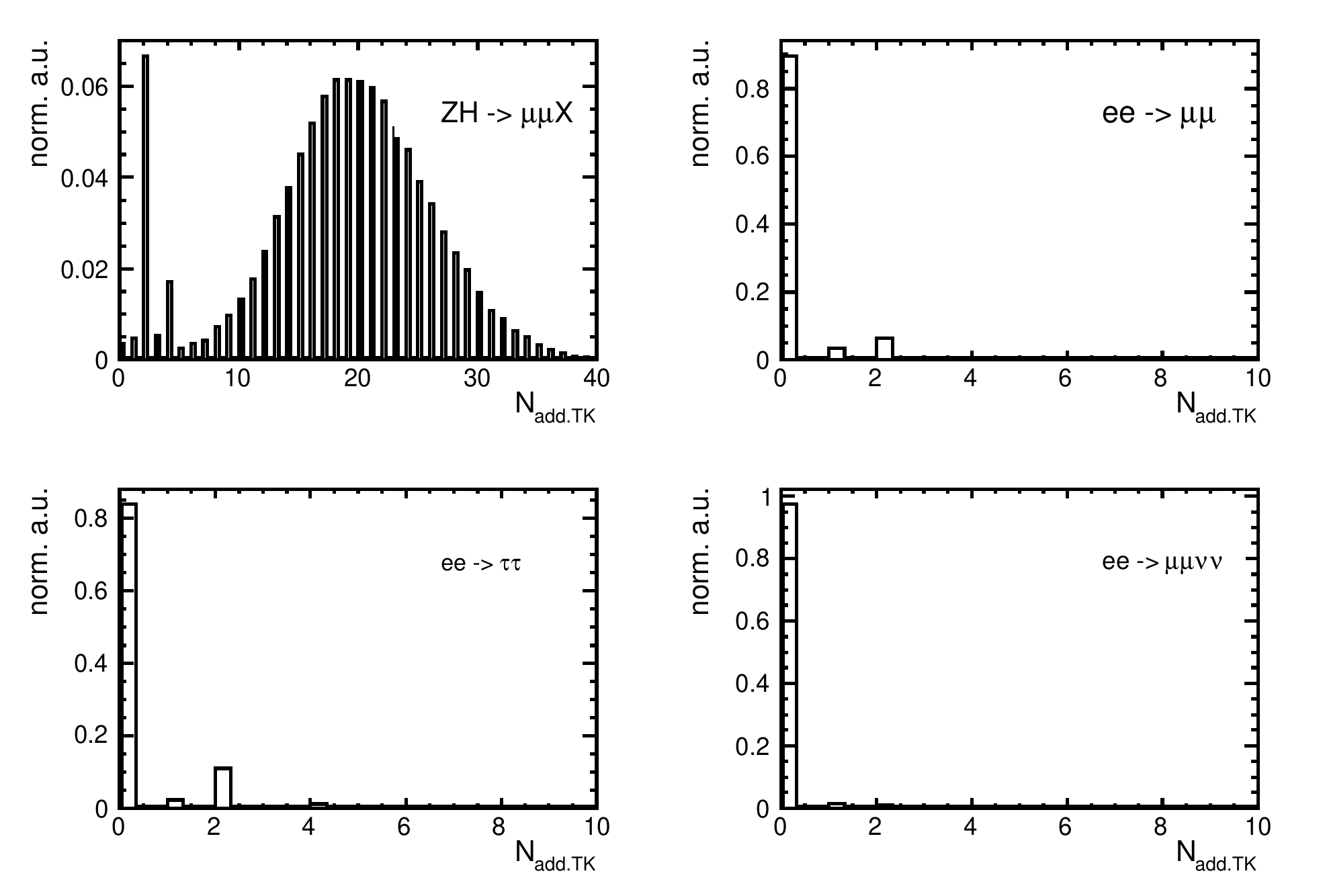}
\caption{Number of additional tracks ($N_{add.TK}$) for $\mu\mu X$, $\mu\mu$, $\tau\tau$ and $\mu\mu\nu\nu$ final states.}
\label{fig:ntks}
\end{figure}

If the analysis of the Higgs-strahlung process is restricted to modes in which the Higgs can solely decay into charged particles as e.g. suggested by the Standard Model, hence introducing a \emph{Model Dependency}, the different track multiplicities can be used for the separation of signal and background events. The Higgs boson decays into oppositely charged particles such that events with less than four tracks can be considered as background. Figure~\ref{fig:ntks} shows the number of reconstructed tracks beside the ones from the di-lepton system for final states of the types $\mu\mu X$, $\mu\mu$, $\tau\tau$ and $\mu\mu\nu\nu$.

As expected, the Higgs-strahlung process leads to a considerable amount of charged particles while processes with a low multiplicity of charged particles also create only a small number of tracks. The distributions tell that a large fraction of events have exactly two additional tracks  beside those of the di-lepton system. The two additional tracks originate
from two sources.
\begin{itemize}
\item Tracks created by charged particles by $\Higgs\rightarrow\tau^+\tau^-$ and the subsequent decays of the $\mathrm{\tau}$-Leptons into charged particles.
\item Tracks created by photon conversion. This photon may be created by initial state radiation. 
\end{itemize}
 
\begin{figure}[!h]
\centering
\includegraphics[width=0.8\textwidth]{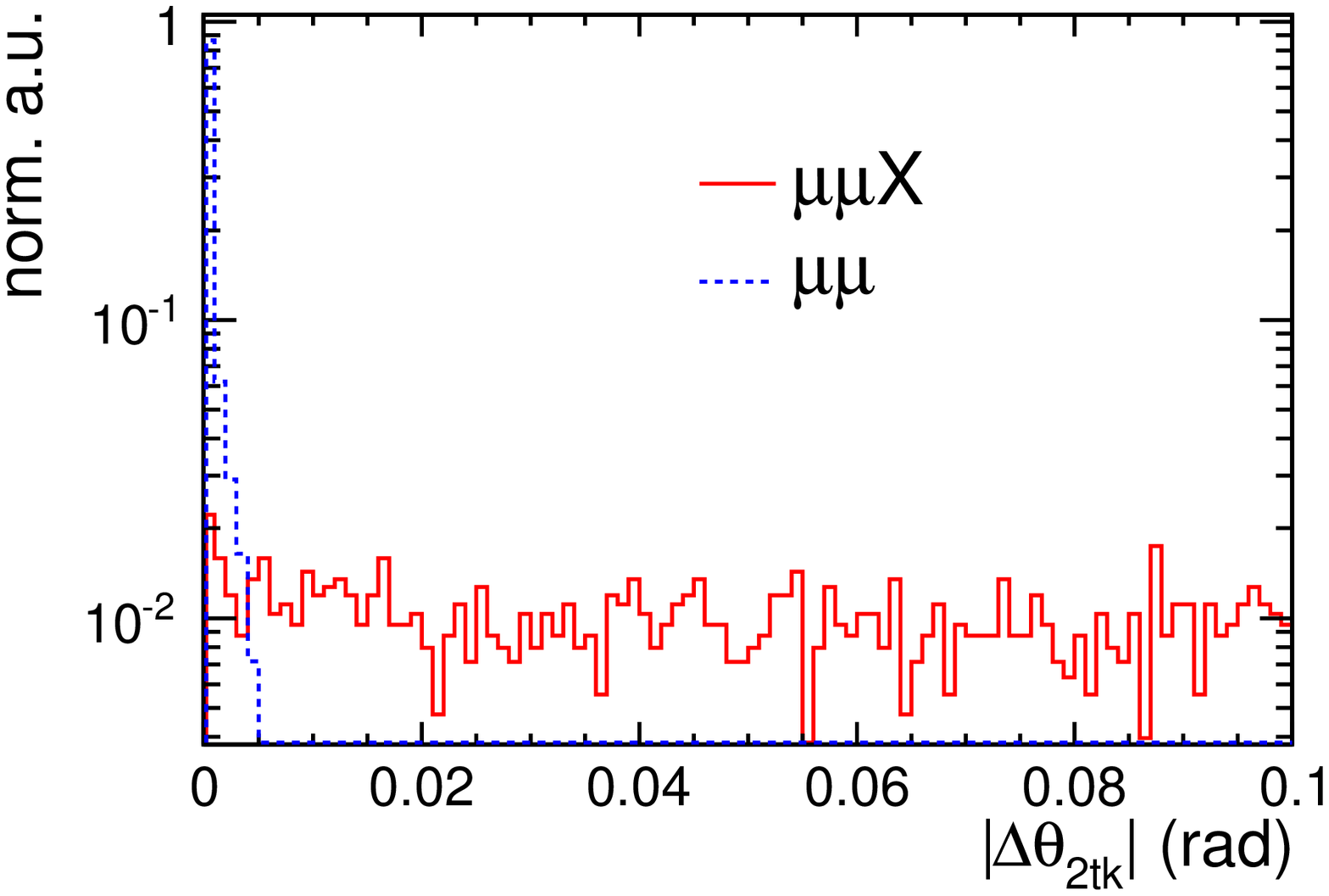}
\caption{Distribution of $\Delta\theta_{2tk}$, which is the $\Delta\theta$ between two additional tracks for $N_{add.TK}=2$, for signal events ($\mu\mu X$)  and background by muon pair production ($\mu\mu$).} 
\label{fig:dtha2tk}
\end{figure}

\begin{figure}[hbt]
\centering
\includegraphics[width=0.8\textwidth]{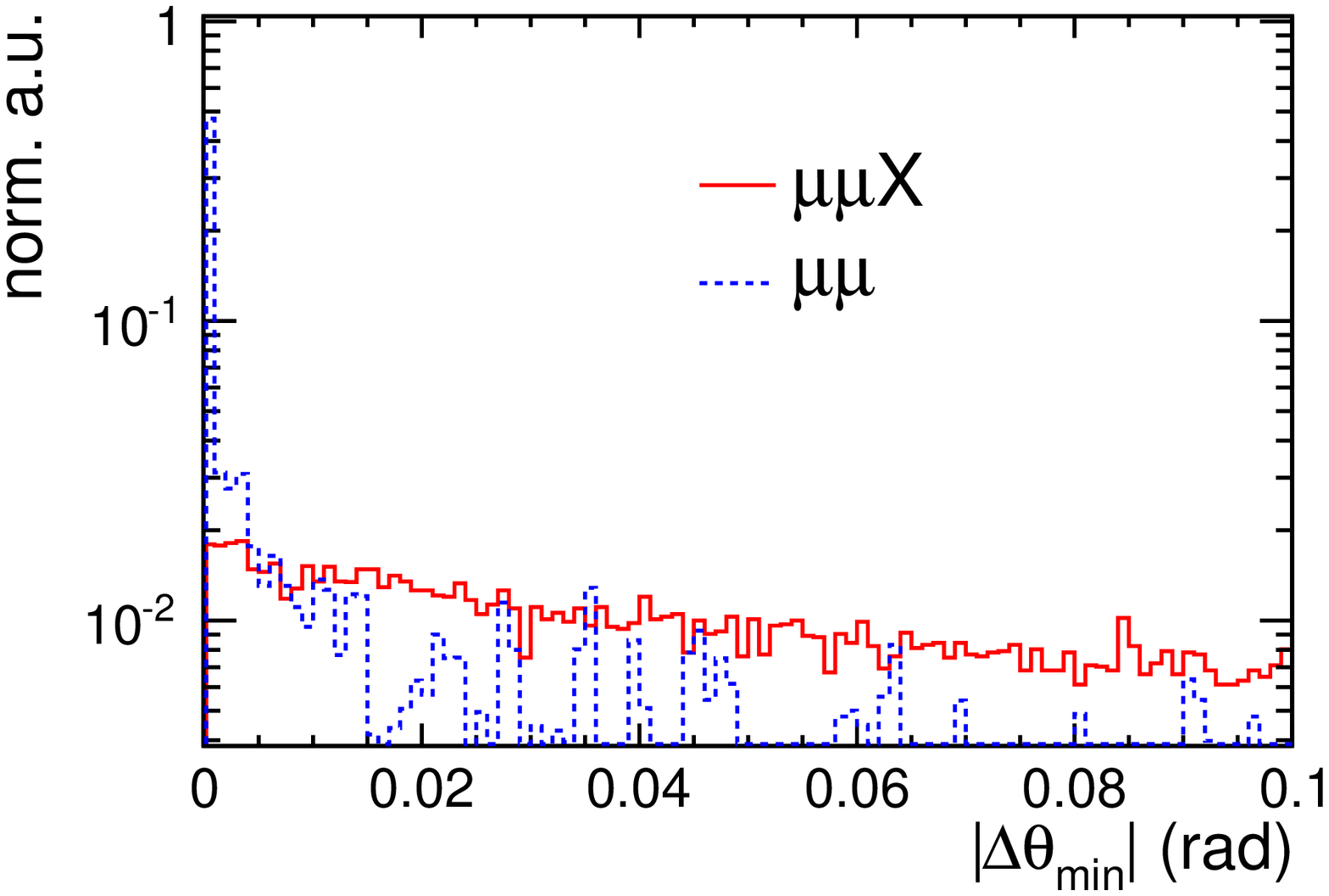}
\caption{Distribution of $\Delta\theta$ between muon candidates and additional tracks, for the signal events ($\mu\mu X$) and background by muon pair creation ($\mu\mu$).} 
\label{fig:dthamin_mu}
\end{figure}

The first type of events need to be kept in the signal as the $\tau$-Leptons constitute an important analyser to determine e.g. quantum numbers like \emph{CP} of the Higgs boson~\cite{Higgswg}. The second type of events can be rejected  by taking into account that the opening angle of $e^+e^-$ pair created by photon conversion is expected to be very small. This is underlined by Figure~\ref{fig:dtha2tk} which shows the angular difference $\Delta\theta_{2tk}$ between the two additional tracks. 
While the signal events result in a flat distribution, the background events exhibit a strong maximum around $\Delta\theta_{2tk}=0$. This observation motivates the a cut  $|\Delta\theta_{2tk}|>0.01$. 
The di-lepton system of a given type might be contaminated from particles of the other type. Therefore, the polar angle of each of the two particles of the di-lepton system is also compared with the polar angle of the additional tracks, defining the observable $\Delta\theta_{min}$ as shown in Figure~\ref{fig:dthamin_mu}. Again, a strong maximum around $\Delta\theta_{min}=0$ can be observed which suggests the cut $\Delta\theta_{min}>0.01$ 

Tables~\ref{tab:rej_mu_sm_lr} through~\ref{tab:rej_el_sm_rl} give the resulting number of events after each cut applied under the assumption that the Higgs boson decays into Standard Model particles. Note, that the cut on the transverse momentum of the di-lepton system has been omitted in order to maximise the number of signal events. The combination of cuts will be later referred as \emph{MD Cuts}. The numbers in the tables show that the cuts introduced for the additional tracks allow for an entire suppression of backgrounds with a small multiplicity of charged particles in the final state. It has to be pointed out that in particular the background from Bhabha events can be removed almost completely. On the other hand at least 50\% of the signal is retained by the cuts.

Again the remaining major background is given by events in which vector bosons pairs are produced. This background is further reduced by a likelihood analysis as described above. The results of this likelihood analysis is also given in Tables~\ref{tab:rej_mu_sm_lr} through~\ref{tab:rej_el_sm_rl}. 
From these tables, it can be deferred that the $f_L$ cuts reject the background from $\Zzero$ pair production by a factor of two, and reduce the signal by only 10\%. At the same time, the background $\mu\mu$, $ee$, $\tau\tau$, $\mu\mu\nu\nu$ and $ee\nu\nu$ is entirely suppressed.

\newpage

\subsection{Tables of Background Rejection}\label{sec:zhana_rejtab}

\begin{table}[!htbp]
\centering 
\begin{tabular}{|l|r r r r r|}
\hline
\multicolumn{6}{|c|}{\bf MI, $\mathbf{\mmX}$, $\mathbf{\polLR}$}  \\ \hline

$N_{evts}$ Remained	&\multicolumn{1}{c}{$\mmX$}	&	$\mpmm$	&	$\tptm$	&	$\mpmm\nu\nu$	&	$\mpmm ff$	\\ \hline \hline
Before any restriction	&	2918	($	100.0	\%$)&	2.6M	&	1.6M	&	111k	&	317k	\\ \hline
+ Lepton ID &&&&&\\
+ Tightened Pre-Cuts	&	2472	($	84.72	\%$)&	9742	&	4582	&	9268	&	8175	\\ \hline
+ $\Ptdl>20\GeV$	&	2408	($	82.50	\%$)&	7862	&	3986	&	8462	&	7222	\\ \hline
+ $\Mdl \in (80,100)\GeV$	&	2292	($	78.54	\%$)&	6299	&	2679	&	5493	&	5658	\\ \hline
+ $acop \in (0.2,3.0)$	&	2148	($	73.61	\%$)&	5182	&	112	&	5179	&	5083	\\ \hline
+ $\Ptbal > 10\GeV$	&	2107	($	72.20	\%$)&	335	&	80	&	4705	&	4706	\\ \hline
+ $\dthattk>0.01$	&	2104	($	72.11	\%$)&	149	&	80	&	4647	&	4676	\\ \hline
+ $\cthamiss<0.99$	&	2046	($	70.09	\%$)&	82	&	80	&	4647	&	3614	\\ \hline
+ $\Mrecoil \in (115, 150)\GeV$	&	2028	($	69.48	\%$)&	75	&	80	&	3642	&	2640	\\ \hline
+ $f_L>0.26$	&	1596	($	54.68	\%$)&	41	&	0	&	1397	&	1125	\\ \hline

\end{tabular}
\caption{Number of events left after each cut for the $\mmX$ channel in the MI Analysis. Fractions of number of events remained of the Higgs-Strahlung process are given inside parentheses, the last one gives the efficiency of signal selection. The polarisation mode is $\polLR$. }
\label{tab:rej_mu_mi_lr}
\end{table}

\begin{table}[!htbp]
\centering 
\begin{tabular}{|l|r r r r r|}
\hline
\multicolumn{6}{|c|}{\bf MI, $\mathbf{\eeX}$, $\mathbf{\polLR}$}  \\ \hline

$N_{evts}$ Remained	&\multicolumn{1}{c}{$\eeX$}&	$\epem$	&	$\tptm$	&	$\epem\nu\nu$	&	$\epem ff$	\\ \hline \hline
Before any restriction	&	3138	($	100.0	\%$)	&	4.3G	&	1.6M	&	147k	&	110k	\\ \hline
+ Lepton ID &&&&&\\
+ Tightened Pre-Cuts	&	2019	($	64.33	\%$)&	43607	&	6422	&	13196	&	12548	\\ \hline
+ $\Ptdl>20\GeV$	&	1962	($	62.50	\%$)&	39152	&	5551	&	12054	&	10583	\\ \hline
+ $\Mdl \in (80,100)\GeV$	&	1755	($	55.93	\%$)&	25501	&	3806	&	7786	&	7509	\\ \hline
+ $acop \in (0.2,3.0)$	&	1645	($	52.41	\%$)&	23228	&	245	&	7239	&	6739	\\ \hline
+ $\Ptbal > 10\GeV$	&	1606	($	51.16	\%$)&	1725	&	157	&	6286	&	5904	\\ \hline
+ $\dthattk>0.01$	&	1603	($	51.09	\%$)&	990	&	157	&	6150	&	5844	\\ \hline
+ $\cthamiss<0.99$	&	1564	($	49.83	\%$)&	679	&	157	&	6149	&	4643	\\ \hline
+ $\Mrecoil \in (115, 150)\GeV$	&	1539	($	49.04	\%$)&	576	&	41	&	4824	&	3335	\\ \hline
+ $f_L>0.28$	&	1153	($	36.74	\%$)&	243	&	29	&	2019	&	1217	\\ \hline

\end{tabular}
\caption{Number of events left after each cut for the $\eeX$ channel in the MI Analysis. Fractions of number of events remained of the Higgs-Strahlung process are given inside parentheses, the last one gives the efficiency of signal selection. The polarisation mode is $\polLR$. }
\label{tab:rej_el_mi_lr}
\end{table}

\begin{table}[!htbp]
\centering 
\begin{tabular}{|l|r r r r r|}
\hline
\multicolumn{6}{|c|}{\bf MI, $\mathbf{\mmX}$, $\mathbf{\polRL}$}  \\ \hline

$N_{evts}$ Remained	&\multicolumn{1}{c}{$\mmX$}&	$\mpmm$	&	$\tptm$	&	$\mpmm\nu\nu$	&	$\mpmm ff$	\\ \hline \hline
Before any restriction	&	1967	($	100.0	\%$)&	2.0M	&	1.2M	&	9k	&	291k	\\ \hline
+ Lepton ID &&&&&\\
+ Tightened Pre-Cuts	&	1667	($	84.73	\%$)&	6696	&	3471	&	1048	&	5324	\\ \hline
+ $\Ptdl>20\GeV$	&	1623	($	82.48	\%$)&	5419	&	3037	&	957	&	4600	\\ \hline
+ $\Mdl \in (80,100)\GeV$	&	1544	($	78.47	\%$)&	4347	&	2092	&	702	&	3530	\\ \hline
+ $acop \in (0.2,3.0)$	&	1448	($	73.60	\%$)&	3592	&	113	&	656	&	3169	\\ \hline
+ $\Ptbal > 10\GeV$	&	1421	($	72.21	\%$)&	229	&	81	&	632	&	2873	\\ \hline
+ $\dthattk>0.01$	&	1419	($	72.10	\%$)&	101	&	81	&	625	&	2851	\\ \hline
+ $\cthamiss<0.99$	&	1379	($	70.10	\%$)&	54	&	81	&	625	&	2065	\\ \hline
+ $\Mrecoil \in (115, 150)\GeV$	&	1367	($	69.49	\%$)&	50	&	81	&	487	&	1506	\\ \hline
+ $f_L>0.19$	&	1165	($	59.20	\%$)&	28	&	0	&	243	&	752	\\ \hline

\end{tabular}
\caption{Number of events left after each cut for the $\mmX$ channel in the MI Analysis. Fractions of number of events remained of the Higgs-Strahlung process are given inside parentheses, the last one gives the efficiency of signal selection. The polarisation mode is $\polRL$. }
\label{tab:rej_mu_mi_rl}
\end{table}

\begin{table}[!htbp]
\centering 
\begin{tabular}{|l|r r r r r|}
\hline
\multicolumn{6}{|c|}{\bf MI, $\mathbf{\eeX}$, $\mathbf{\polRL}$}  \\ \hline

$N_{evts}$ Remained	&\multicolumn{1}{c}{$\eeX$}&	$\epem$	&	$\tptm$	&	$\epem\nu\nu$	&	$\epem ff$	\\ \hline \hline
Before any restriction	&	2107	($	100.0	\%$)&	4.3G	&	1.2M	&	17k	&	1.1M	\\ \hline
+ Lepton ID &&&&&\\
+ Tightened Pre-Cuts	&	1352	($	64.16	\%$)&	40896	&	5257	&	1469	&	10198	\\ \hline
+ $\Ptdl>20\GeV$	&	1313	($	62.33	\%$)&	36742	&	4546	&	1351	&	8430	\\ \hline
+ $\Mdl \in (80,100)\GeV$	&	1177	($	55.88	\%$)&	23993	&	3051	&	943	&	5909	\\ \hline
+ $acop \in (0.2,3.0)$	&	1103	($	52.36	\%$)&	21846	&	107	&	881	&	5266	\\ \hline
+ $\Ptbal > 10\GeV$	&	1077	($	51.11	\%$)&	1612	&	92	&	805	&	4517	\\ \hline
+ $\dthattk>0.01$	&	1076	($	51.05	\%$)&	927	&	92	&	799	&	4465	\\ \hline
+ $\cthamiss<0.99$	&	1050	($	49.82	\%$)&	638	&	92	&	799	&	3484	\\ \hline
+ $\Mrecoil \in (115, 150)\GeV$	&	1033	($	49.04	\%$)&	539	&	12	&	586	&	2521	\\ \hline
+ $f_L>0.16$	&	909	($	43.14	\%$)&	326	&	4	&	368	&	1294	\\ \hline

\end{tabular}
\caption{Number of events left after each cut for the $\eeX$ channel in the MI Analysis. Fractions of number of events remained of the Higgs-Strahlung process are given inside parentheses, the last one gives the efficiency of signal selection. The polarisation mode is $\polRL$. }
\label{tab:rej_el_mi_rl}
\end{table}

\begin{table}[!htbp]
\centering 
\begin{tabular}{|l|r r r r r|}
\hline
\multicolumn{6}{|c|}{\bf MD, $\mathbf{\mmX}$, $\mathbf{\polLR}$}  \\ \hline

$N_{evts}$ Remained	&\multicolumn{1}{c}{$\mmX$}&	$\mpmm$	&	$\tptm$	&	$\mpmm\nu\nu$	&	$\mpmm ff$	\\ \hline \hline
Before any restriction	&	2918	($	100.0	\%$)&	2.6M	&	1.6M	&	111k	&	317k	\\ \hline
+ Lepton ID &&&&&\\
+ Tightened Pre-Cuts	&	2472	($	84.72	\%$)&	9742	&	4582	&	9268	&	8175	\\ \hline
+ $\Naddtks>1$	&	2453	($	84.05	\%$)&	604	&	842	&	145	&	6321	\\ \hline
+ $\dthattk>0.01$	&	2449	($	83.91	\%$)&	63	&	816	&	14	&	6254	\\ \hline
+ $\dthamin>0.01$	&	2417	($	82.81	\%$)&	38	&	261	&	1	&	5711	\\ \hline
+ $acop \in (0.2,3.0)$	&	2256	($	77.29	\%$)&	32	&	0	&	1	&	5051	\\ \hline
+ $\cthamiss<0.99$	&	2189	($	75.00	\%$)&	16	&	0	&	1	&	3843	\\ \hline
+ $\Mrecoil \in (115, 150)\GeV$	&	2154	($	73.81	\%$)&	15	&	0	&	1	&	2830	\\ \hline
+ $f_L>0.17$	&	1911	($	65.49	\%$)&	11	&	0	&	0	&	1387	\\ \hline

\end{tabular}
\caption{Number of events left after each cut for the $\mmX$ channel in the MD Analysis. Fractions of number of events remained of the Higgs-Strahlung process are given inside parentheses, the last one gives the efficiency of signal selection. The polarisation mode is $\polLR$. }
\label{tab:rej_mu_sm_lr}
\end{table}

\begin{table}[!htbp]
\centering 
\begin{tabular}{|l|r r r r r|}
\hline
\multicolumn{6}{|c|}{\bf MD, $\mathbf{\eeX}$, $\mathbf{\polLR}$}  \\ \hline

$N_{evts}$ Remained	&\multicolumn{1}{c}{$\eeX$}&	$\epem$	&	$\tptm$	&	$\epem\nu\nu$	&	$\epem ff$	\\ \hline \hline
Before any restriction	&	3138	($	100.0	\%$)&	4.3G	&	1.6M	&	147k	&	110k	\\ \hline
+ Lepton ID &&&&&\\
+ Tightened Pre-Cuts	&	2019	($	64.33	\%$)&	43607	&	6422	&	13196	&	12548	\\ \hline
+ $\Naddtks>1$	&	2004	($	63.87	\%$)&	3136	&	1740	&	374	&	10202	\\ \hline
+ $\dthattk>0.01$	&	2001	($	63.77	\%$)&	655	&	1073	&	79	&	10095	\\ \hline
+ $\dthamin>0.01$	&	1969	($	62.75	\%$)&	155	&	128	&	6	&	9271	\\ \hline
+ $acop \in (0.2,3.0)$	&	1840	($	58.62	\%$)&	134	&	0	&	6	&	8366	\\ \hline
+ $\cthamiss<0.99$	&	1792	($	57.11	\%$)&	91	&	0	&	6	&	6696	\\ \hline
+ $\Mrecoil \in (115, 150)\GeV$	&	1731	($	55.16	\%$)&	73	&	0	&	1	&	4950	\\ \hline
+ $f_L>0.27$	&	1378	($	43.90	\%$)&	27	&	0	&	0	&	1652	\\ \hline

\end{tabular}
\caption{Number of events left after each cut for the $\eeX$ channel in the MD Analysis. Fractions of number of events remained of the Higgs-Strahlung process are given inside parentheses, the last one gives the efficiency of signal selection. The polarisation mode is $\polLR$. }
\label{tab:rej_el_sm_lr}
\end{table}

\begin{table}[!htbp]
\centering 
\begin{tabular}{|l|r r r r r|}
\hline
\multicolumn{6}{|c|}{\bf MD, $\mathbf{\mmX}$, $\mathbf{\polRL}$}  \\ \hline

$N_{evts}$ Remained	&\multicolumn{1}{c}{$\mmX$}&	$\mpmm$	&	$\tptm$	&	$\mpmm\nu\nu$	&	$\mpmm ff$	\\ \hline \hline
Before any restriction	&	1967	($	100.0	\%$)&	2.0M	&	1.2M	&	9k	&	291k	\\ \hline
+ Lepton ID &&&&&\\
+ Tightened Pre-Cuts	&	1667	($	84.73	\%$)&	6696	&	3471	&	1048	&	5324	\\ \hline
+ $\Naddtks>1$	&	1654	($	84.07	\%$)&	415	&	391	&	9	&	4160	\\ \hline
+ $\dthattk>0.01$	&	1651	($	83.93	\%$)&	41	&	379	&	0	&	4108	\\ \hline
+ $\dthamin>0.01$	&	1629	($	82.81	\%$)&	22	&	105	&	0	&	3739	\\ \hline
+ $acop \in (0.2,3.0)$	&	1522	($	77.34	\%$)&	20	&	0	&	0	&	3312	\\ \hline
+ $\cthamiss<0.99$	&	1476	($	75.03	\%$)&	11	&	0	&	0	&	2438	\\ \hline
+ $\Mrecoil \in (115, 150)\GeV$	&	1453	($	73.85	\%$)&	10	&	0	&	0	&	1803	\\ \hline
+ $f_L>0.17$	&	1289	($	65.53	\%$)&	8	&	0	&	0	&	875	\\ \hline

\end{tabular}
\caption{Number of events left after each cut for the $\mmX$ channel in the MD Analysis. Fractions of number of events remained of the Higgs-Strahlung process are given inside parentheses, the last one gives the efficiency of signal selection. The polarisation mode is $\polRL$. }
\label{tab:rej_mu_sm_rl}
\end{table}

\begin{table}[!htbp]
\centering 
\begin{tabular}{|l|r r r r r|}
\hline
\multicolumn{6}{|c|}{\bf MD, $\mathbf{\eeX}$, $\mathbf{\polRL}$}  \\ \hline

$N_{evts}$ Remained	&\multicolumn{1}{c}{$\eeX$}&	$\epem$	&	$\tptm$	&	$\epem\nu\nu$	&	$\epem ff$	\\ \hline \hline
Before any restriction	&	2107	($	100.0	\%$)&	4.3G	&	1.2M	&	17k	&	1.1M	\\ \hline
+ Lepton ID &&&&&\\
+ Tightened Pre-Cuts	&	1352	($	64.16	\%$)&	40896	&	5257	&	1469	&	10198	\\ \hline
+ $\Naddtks>1$	&	1342	($	63.69	\%$)&	2935	&	1500	&	22	&	8227	\\ \hline
+ $\dthattk>0.01$	&	1340	($	63.60	\%$)&	617	&	859	&	4	&	8133	\\ \hline
+ $\dthamin>0.01$	&	1319	($	62.59	\%$)&	146	&	57	&	0	&	7388	\\ \hline
+ $acop \in (0.2,3.0)$	&	1232	($	58.47	\%$)&	125	&	0	&	0	&	6651	\\ \hline
+ $\cthamiss<0.99$	&	1201	($	57.00	\%$)&	84	&	0	&	0	&	5265	\\ \hline
+ $\Mrecoil \in (115, 150)\GeV$	&	1161	($	55.10	\%$)&	67	&	0	&	0	&	3886	\\ \hline
+ $f_L>0.32$	&	889	($	42.20	\%$)&	20	&	0	&	0	&	1119	\\ \hline

\end{tabular}
\caption{Number of events left after each cut for the $\eeX$ channel in the MI Analysis. Fractions of number of events remained of the Higgs-Strahlung process are given inside parentheses, the last one gives the efficiency of signal selection. The polarisation mode is $\polRL$. }
\label{tab:rej_el_sm_rl}
\end{table}

%\subsection{Background from Photon Scattering}
%Check Kazutoshis analysis ....
\clearpage
\section{Extraction of Higgs Mass and the Higgs production Cross Section}
 
In the previous section the criteria to select the signal events and to suppress the background from various sources have been introduced and applied. The remaining spectra are a superposition of signal and background events convoluted with beam effects. In the following, the relevant observables as the Higgs boson mass $\mH$ and the total Higgs-strahlung cross section $\sigma$  are extracted. Note in passing, that the results for the $eeX$-channel will always contain a small admixture of the $\Zzero\Zzero$ fusion process. 

As indicated above, the resulting spectrum is composed by several components. This motivates to approximate this spectrum in a non-parametric way using a Kernel Estimation as introduced in~\cite{cranmer1} and applied in~\cite{cranmer2, opal}. In order to reduce the effort 
of finding a parent function using either the already simulated data set or by simulating another  independent set of data a so-called {\it Simplified Kernel Estimation} is proposed.

The signal spectrum is approximated by the following function:

\begin{equation}
F_{S}(x)=\frac{1}{N}\sum^{m}_{j=1}n_{j}G(x;t_{j};h_{j})
\end{equation}
with
\begin{equation}
h_{j}=\biggl(\frac{4}{3}\biggr)^{1/5}N^{-1/5}\Delta x\sqrt{\frac{N}{n_j}},
\end{equation}

Here $G$ is a Gaussian with the parameters $\mu=t_j$ where $t_j$ is the center of the $jth$ bin of a histogram with $m$ bins and $\sigma=h_j$ where $h_j$ is the {\it smoothing parameter} of bandwidth of the individual Gaussians placed around the bin centers. The parameter $\Delta x$ is assumed to the
the standard deviation in each bin and $\frac{n_j}{\Delta x N}$ is an estimate for the parent distribution. By the transformation $x \rightarrow x'=x-m_H$, the approximated function becomes sensitive to the
value of the Higgs-Mass.

The background is approximated by a second order Chebyshev polynomial. By this statistical fluctuations in the remaining background events are smoothened. Using this polynomial as input the background is generated again with 40 times higher statistics. Therefore, statistical uncertainties are 
nearly excluded. The combination of signal and background is finally fitted by the sum of the signal and the background functions. 

\begin{figure}[!h]
\centering
\includegraphics[width=0.99\textwidth]{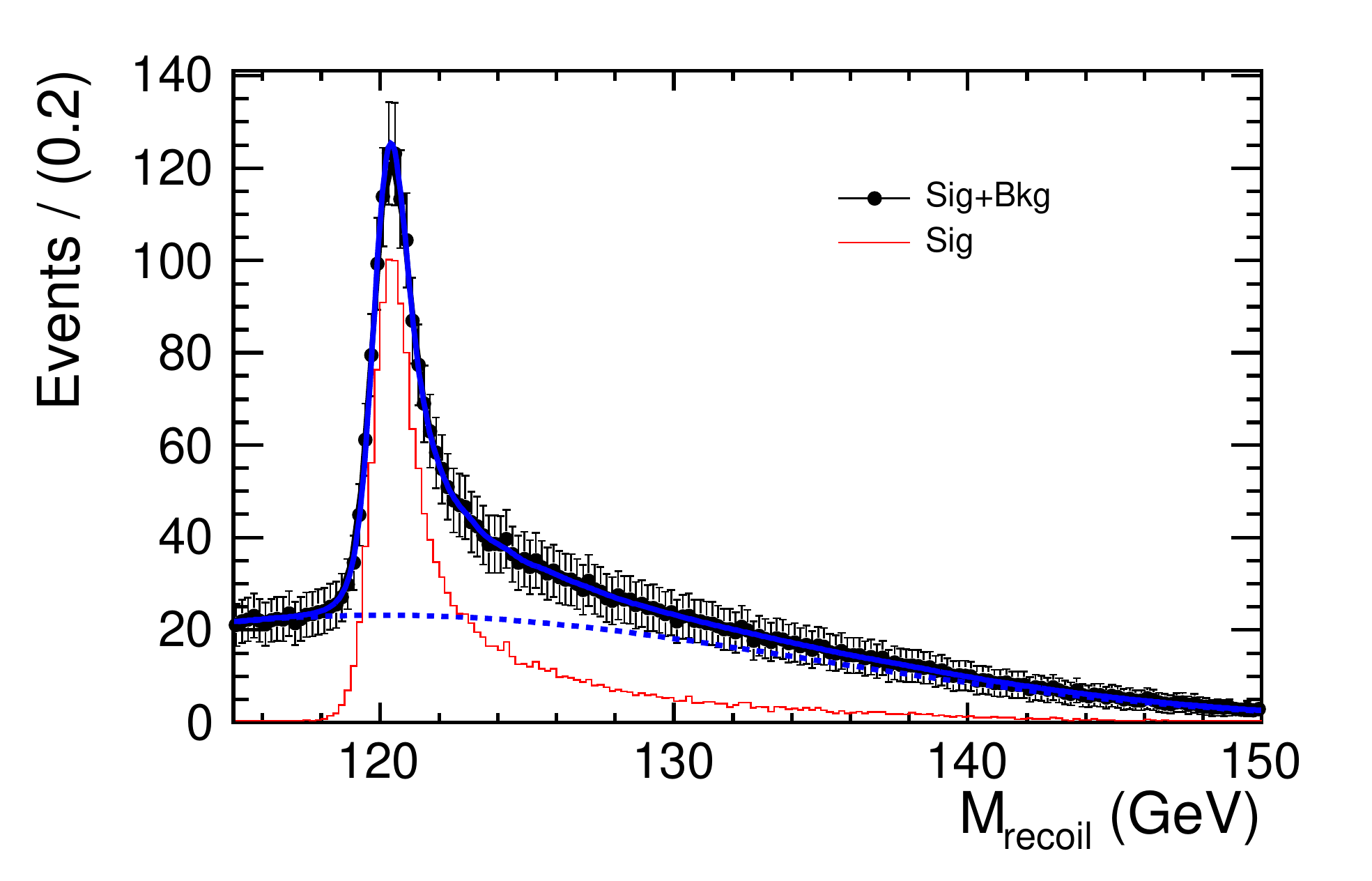}
\includegraphics[width=0.99\textwidth]{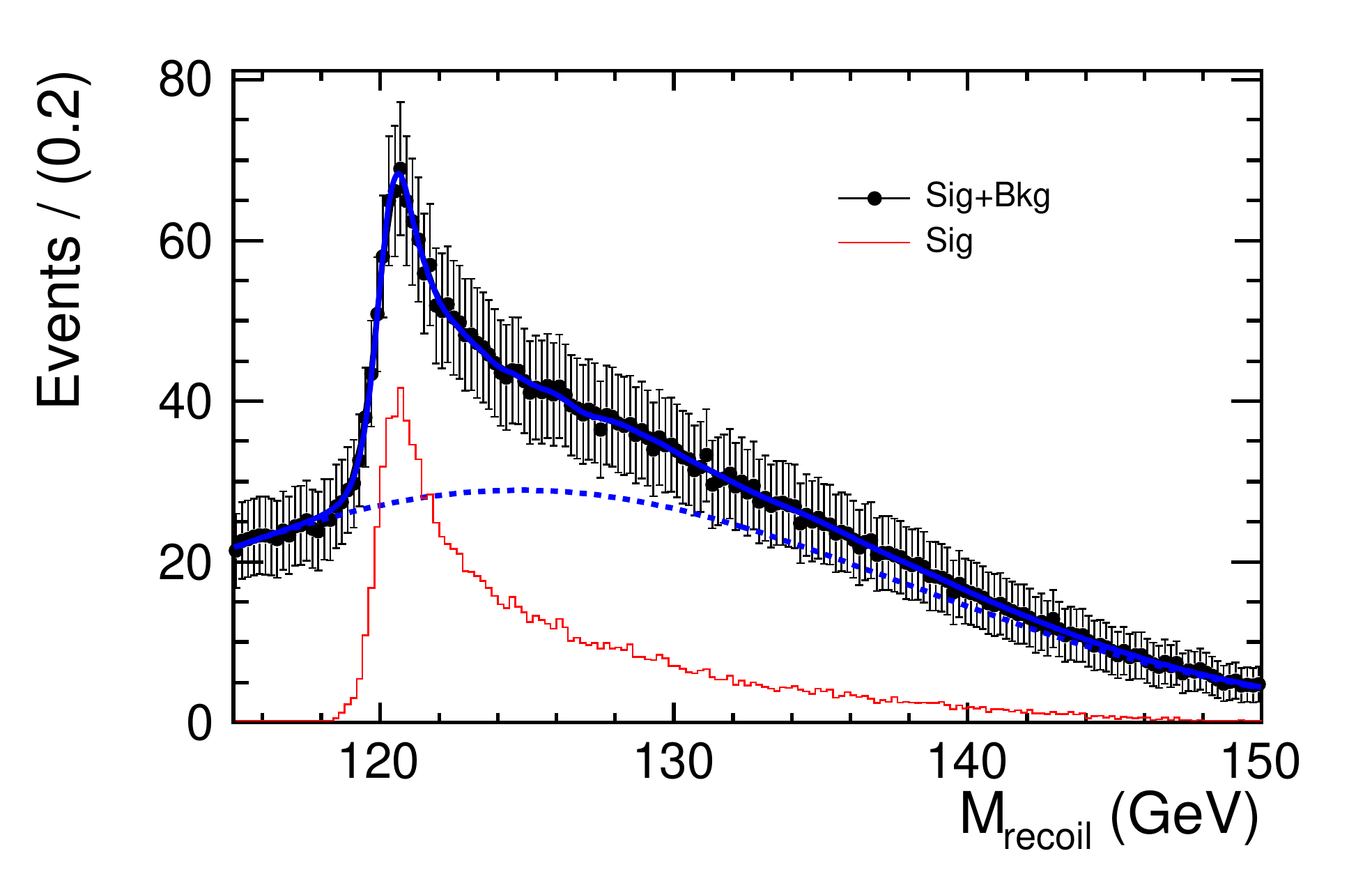}
\caption{Reconstructed Higgs mass spectrum together with the sum of underlying background for the \emph{Model Independent Analysis} for the $\mu\mu X$-channel (top) and $eeX$-channel (bottom).  The polarisation mode is $\eminus_L\eplus_R$.  The lines show the fits using the Simplified Kernel Estimation fitting formula to the signal and a polynomial of second order to the background as explained in the text.} 
\label{fig:fit_mi_lr_ke}
\end{figure}

\begin{figure}[h]
\centering
\includegraphics[width=0.99\textwidth]{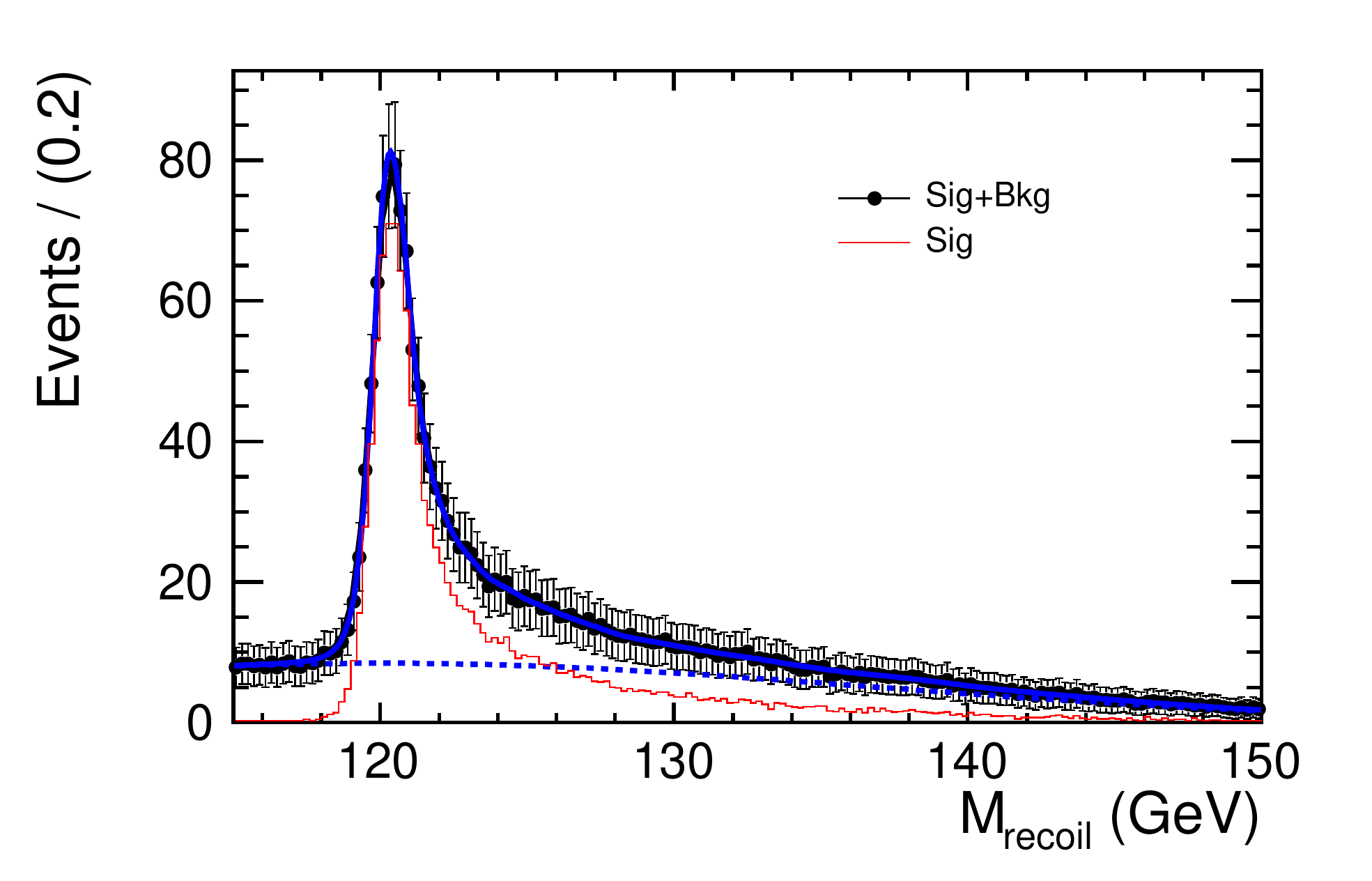}
\includegraphics[width=0.99\textwidth]{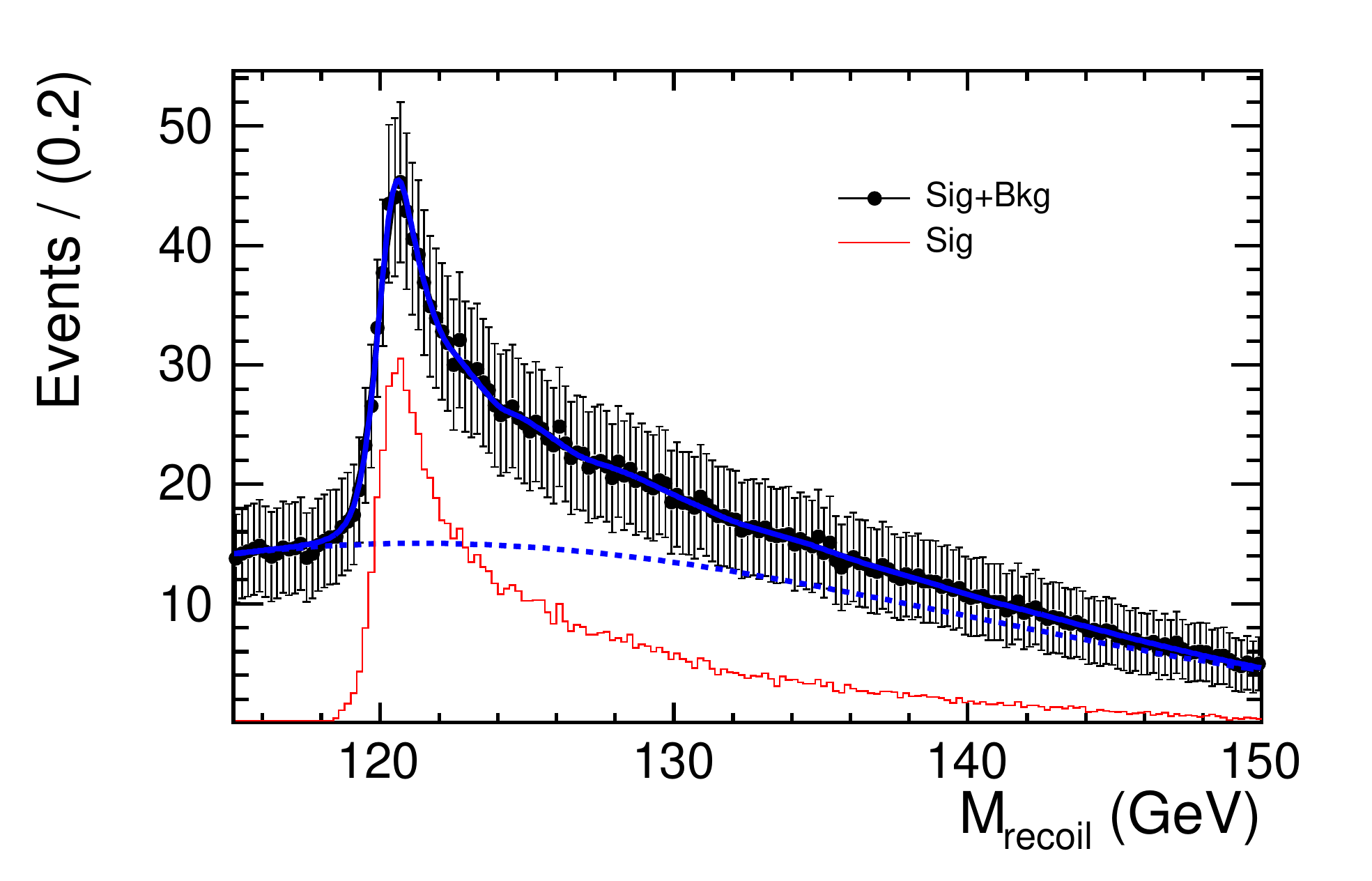}
\caption{
Reconstructed Higgs mass spectrum together with the sum of underlying background for the \emph{Model Independent Analysis} for the $\mu\mu X$-channel (top) and $eeX$-channel (bottom).  The polarisation mode is $\eminus_R\eplus_L$.  The lines show the fits using
 the Simplified Kernel Estimation fitting formula 
 to the signal and a polynomial of second order to the background as explained in the text.} 
\label{fig:fit_mi_rl_ke}
\end{figure}

\begin{figure}[h]
\centering
\includegraphics[width=0.99\textwidth]{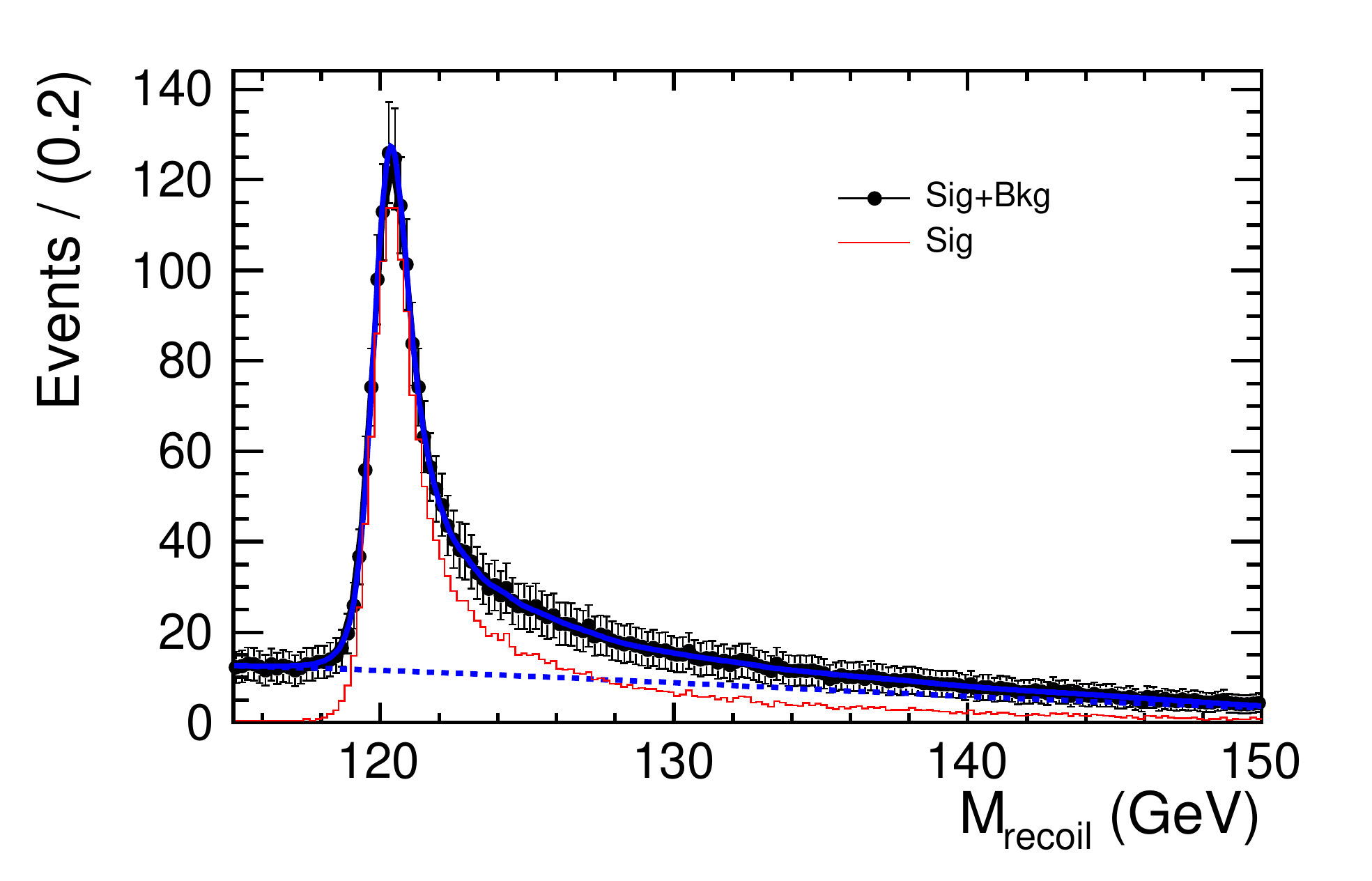}
\includegraphics[width=0.99\textwidth]{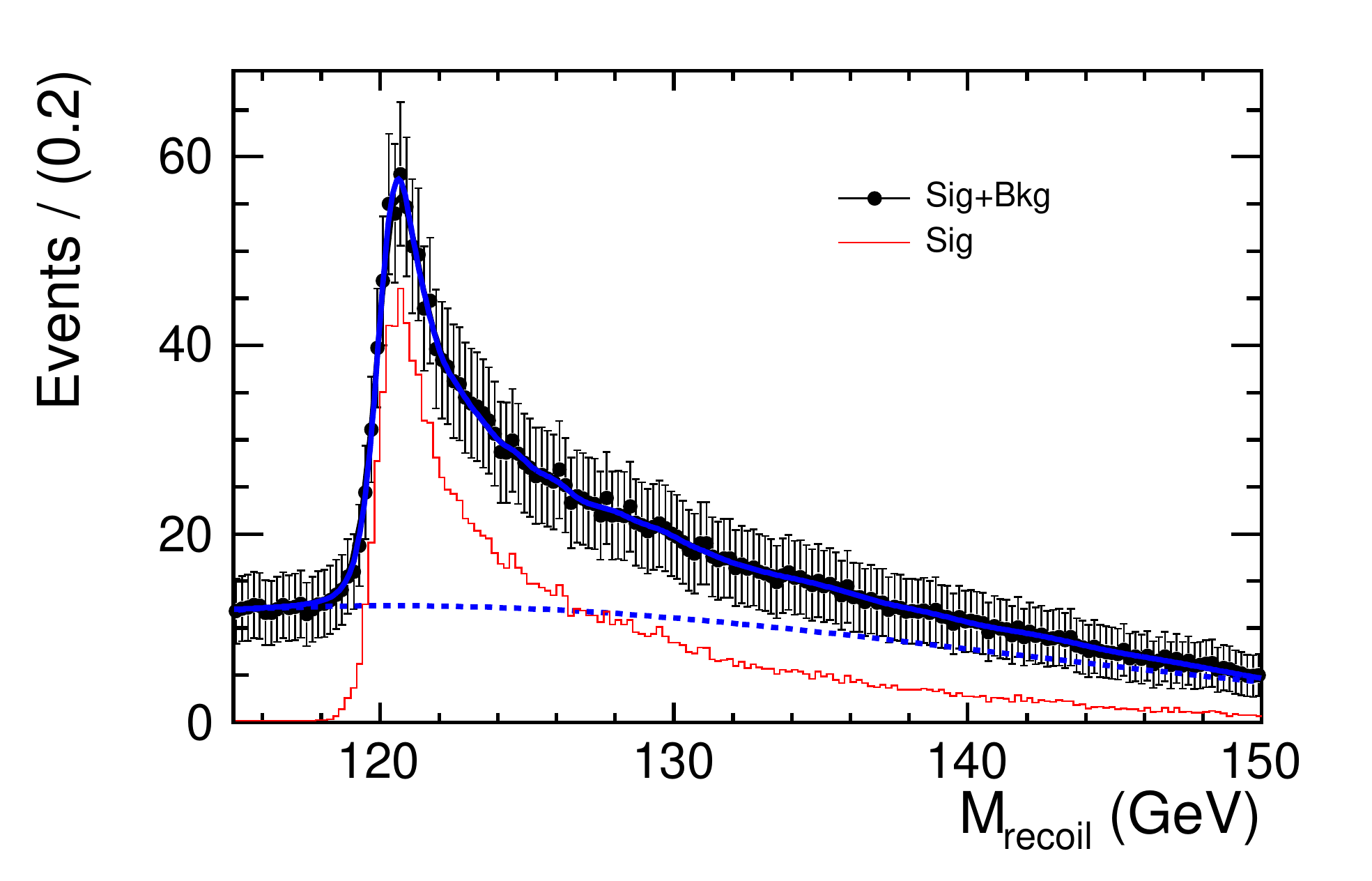}
\caption{
Reconstructed Higgs mass spectrum together with the sum of underlying background for the \emph{Model Dependent Analysis} for the $\mu\mu X$-channel (top) and $eeX$-channel (bottom).  The polarisation mode is $\eminus_L\eplus_R$.  The lines show the fits using 
the Simplified Kernel Estimation fitting formula to the signal and a polynomial of second order to the background as explained in the text. } 
\label{fig:fit_sm_lr_ke}
\end{figure}

\begin{figure}[h]
\centering
\includegraphics[width=0.99\textwidth]{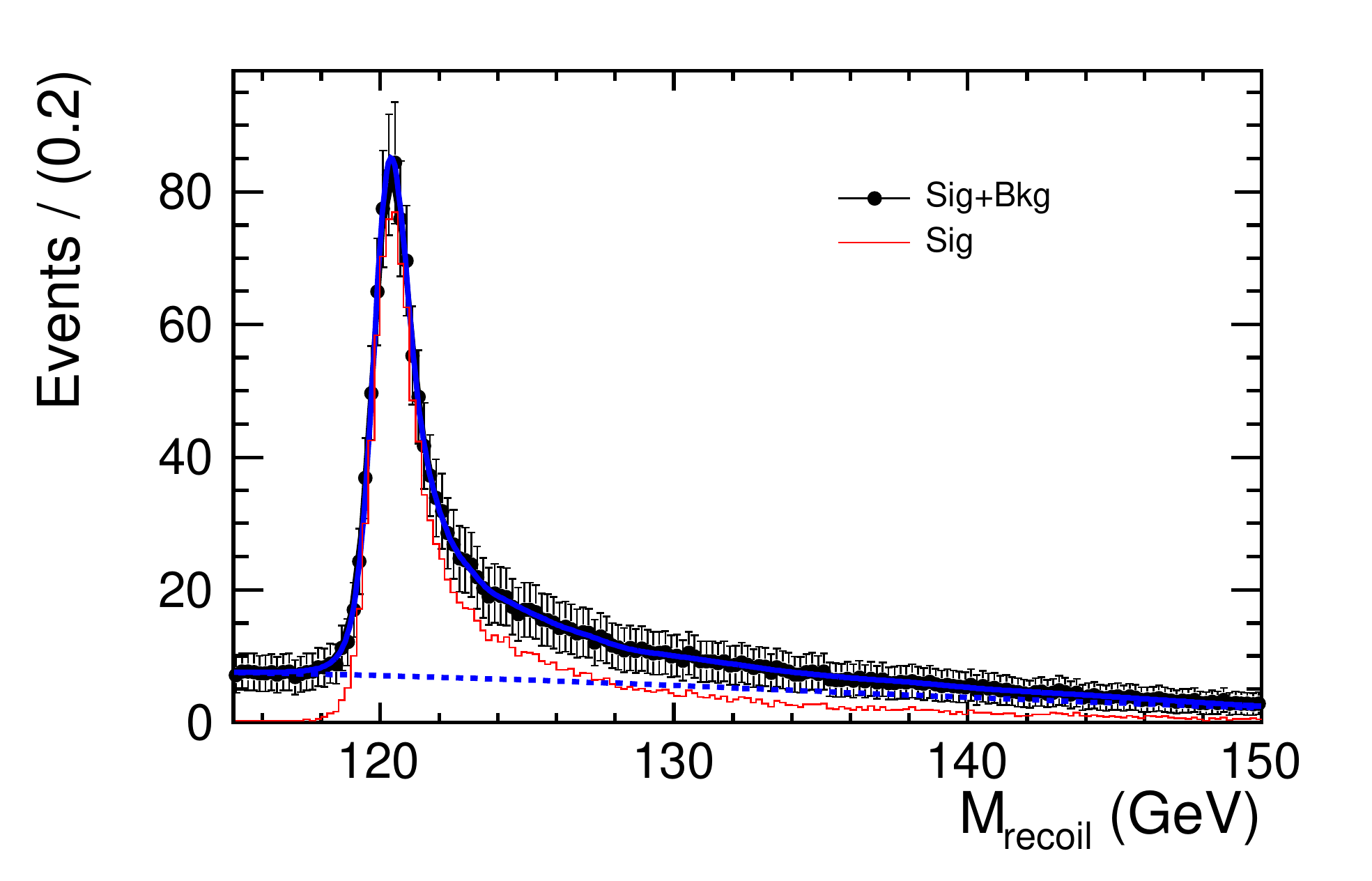}
\includegraphics[width=0.99\textwidth]{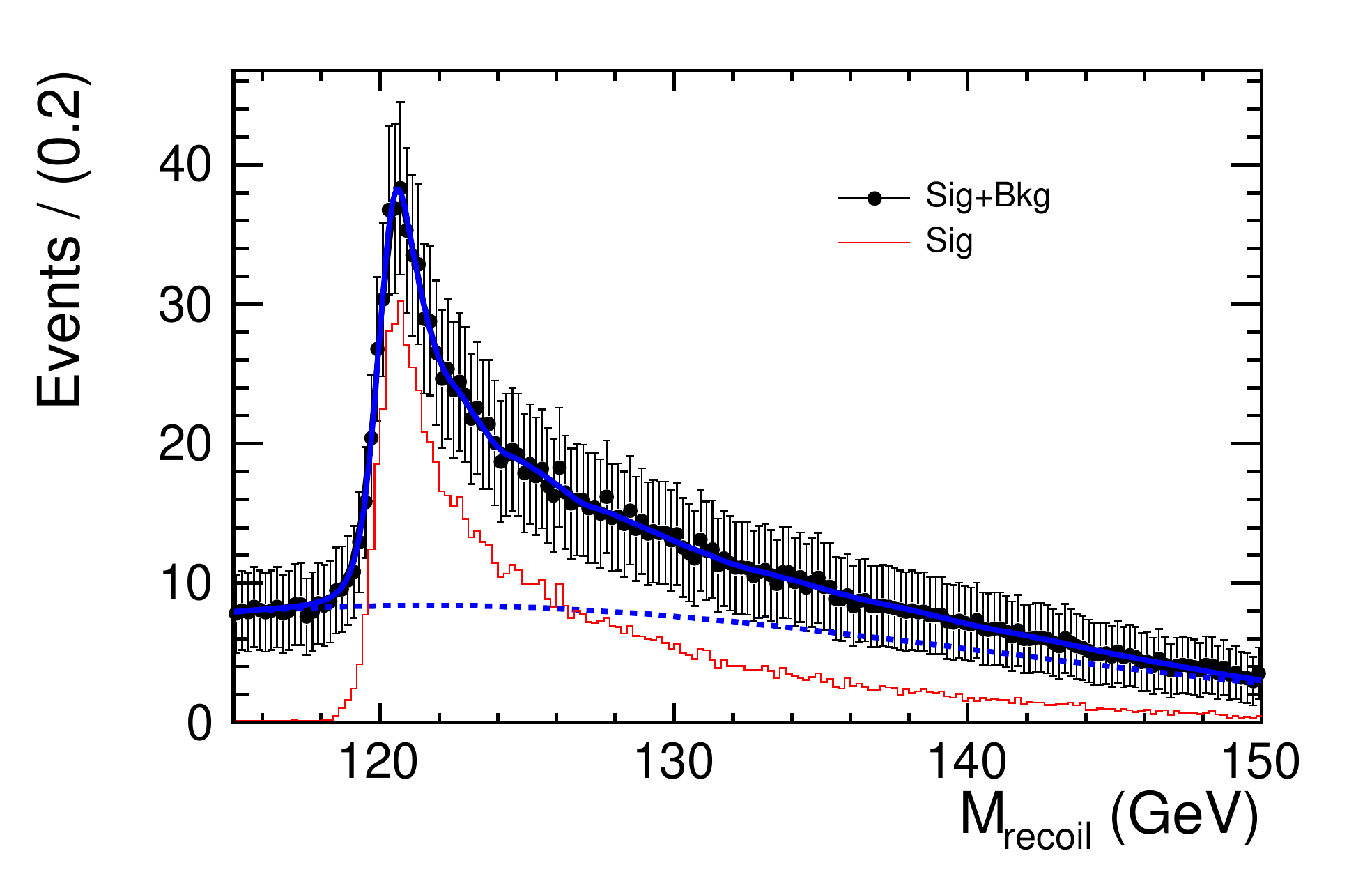}
\caption{Reconstructed Higgs mass spectrum together with the sum of underlying background for the \emph{Model Dependent Analysis} for the $\mu\mu X$-channel (top) and $eeX$-channel (bottom).  The polarisation mode is $\eminus_R\eplus_L$.  The lines show the fits using  the Simplified Kernel Estimation fitting formula to the signal and a polynomial of second order to the background as explained in the text. } 
\label{fig:fit_sm_rl_ke}
\end{figure}

The simulated signal sample is separated into two sets of data. One of them, the \emph{Reference Sample}, is employed to determine all fit parameters except the normalisation $N$ of the signal signal and the actual Higgs mass, $\mH$. The normalisation $N$ and the Higgs mass $\mH$ enter as free parameters of the fit to the second sample, the \emph{Result Sample}. The spectra of the Result Sample, scaled to a luminosity of $\mathcal{L}$=250\,$\invfb$, including the defined fitting function are displayed in Figures~\ref{fig:fit_mi_lr_ke} and~\ref{fig:fit_mi_rl_ke} for the Model Independent Analysis and in Figures~\ref{fig:fit_sm_lr_ke} to~\ref{fig:fit_sm_rl_ke} for the Model Dependent Analysis. The fit based on the Kernel estimation for the signal part  describes the shape of the mass spectra very well and are therefore suited for the extraction of the relevant parameters which are listed in Table~\ref{tab:rst_ke_mi} for the Model Independent Analysis and in Table~\ref{tab:rst_ke_sm} for the Model Dependent Analysis. In~\cite{phd_hengne} alternative fit methods are discussed which lead to nearly identical results.

% KE MI
\begin{table}[phtb]
\centering
\begin{tabular}{|c|l|l|l|}
\hline
Pol. &  Ch. &   $\mH$ (GeV) &                           $\rm{\sigma}$ (fb) \\ \hline                                            
$\polRL$ &      $\mmX$ &        120.006 $\pm$(  0.039   )&      7.89    $\pm$   0.28    (       3.55    \%)     \\ \cline{2-4}
$\mathcal{L}=250\ \rm{fb^{-1}}$ &       $\eeX$ &        120.005 $\pm$(  0.092   )&      8.46    $\pm$   0.43    (       5.08    \%)     \\ \cline{2-4}
&       merged &        120.006 $\pm$(  0.036   )&      8.06    $\pm$   0.23    (       2.91    \%)     \\ \hline
$\polLR$ &      $\mmX$ &        120.008 $\pm$(  0.037   )&      11.70   $\pm$   0.39    (       3.33    \%)     \\ \cline{2-4}
$\mathcal{L}=250\ \rm{fb^{-1}}$ &       $\eeX$ &        119.998 $\pm$(  0.085   )&      12.61   $\pm$   0.62    (       4.92    \%)     \\ \cline{2-4}
&       merged &        120.006 $\pm$(  0.034   )&      11.96   $\pm$   0.33    (       2.76    \%)     \\ \hline
%Joint &        $\mmX$ &        120.007 $\pm$(  0.038   )&                                                      \\ \cline{2-4}
%$\mathcal{L}=$ $125\ \rm{fb^{-1}}(\polLR)$ &   $\eeX$ &        120.001 $\pm$(  0.088   )&                                                      \\ \cline{2-4}
%$\ \ \ \ + 125\ \rm{fb^{-1}}(\polRL) $&        merged &        120.006 $\pm$(  0.035   )&                                                      \\ \hline
\end{tabular}
\caption{Resulting Higgs mass $\mH$ and cross section $\sigma$ of the \emph{MI Analysis} using \emph{Kernel Estimation}.}
\label{tab:rst_ke_mi}
\end{table}

%KE SM
\begin{table}[phtb]
\centering
\begin{tabular}{|c|l|l|l|}
\hline
Pol. &  Ch. &   $\mH$ (GeV) &                           $\rm{\sigma}$ (fb) \\ \hline                                            
$\polRL$ &      $\mmX$ &        120.008 $\pm$   0.037   &       7.88    $\pm$   0.26    (       3.30    \%)     \\ \cline{2-4}
$\mathcal{L}=250\ \rm{fb^{-1}}$ &       $\eeX$ &        120.001 $\pm$   0.081   &       8.46    $\pm$   0.38    (       4.49    \%)     \\ \cline{2-4}
&       merged &        120.007 $\pm$   0.034   &       8.06    $\pm$   0.21    (       2.66    \%)     \\ \hline
$\polLR$ &      $\mmX$ &        120.009 $\pm$   0.031   &       11.68   $\pm$   0.32    (       2.74    \%)     \\ \cline{2-4}
$\mathcal{L}=250\ \rm{fb^{-1}}$ &       $\eeX$ &        120.007 $\pm$   0.065   &       12.58   $\pm$   0.46    (       3.66    \%)     \\ \cline{2-4}
&       merged &        120.009 $\pm$   0.028   &       11.97   $\pm$   0.26    (       2.19    \%)     \\ \hline
%Joint &        $\mmX$ &        120.009 $\pm$   0.034   &                                                       \\ \cline{2-4}
%$\mathcal{L}=$ $125\ \rm{fb^{-1}}(\polLR)$ &   $\eeX$ &        120.005 $\pm$   0.072   &                                                       \\ \cline{2-4}
%$\ \ \ \ + 125\ \rm{fb^{-1}}(\polRL) $&        merged &        120.008 $\pm$   0.030   &                                                       \\ \hline

\end{tabular}
\caption{Resulting Higgs mass $\mH$ and cross section $\sigma$ of the \emph{MD Analysis} using \emph{Kernel Estimation}.}
\label{tab:rst_ke_sm}
\end{table}

\subsection{Discussion of the Results}
 
The Higgs mass can be determined to a precision of the order of 0.03\% when the $eeX$ channel and the $\mu\mu X$  are combined. Regarding the individual results, it can be deferred that the precision in the $\mu\mu X$ channel is more than two times smaller than that of the $eeX$ Channel
This increase of the error is mainly induced by bremsstrahlung of the electrons in the detector material. This can be concluded by comparing the results between Tables~\ref{tab:rst_ke_mi} and~\ref{tab:rst_ke_sm} as for the latter the background of Bhabha scattering is suppressed  while the difference in the precision between the two decay modes of the $\Zzero$ bosons remains the same. The precision on the cross section is less sensitive to this experimental drawback as it is derived within a basically arbitrary mass window.  The derived values for the Model Dependent Analysis are consistently slightly more precise. The relatively small difference of the results confirms that the methods employed for background suppression in the Model Independant Analysis are already very efficient.

\begin{figure}[!h]
\centering
\includegraphics[width=0.5\textwidth]{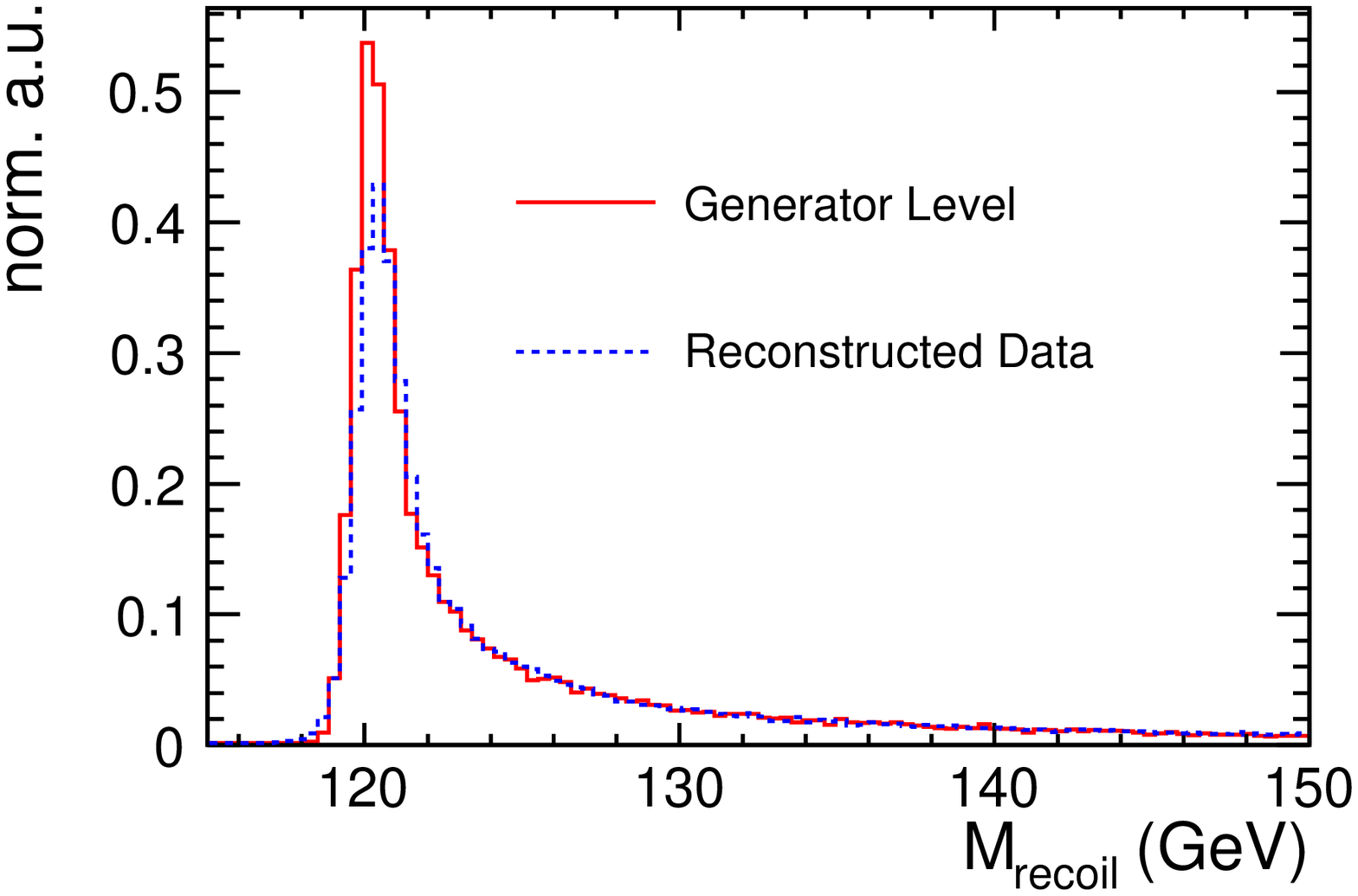}%
\includegraphics[width=0.5\textwidth]{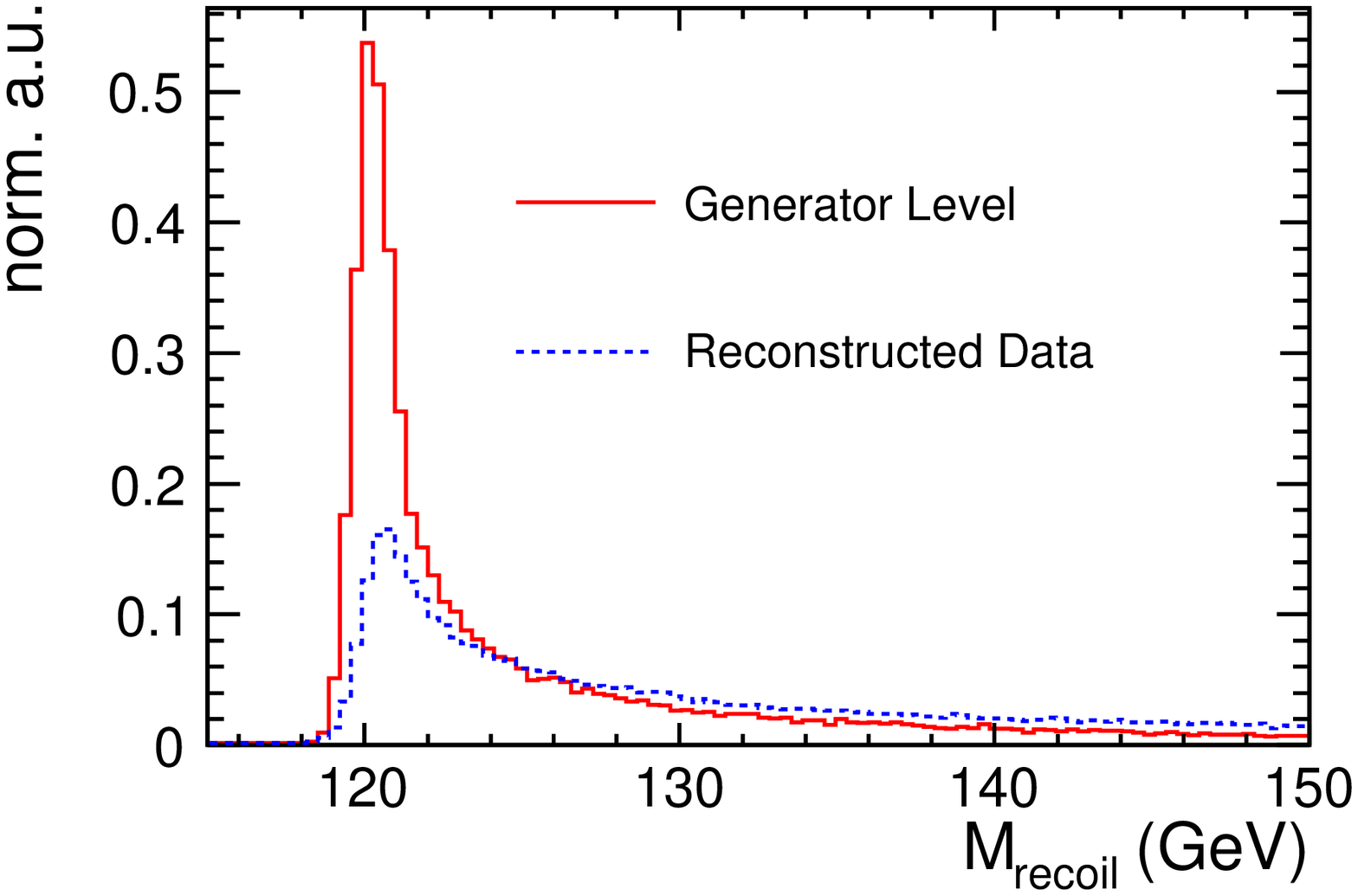}
\caption{Comparisons of recoil mass spectra in generator level and after full simulation, for the $\mu\mu X$-channel (left) and the $eeX$-channel (right).  } 
\label{fig:mh_comp}
\end{figure}

\begin{table}[!h]
\centering 
\begin{tabular}{|c|c|c|c|}
\hline
& $\Delta M_{tot.}$ (MeV) & $\Delta M_{mac.}$ (MeV) & $\Delta M_{dec.}$ (MeV) \\
\hline
$\mu\mu X$ & 650 & 560 & 330 \\
\hline
$ee X$ & 750 & 560 & 500 \\
\hline
\end{tabular}
\caption{Mass Resolution with contributions by machine ($\Delta M_{mac.}$) and detector ($\Delta M_{det.}$) separated. }
\label{tab:mass_resol}
\end{table}

The width of the Higgs boson mass as shown before is mainly given by a convolution of detector uncertainties and uncertainties on the energy of the incoming beams. Uncertainties on the energy of the incoming beams are imposed by accelerator components such as the initial linac, the damping rings or, in the case of electrons, by a tentative undulator in the electron beam line. Another source of uncertainty is the beamstrahlung when particles of a beam bunch interact in the electromagnetic field of the opposite one. The Figure~\ref{fig:mh_comp} shows the Higgs mass spectrum before and after full detector simulation. The detector response leads only to small additional widening of the maximum of the recoil mass distribution. Using a Gaussian fit to the left side of the recoil mass distribution, the width before detector simulation can be quantified to be 560\,MeV while it increases to 650\,MeV for the $\mu\mu X$ channel after detector simulation, see Table~\ref{tab:mass_resol}. For the given configuration, the uncertainty on the incoming beams remains the dominant contribution to the observed width even for the $eeX$ channel. 

\subsubsection*{On the Control of Systematic Errors}
Naturally, the measurement of the Higgs mass is sensitive to the calibration of the detector and the beam energies as the Higgs mass is directly computed from the four momenta of the particles composing the di-lepton system and the centre-of-mass energy. Both uncertainties can be controlled by the measurement of the $\epem \rightarrow\Zzero\Zzero$ process as the $\Zzero$ mass is known to a few MeV and the cross section for $\Zzero$ pair production is approximately 40 times higher than that of the Higgs-strahlungs process. Once the detector is calibrated the Higgs-strahlung process can be used to determine, within reasonable limits,  arbitrary Higgs masses.  The algorithms presented in this note are also suited for the quantification of the systematic error.  Note, that the systematic error of the cross section determination might be easier to control by using a smaller set of cut variables than those presented above. Such a set could comprise only the invariant mass and the transverse momentum of the di-lepton system, $M_{dl}$, $P_{Tdl}$ or, in case of the model dependant analysis, the number of additional tracks, $N_{add, TK}$.  The expected increase of the statistical error is only about 10\%.

\subsubsection{Recovery of Bremsstrahlungs Photons}
\label{brems}
The lower precision obtained in the $eeX$-Channel is due to the Bremsstrahlung of the
final state electrons in the passive material of the detector. In the following an attempt is made
to improve the precision in that channel by recovering the bremsstrahlungs photons~\cite{brems}. The four momenta of the selected electrons are combined with those of photons which have a small angular distance to the electrons. If these combined objects together with the corresponding other electron candidate form the
$\Zzero$ mass, they are included in the $\Zzero$ reconstruction. The inclusion of low energetic photons leads to a penalty in the momentum reconstruction due to the poor energy resolution of the electromagnetic calorimeter for low energetic particles. This drawback might get counterbalanced by the gain in statistics due to the described recovery of the energy loss.
  
\begin{figure}[hbt]
\centering
\includegraphics[width=0.9\textwidth]{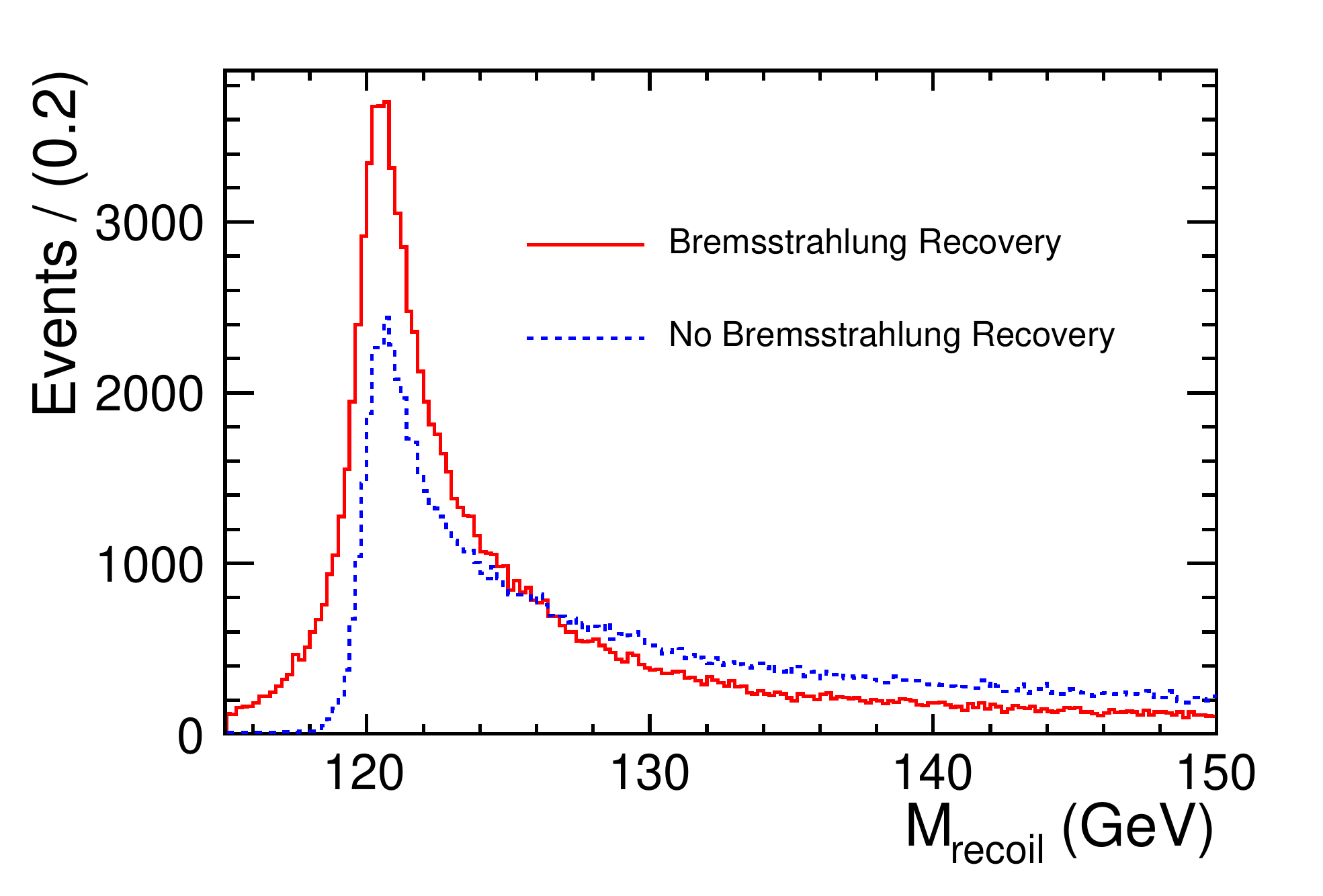}
\caption{Comparison of the Higgs recoil mass distributions of $\eeX$ channel with and without the bremsstrahlung photons recovery.} 
\label{fig:mh_br_nobr}
\end{figure}

\begin{figure}[!h]
\centering
\includegraphics[width=0.99\textwidth]{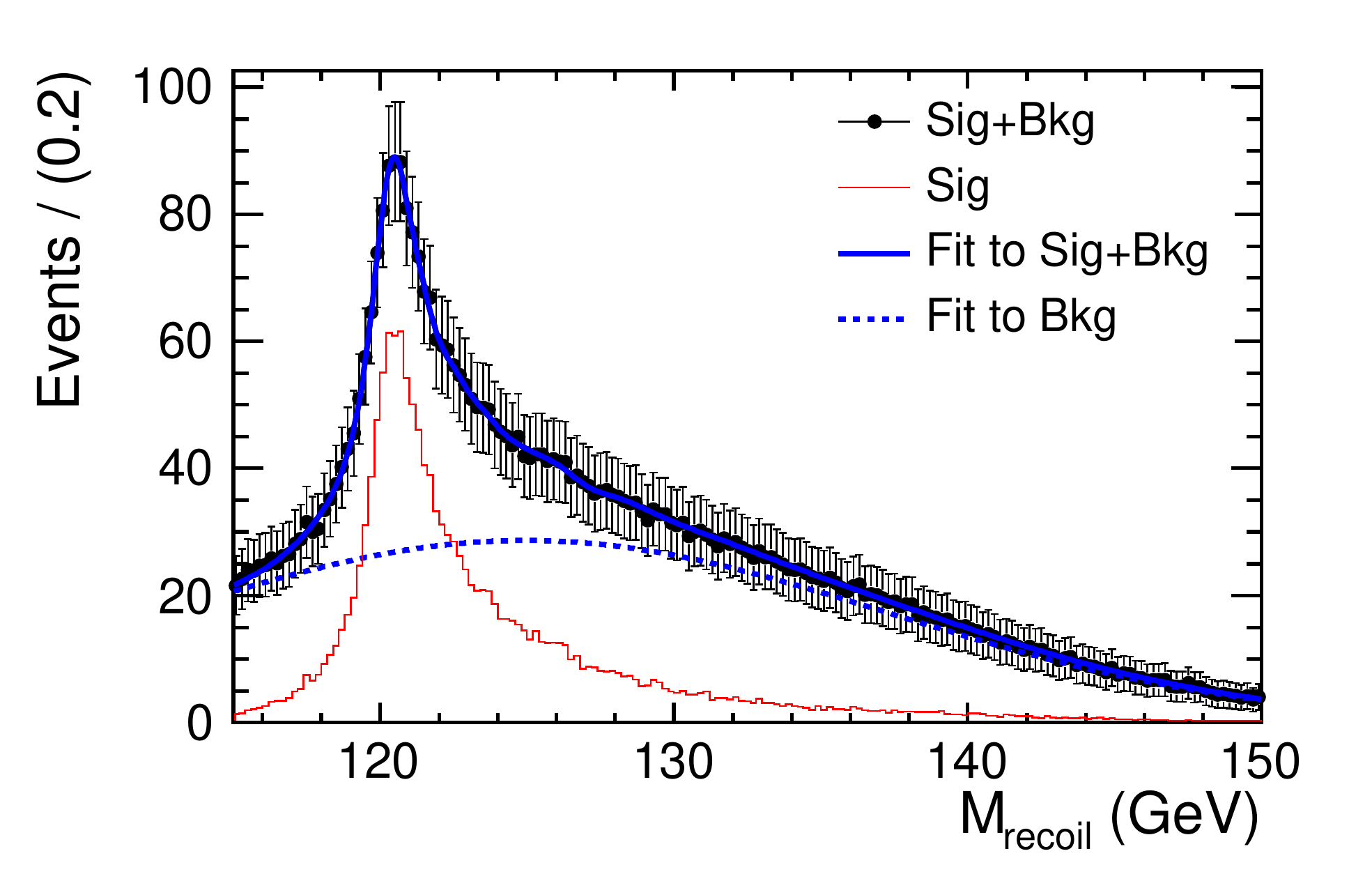}
\includegraphics[width=0.99\textwidth]{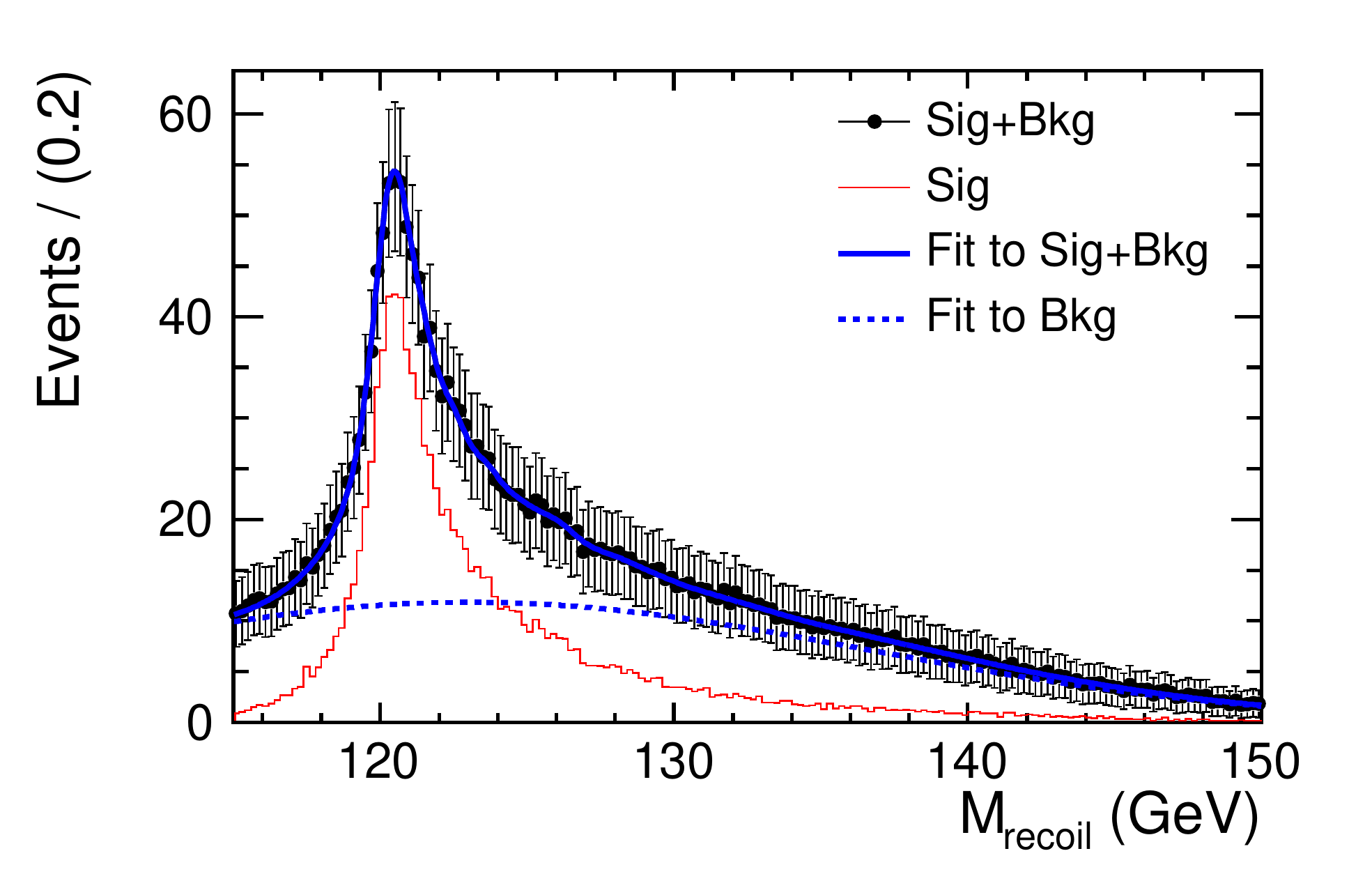}
\caption{Reconstructed Higgs mass spectrum \emph{after the recovery of Bremsstrahlungs photons} together with the sum of underlying background for the \emph{Model Independent Analysis} for $eeX$-channel.  The polarisation mode is $\eminus_L\eplus_R$ (top) and $\eminus_R\eplus_L$ (bottom).  The lines show the fits using the Simplified Kernel Estimation fitting formula to the signal and a polynomial of second order to the background as explained in the text.} 
\label{fig:fit_el_mi_zf_ke}
\end{figure}

\begin{figure}[!h]
\centering
\includegraphics[width=0.99\textwidth]{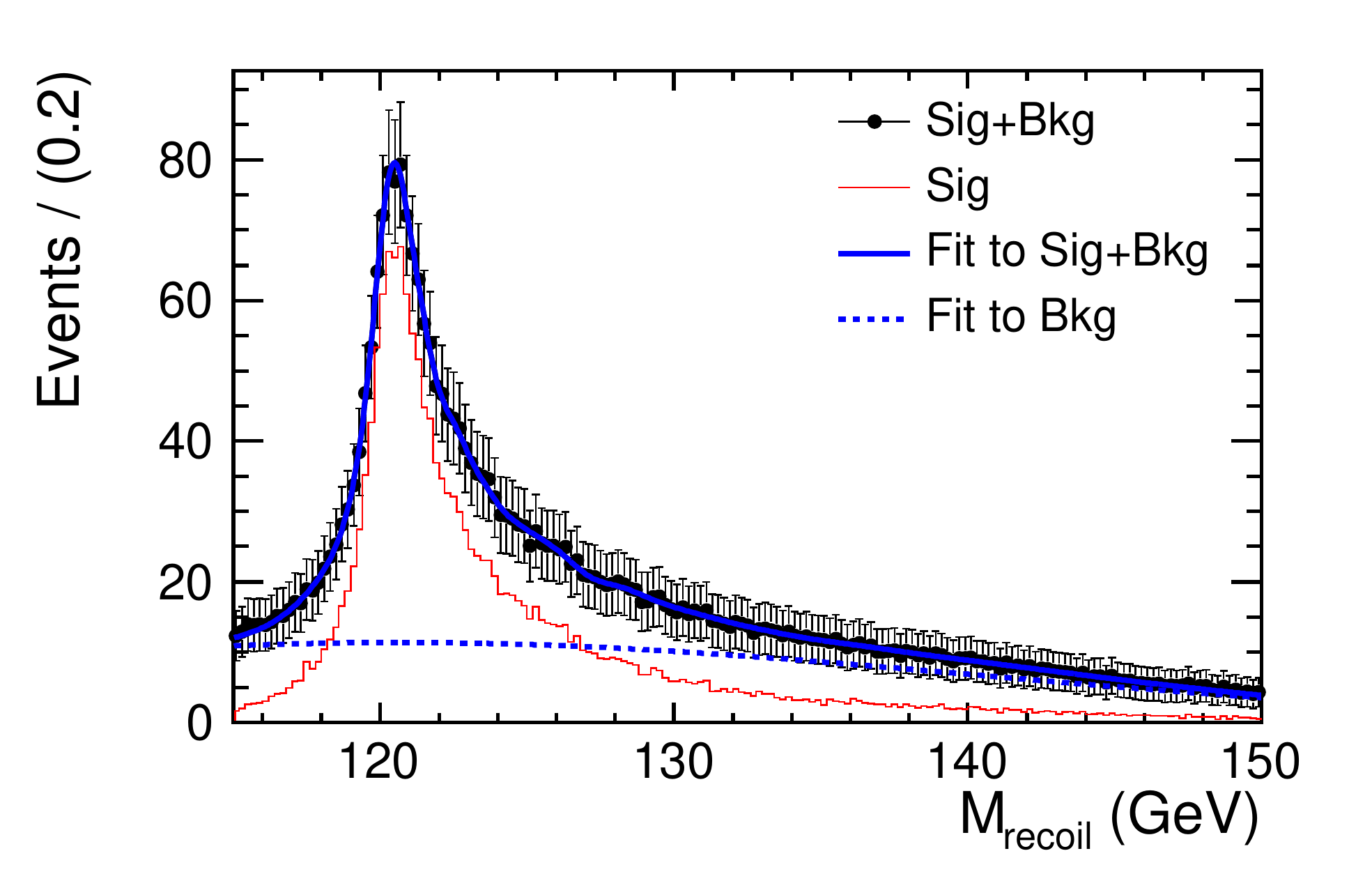}
\includegraphics[width=0.99\textwidth]{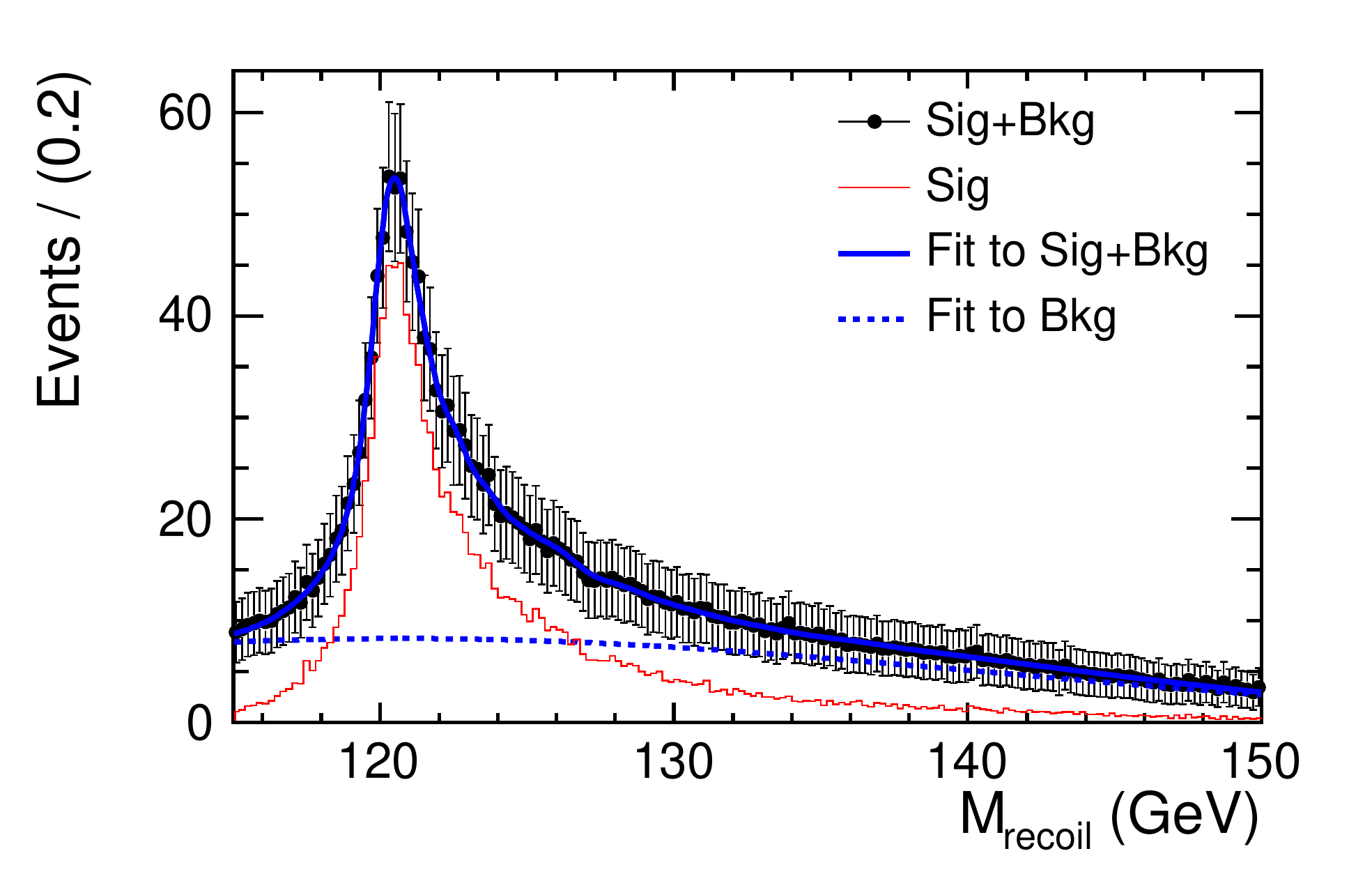}
\caption{Reconstructed Higgs mass spectrum \emph{after the recovery of Bremsstrahlungs photons} together with the sum of underlying background for the \emph{Model Dependent Analysis} for $eeX$-channel.  The polarisation mode is $\eminus_L\eplus_R$ (top) and $\eminus_R\eplus_L$ (bottom).  The lines show the fits using the Simplified Kernel Estimation fitting formula to the signal and a polynomial of second order to the background as explained in the text.} 
\label{fig:fit_el_sm_zf_ke}
\end{figure}

Figure~\ref{fig:mh_br_nobr} shows the recoil mass spectrum after the recovery of the Bremsstrahlungs photons. The worse resolution around the mass maximum is clearly visible. The corresponding results are given Tables~\ref{tab:sel_el_br} through~\ref{tab:zfinder_rst} and the fitted spectra in Figures~\ref{fig:fit_el_mi_zf_ke} through~\ref{fig:fit_el_sm_zf_ke}. The numbers show that the mass resolution is improved by 10\% while the precision in the cross section is improved by 20\%. The cross section benefits directly from the gain in statistics while the determination of the recoil mass suffers from the reduced momentum resolution. 

\begin{table}[htb]
\centering
\begin{tabular}{|l|c|l|l|l|}

\hline
Ana. &  Pol.                            & Ch.                   &       S (\%)                  &       B               \\ \hline                                     
                
MI      &       $\polRL$        &       $\eeX$  &       1029 (48.84\%)          &       1408            \\ \cline{2-5}
        &       $\polLR$        &       $\eeX$  &       1491 (41.51\%)          &       3394            \\ \hline
MD      &       $\polRL$        &       $\eeX$  &       1152 (54.66\%)          &       1114            \\ \cline{2-5}
        &       $\polLR$        &       $\eeX$  &       1724 (54.94\%)          &       1513            \\  
\hline
\end{tabular}
\caption{Resulting Number of Signal (S) and Number of Background (B), and the efficiencies of signal selection (in the parentheses) after background rejection, for $\
eeX$ channel with \emph{Bremsstrahlung Photons Recovery}}
\label{tab:sel_el_br}
\end{table}

\begin{table}[htb]
\centering
\begin{tabular}{|c|l|l|l|}
\hline
Ana. &  Pol. &  $\mH$ (GeV) &                           $\rm{\sigma}$ (fb) \\ \hline                                            
MI &    $\polRL$ &      120.003 $\pm$   0.081   &       8.41    $\pm$   0.36    (       4.28    \%)     \\ \cline{2-4}
&       $\polLR$ &      119.997 $\pm$   0.073   &       12.52   $\pm$   0.49    (       3.91    \%)     \\ \hline
MD&     $\polRL$ &      119.999 $\pm$   0.074   &       8.41    $\pm$   0.31    (       3.69    \%)     \\ \cline{2-4}
&       $\polLR$ &      120.001 $\pm$   0.060   &       12.51   $\pm$   0.38    (       3.04    \%)     \\ \hline
\end{tabular}
\caption{Resulting Higgs mass $\mH$ and cross section $\sigma$ for the \emph{Model Independent Analysis} and \emph{Model Dependent Analysis} of the Higgs-Strahlung pr
ocess of $\eeX$ channel with \emph{Bremsstrahlung photons Recovery}.}
\label{tab:zfinder_rst}
\end{table}

\section{Conclusion and Outlook}
 
Using mainly the Higgs-strahlung process with the $\Zzero$ boson decaying leptonically and a Higgs boson mass of 120\,GeV as input, the current design of the ILD detector promises to determine the mass of the Higgs boson to a precision of the order of 30\,MeV. According to~\cite{lal230} and references therein, such a precision renders sensitivity to effects from super-symmetric extensions to the Standard Model. Assuming a heavier Higgs, the precision might allow also for the determination of the Higgs boson mass width at centre-of-mass energies higher than 250\,GeV. Staying with small Higgs masses, it has been demonstrated semi-analytically~\cite{lal230} and confirmed with full simulation studies~\cite{hengne} that the precision can be further increased by working at a centre-of-mass energy close to the $\Higgs\Zzero$ production threshold, i.e. at $\roots$=230\,GeV. In the present study, the cross section and therefore the coupling strength at the $\Higgs\Zzero\Zzero$ vertex is determined to a precision of the order of 2-3\% which might already be sufficient to get sensitive to contributions to this coupling from physics beyond the Standard Model. 

The signal to background ratio can be enhanced to a value of at least 30\% even if the cross sections of the background processes are several orders of magnitudes higher. Note, that this ratio is way higher in the region around the signal maximum. The background suppression exploits the considerable capabilities of track recognition as allowed by the current design of the ILD detector. The precision of the measurement can be improved by a better muon recognition by e.g. including a muon system in the analysis which has not been done so far. The precision obtained in the branch in which the $\Zzero$ boson decays into electrons might gain considerably from a revision of the amount of passive material in the detector. Both decay modes may gain also from an exploitation of the
particle identification of the ILD detector by means of a $dE/dx$ measurement in the TPC. For this, further studies are needed. A future study clearly will have to quantify the systematic uncertainties and to identify those which have the
largest influence on the systematic errors. This would give important directions on the final detector layout and
the precision needed for e.g. alignment systems. For such a study realistic inputs on e.g. the uncertainty of drift times in the TPC or residual misalignments after detector movements are needed.

The analysis has proven that the results are sensitive to details of the accelerator configuration. Using the set of parameters  as has been chosen for the SLAC samples which in turn correspond to the current best knowledge of the beam parameters, approximately half of the statistical error is generated by uncertainties caused by beamstrahlung and the energy spread of the incoming beams. The Higgs-strahlung process constitutes an important benchmark for the optimisation of the accelerator performance.

\newpage

\newpage

\newpage

\newpage

\newpage

\newpage

\clearpage 
\newpage
\begin{footnotesize}
% IF YOU DO NOT USE BIBTEX, USE THE FOLLOWING SAMPLE SCHEME FOR THE REFERENCES
% ----------------------------------------------------------------------------

% ----------------------------------------------------------------------------

% IF YOU USE BIBTEX,
% - DELETE THE TEXT BETWEEN THE TWO ABOVE DASHED LINES
% - UNCOMMENT THE NEXT TWO LINES AND REPLACE 'Name_Of_Your_BibFile'

%\bibliographystyle{unsrt}
%\bibliography{Name_Of_Your_BibFile}
% example of Name_Of_Your_BibFile.bib
% @Article{Turcato:2006ch,
%      author    = "Turcato, M.",
%  collaboration = "ZEUS and H1",
%      title     = "Lepton flavour violation and charmonium physics at HERA",
%      journal   = "Nucl. Phys. Proc. Suppl.",
%      volume    = "162",
%      year      = "2006", 
%      pages     = "283-287",
%      SLACcitation  = "%%CITATION = NUPHZ,162,283;%%"
% }
% 
% @Unpublished{Gogitidze:2007du,
%      author    = "Gogitidze, N.",
%  collaboration = "H1", 
%      title     = "Prompt photons and particle momentum distributions at
%                   HERA", 
%      year      = "2007",
%      note    = "hep-ex/0701033",
%      SLACcitation  = "%%CITATION = HEP-EX 0701033;%%"
% }

\end{footnotesize}

% ****************************************************************************
% END OF BIBLIOGRAPHY AREA
% ****************************************************************************

\end{document}